\documentclass[12pt,a4paper,openright,twoside]{report}
\usepackage[english]{babel}
\usepackage{pdfpages}
\usepackage[latin1]{inputenc}
\usepackage{fancyhdr}
\usepackage{amsthm,bm,mathrsfs,stmaryrd}

\usepackage{indentfirst}
\usepackage{graphicx}
\usepackage{newlfont}
\usepackage{amssymb}
\usepackage{amsmath}
\usepackage{latexsym}
\usepackage{amsthm}
\renewcommand{\chaptermark}[1]{\markboth{\thechapter.\ #1}{}}

\usepackage{latexsym}
\usepackage{verbatim}
\usepackage{mathrsfs}
\usepackage{amssymb,amsmath}
\usepackage{amsthm}
\usepackage{amscd}
\usepackage{url}
\usepackage{slashed}
\usepackage[latin1]{inputenc}
\usepackage{amsfonts}
\usepackage{bbm}
\usepackage{mathrsfs}

\DeclareMathAlphabet{\mathpzc}{OT1}{pzc}{m}{it}

\renewcommand{\theequation}{\thesection.\arabic{equation}} \csname
@addtoreset\endcsname{equation}{section}

\newcommand{\mt}[1]{$\mathop{#1}$}

\newcommand{\ra}{\rangle}

\newcommand{\st}{\scriptstyle}
\newcommand{\sst}{\scriptscriptstyle}

\newcommand{\nn}{\nonumber\\}

\newcommand{\edf}{&:=&}

\def\gz0{\gamma^{0}}

\def\eq{&=&}

\def\a{\alpha}
\def\b{\beta}
\def\g{\gamma}

\def\d{\delta}
\def\D{\Delta}
\def\e{\epsilon}

\def\l{\lambda}
\def\L{\Lambda}
\def\m{\mu}
\def\n{\nu}

\def\s{\sigma}

\def\t{\tau}

\def\o{\omega}
\def\O{\Omega}

\def\cA{{\cal A}}
\def\cB{{\cal B}}
\def\cC{{\cal C}}
\def\cD{{\cal D}}
\def\cE{{\cal E}}

\def\cG{{\cal G}}
\def\cH{{\cal H}}

\def\cK{{\cal K}}
\def\cL{{\cal L}}

\def\cO{{\cal O}}
\def\cP{{\cal P}}

\def\cR{{\cal R}}

\def\cV{{\cal V}}

\def\cZ{{\cal Z}}

\def\be{\begin{equation}}
\def\ee{\end{equation}}
\def\ba{\begin{eqnarray}}
\def\ea{\end{eqnarray}}
\def\bec{\begin{center}}
\def\ec{\end{center}}

\def\12{\frac{1}{2}}

\def\eq{&=&}

\def\ra{\rightarrow}

\begin{document}

\includepdfmerge[pages={2},pagecommand={\thispagestyle{empty}}]{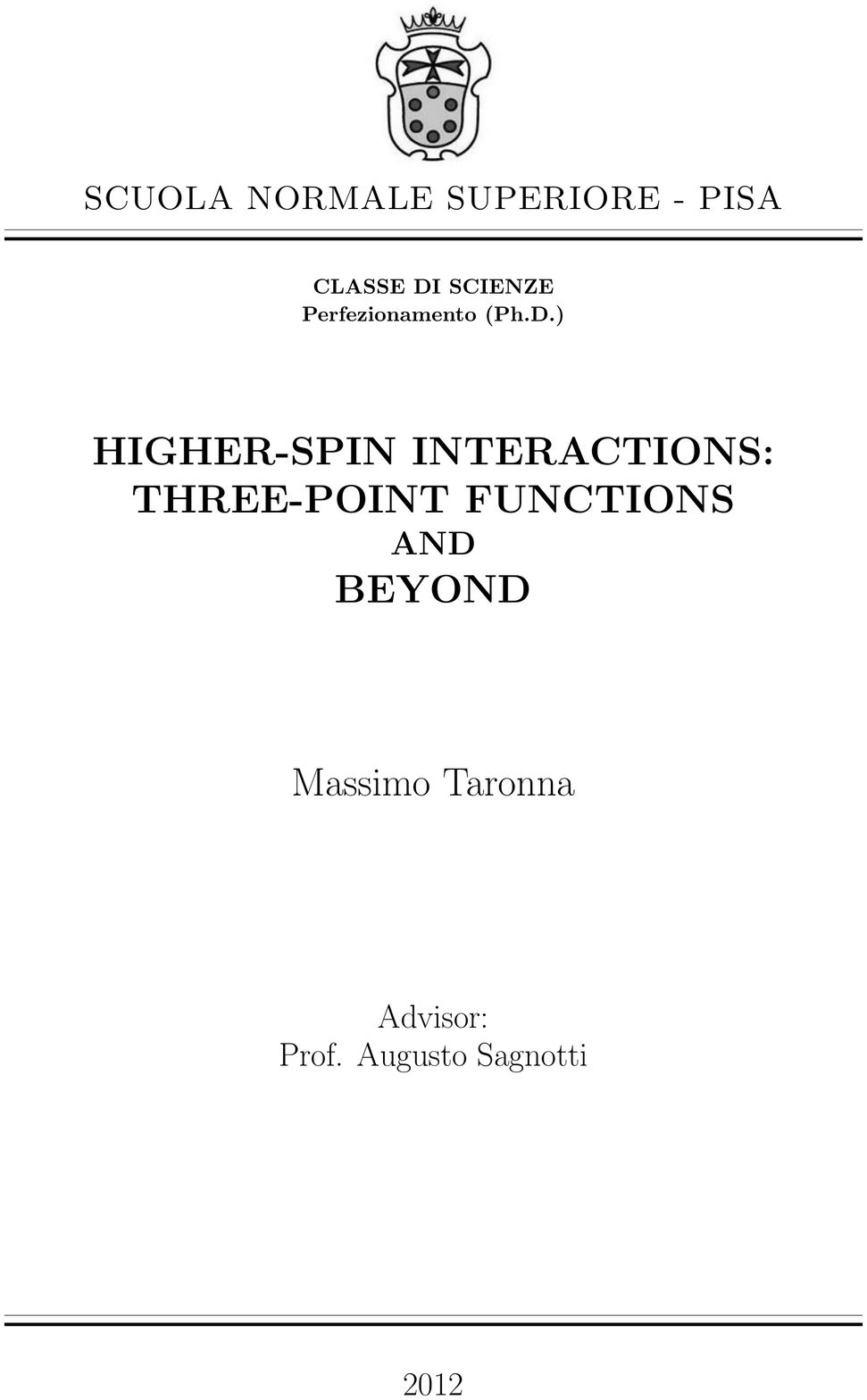}
\newpage                                
\clearpage{\pagestyle{empty}\cleardoublepage}
\begin{titlepage}                       
%
\thispagestyle{empty}                   
\topmargin=6.5cm                        
\raggedleft                             
\large                                
\em                                     
Alla mia famiglia e ad Aurora\ldots
\newpage                                
\clearpage{\pagestyle{empty}\cleardoublepage}
\thispagestyle{empty}                   
\topmargin=6.5cm                        
\raggedleft                             
\em                                     
\em \small{``My work always tried to unite the truth with the beautiful, but when I had to choose one or the other, I usually chose the beautiful\ldots''\\
H. Weyl}
\newpage                                
%

\clearpage{\pagestyle{empty}\cleardoublepage}
\end{titlepage}
\pagenumbering{roman}                   

\clearpage{\pagestyle{empty}\cleardoublepage}
\chapter*{Acknowledgments}
\thispagestyle{empty}

This Thesis summarizes most of the research that I have done during my PhD course at Scuola Normale Superiore. First of all my gratitude goes to this small but beautiful Institution that gave me a huge support from the very beginning of my studies, when I arrived in Pisa as an undergraduate student. These years here have made my passion for Physics grow stronger and stronger, continuously making me face beautiful problems to solve, or just to think about, in several branches of Science. I am especially grateful to my advisor, Augusto Sagnotti, for his constant support during this time (first when I was an undergraduate student and later when I was a graduate student) and for having introduced me to this fascinating research field, encouraging me from the very beginning to think about any research problem in my own way. I really owe him a lot of what I am now as a scientist, and he will always be of great inspiration for me. I am also grateful to my collaborators and to other colleagues for many stimulating discussions. They created a fantastic environment around me to make me grow.
I should also thank my parents Gabriella and Salvatore, my brother Luca, Aurora and my friends for their support, love and friendship, and especially my parents, because it is thanks to them that I have discovered my passion for Physics.

\vspace{5pt}

This Thesis is dedicated to them.

\clearpage{\pagestyle{empty}\cleardoublepage}
\tableofcontents                        
\rhead[\fancyplain{}{\bfseries\leftmark}]{\fancyplain{}{\bfseries\thepage}}
\lhead[\fancyplain{}{\bfseries\thepage}]{\fancyplain{}{\bfseries
Index}}
\clearpage{\pagestyle{empty}\cleardoublepage}
\listoffigures                          
\clearpage{\pagestyle{empty}\cleardoublepage}

\pagenumbering{arabic}                  


\chapter*{Introduction}                 
\rhead[\fancyplain{}{\bfseries
INTRODUCTION}]{\fancyplain{}{\bfseries\thepage}}
\lhead[\fancyplain{}{\bfseries\thepage}]{\fancyplain{}{\bfseries
INTRODUCTION}}
\addcontentsline{toc}{chapter}{Introduction}

\renewcommand{\theequation}{\arabic{equation}}

\vspace{30pt}
After decades of intense efforts, Higher-Spin (HS) theories still present several unanswered questions as well as many open problems that along the years have highlighted the many difficulties of the subject and, most importantly, the limitations of the well known frameworks and techniques that have been successfully applied to their lower-spin counterparts. For these reasons HS theories have been over the last decades an intense field of research that has attracted an increasing attention, starting from the works of the late 80's by Fradkin and Vasiliev (FV) \cite{Fradkin:1986qy} and Vasiliev \cite{Vasiliev:1988sa,Vasiliev:1989yr,Vasiliev:1990en,Vasiliev:1990vu,Vasiliev:1990bu,Vasiliev:1992av,Vasiliev:1995dn,Vasiliev:2004qz,Vasiliev:1999ba,Vasiliev:2005zu} that opened the way to the first classically consistent examples of non-abelian interactions of this type.

However, we are still far from a satisfactory understanding of the problem, so much so that only in the last few years a reasonable understanding of the \emph{free} HS theory has been attained\footnote{For some recent reviews of HS gauge theories, see e.g. the proceeding \cite{SolvayHS} (which includes contributions \cite{Bianchi:2005yh, Francia:2005bv, Bouatta:2004kk, Bekaert:2005vh, Sagnotti:2005ns}) and \cite{Sorokin:2004ie, Francia:2006hp,Fotopoulos:2008ka,Campoleoni:2009je,Bekaert:2010hw,Campoleoni:2011tn,Sagnotti:2011qp}.
}. Two distinct approaches have come to terms with a problem that, in some respects, dates back to the early days of Quantum Field Theory (QFT) (see, e.g.~\cite{Majorana:1932rj,Dirac:1936tg,Fierz:1939ix,Rarita:1941mf,Wigner:1939cj,Bargmann:1948ck}). The first is a ``metric-like'' approach, initiated in the works of Hagen and Singh \cite{Hagen:1982ez}, Fronsdal \cite{Fronsdal:1978rb} and de Wit and Freedman \cite{Wit:1979pe}, and reconsidered more recently by Francia and Sagnotti \cite{Francia:2002aa,Francia:2002pt,Francia:2007ee}. In their works the authors of \cite{Francia:2002aa} proposed a \emph{geometric} reinterpretation of the free-field equations that can be expressed as
\be
\frac{1}{\square^{\,n}}\ \partial\cdot{\cR^{\,[n]}}_{;\m_1\ldots\m_{2n+1}}\,=\,0\,,
\ee
for odd spins $s\,=\,2n\,+\,1$, and
\be
\frac{1}{\square^{\,n-1}}\ {\cR^{\,[n]}}_{;\m_1\ldots\m_{2n}}\,=\,0\,,
\ee
for even spins $s\,=\,2n$, together with the related \emph{minimal} Lagrangian formulation \cite{Francia:2005bu,Francia:2007qt}, that rests for any $s$ on at most two additional fields and simplifies the previous BRST (Becchi, Rouet, Stora, Tyutin) constructions of \cite{Pashnev:1997rm,Pashnev:1998ti,Burdik:2000kj,Buchbinder:2001bs,Buchbinder:2002ry,Bekaert:2003uc,Buchbinder:2004gp}. This form of the free equations includes the three familiar lower-spin examples, given by the linearized Einstein equations $(s\,=\,2)$, by the Maxwell equations $(s\,=\,1)$ and formally also by the Klein-Gordon equation $(s\,=\,0)$, together with \emph{non-local} equations for spin larger than two. One can thus have an intuition, even if restricted to a single spin at a time, of possible generalizations of the geometric framework of Maxwell theory and Einstein gravity to HS, pointing also to a possible key role of non-localities, that may be more and more fundamental at the interacting level, together with an eventual reconsideration of QFT from a more general perspective. More recently, these results were also generalized to reducible HS \emph{free} fields, starting from the ``triplet'' system \cite{Heanault:book,Bonelli:2003zu,Sagnotti:2003qa} and recovering similar interesting \emph{non-local} structures \cite{Francia:2010ap,Francia:2010qp}. The constructions for symmetric (spinor)-tensors that we have just outlined afford also interesting generalizations to the case of mixed-symmetry fields of the type $\phi_{\,\m_1\ldots\m_{s_1};\n_1\ldots\n_{s_2}\ldots}$, whose non-local geometric equations were first proposed in \cite{Bekaert:2002dt,Bekaert:2003az,deMedeiros:2003dc}, while the Lagrangian formulation was initiated with the pioneering works of Curtright and Labastida \cite{Curtright:1979uz,Curtright:1980yk,Aulakh:1986cb,Ouvry:1986dv,Labastida:1986gy,Siegel:1986zi,Siegel:1986de,Labastida:1986ft,Labastida:1987kw} and was completed only recently \cite{Campoleoni:2008jq,Campoleoni:2009gs,Campoleoni:2012th}. A second kind of approach, the ``frame-like'' one, was developed mostly by Vasiliev and collaborators \cite{Vasiliev:1988sa,Vasiliev:1989yr,Vasiliev:1990en,Vasiliev:1990vu,Vasiliev:1990bu,Vasiliev:1992av,Vasiliev:1995dn,Vasiliev:2004qz,Vasiliev:1999ba,Vasiliev:2005zu} generalizing the Cartan-Weyl framework to HS, and led to the Vasiliev system. Despite the remarkable success of Vasiliev's approach, only recently has it been possible to arrive at a covariant description of all bosonic flat space cubic interactions in \cite{Manvelyan:2009vy,Manvelyan:2010wp,Manvelyan:2010jr} by purely field theoretical methods. At the same time, starting from a String Theory vantage point, all consistent cubic interactions involving bosonic and fermionic fields were obtained in \cite{Taronna:2010qq,Sagnotti:2010at}\footnote{Further off-shell completions were presented in \cite{Fotopoulos:2010ay,Manvelyan:2010je,Manvelyan:2010jf,Mkrtchyan:2010pp}.}. These results were then extended, for what concerns the part of the vertices that is not proportional to traces and divergences of the fields, to constant-curvature backgrounds in \cite{Joung:2011ww,Joung:2012rv}\footnote{See also \cite{Joung:2012fv} for a review of the general ideas and results.} while the formalism was pushed forward to higher orders in the number of fields in \cite{Taronna:2011kt}\footnote{See also the appendix of \cite{Sagnotti:2010at} and \cite{Dempster:2012vw} for a related analysis of the quartic interactions and for further discussions.} identifying a class of higher-order flat vertices. This extended previous results, including the works of the 80's by Bengtsson, Bengtsson and Brink \cite{Bengtsson:1983pd,Bengtsson:1983pg} and the important works of Metsaev \cite{Metsaev:1993ap,Metsaev:1993mj,Metsaev:1995re,Metsaev:1997nj,Metsaev:1999gw,Metsaev:2003cu,Metsaev:2005ar,Metsaev:2006ui,Metsaev:2007rn}, in the light-cone formulation, and the works of Berends, Burgers and van Dam \cite{Berends:1979kg,Berends:1984rq,Berends:1985xx} in a covariant formulation that were then reconsidered and extended by Boulanger and others in\footnote{See also \cite{Alkalaev:2010af,Boulanger:2011qt,Boulanger:2011se,Zinoviev:2011fv,Metsaev:2012uy,Henneaux:2012wg} for more recent results related to mixed-symmetry fields and fermions.} \cite{Boulanger:2005br,Buchbinder:2006eq,Zinoviev:2006im,Boulanger:2006gr,Fotopoulos:2007nm,Fotopoulos:2007yq,Zinoviev:2008ck,Boulanger:2008tg,Zinoviev:2008jz,Zinoviev:2008ze,Zinoviev:2009hu,Bekaert:2010hp,Zinoviev:2010cr,Bekaert:2010hk} and were recently exploited in the interesting work of Bekaert, Joung and Mourad \cite{Bekaert:2009ud}.

This Thesis is aimed at reviewing our current understanding of HS interactions along the lines of the original contributions of the author \cite{Taronna:2010qq,Sagnotti:2010at,Taronna:2011kt,Joung:2011ww,Joung:2012rv,Joung:2012fv} and, more in detail, within the variant of the ambient-space formalism developed originally by the author in\footnote{See also \cite{Bars:2010xi} and references therein for similar ideas in the framework of two-time physics.} \cite{Joung:2011ww,Joung:2012rv}. Moreover, we shall push forward the idea that string results and their structure may give new insights on field theory properties that manifest themselves when looking at HS fields. This motivates a closer relation between ST and HS gauge theories that resonates with the long-held feeling that ST draws its origin from a \emph{generalized} Higgs effect responsible for its massive excitations (see e.g. \cite{Gross:1987kza,Gross:1987ar,Gross:1988ue,Amati:1987wq,Amati:1987uf,Amati:1988tn,Bianchi:2003wx,Beisert:2004di}). The crux of the matter has long been to construct a consistent deformation of the free system at the quartic order. It is indeed this the case in which the HS program has encountered along the years a barrier, both in the metric-like formulation and in the frame-like one, in which Vasiliev's system unfortunately does not provide a transparent answer\footnote{See \cite{Sezgin:2002ru,Sezgin:2003pt,Boulanger:2008tg} for a discussion of the the general strategy in order to extract the couplings starting from the Vasiliev system together with some explicit results for the scalar couplings. See also \cite{Boulanger:2011dd} for a proposal of an action principle for Vasiliev's system.}. Indeed, only recently in \cite{Boulanger:2008tg} the chain of higher-derivative terms found in the work of FV\footnote{See \cite{Vasiliev:2001wa,Alkalaev:2002rq,Vasiliev:2011xf} for the extension of the FV construction beyond $4$ dimensions.} \cite{Fradkin:1986qy} and weighted by inverse powers of the cosmological constant $\L$ was recognized to be related, in the case of the gravitational coupling, to a higher-derivative \emph{seed} (called in this way after \cite{Sagnotti:talk09}). The latter is nicely associated to the simpler flat space cubic vertex, whose exact structure can be recovered, in a suitable scaling limit, wiping out the lower members of the tail. Moreover, even the origin of the spin-$2$ excitation present in the Vasiliev system is still unclear from a field theory perspective, since it can be dressed with Chan-Paton factors like any excitation belonging to the open bosonic string. This would make the ``graviton\footnote{We call it \emph{graviton} here with a little abuse of language since it admits colors.}'' colored, in contrast with standard field theory results pointing out inconsistencies of this kind of option \cite{Boulanger:2000rq} (strictly speaking with finitely many fields). To reiterate, at the quartic order a number of difficulties have piled up along the years, starting from the no-go results \cite{Weinberg:1964ew,Coleman:1967ad,Haag:1974qh,Weinberg:1980kq,Aragone:1979hx,Aragone:1979bm,Damour:1987fp,Porrati:2008rm,Porrati:2008gv,Porrati:2008ha,Porrati:2009bs}, up to the inconsistency pointed out in \cite{Bekaert:2010hp} for the Berends-Burgers-van Dam cubic coupling of spin-$3$ fields (for a recent review see for instance \cite{Bekaert:2010hw} and references therein). Four-point functions of HS fields have been (and still are) somehow the most intriguing source of difficulties. Here we shall review these questions discussing the role of Lagrangian \emph{non-localities} from a more general perspective while keeping in mind the well known breakdown of unitarity and causality that seems unavoidably
to accompany them (see e.g. \cite{Fradkin:1986qy,Simon:1990ic,Bekaert:2010hw,Adams:2006sv}). As we have anticipated, already at the quadratic level some non-localities naturally arise as soon as massless HS particles are considered while at higher orders they may reflect very peculiar and subtle aspects of the corresponding tree-level amplitudes that may clash with the factorization property usually assumed in the framework of the S-matrix theory.

From a more general perspective concepts like \emph{locality} or \emph{unitarity} have proved to be still poorly understood for HS field theories, and in general in theories of Quantum Gravity. An example is the open problem of reconciling unitarity with black hole entropy (see e.g. \cite{Giddings:2006sj}), where the only indirect argument that a unified description should exist comes from the famous AdS/CFT correspondence\footnote{See \cite{Vasiliev:2012vf} for a reinterpretation of the Holographic principle from the point of view of the Unfolding formulation.} \cite{Maldacena:1997re}.
The limitations as well as the comprehension of these concepts deserve further studies, especially whenever HS fields are taken into account. Moreover, it is worth stressing that those concepts are very deeply related to fundamental questions that combine the long standing problems of understanding Gravity and Quantum Mechanics.

Over the years there have been a number of results pointing to the aforementioned direction, one of these being the AdS/CFT correspondence itself. The latter gives indeed an interesting hint on the very nature of the full theory of Quantum Gravity and strongly suggests the need of introducing HS excitations in order to arrive at a systematic analysis of a quantum space-time geometry. The reason is very simple and is related to the fact that many CFT admit primary operators of arbitrary spin, be they related to conserved currents or not. Most of the time such operators can be indeed constructed by considering derivatives of the fundamental fields, as for instance
\be
J^{\m_1\ldots\m_s}\,\sim\,\partial^{\,\m_1}\ldots\partial^{\,\m_k}\phi\,\partial^{\,\m_{k+1}}\ldots\partial^{\,\m_s}\phi\,,
\ee
where the space time indices are carried by the derivatives, or as in ST:
\be
J^{\m_1\ldots\m_s}\,\sim\,\partial X^{\,\m_1}\ldots \partial X^{\,\m_s}\,.
\ee
where the space time indices are carried by the fields themselves. From the space-time point of view this picture may correspond to various phases of some underlying HS field theory whose observables, Ward Identities and global symmetries can be captured by some CFT.

Even if the previous observation is centered on the AdS/CFT correspondence, the very origin of these motivations is to be found in String Theory, that is regarded as a promising scheme for the Fundamental Interactions, leading naturally to a ultraviolet completion of classical gravity. Moreover it is an example of a theory with an infinite number of propagating degrees of freedom, which makes very hard to imagine how the fundamental concepts mentioned above can be enforced systematically or even translated in terms of a fully background independent scenario.

Surely enough, the whole net of string dualities together with the remarkable M-theory picture have led to a number of important results, nonetheless fundamental questions like those mentioned above about a background independent description highlight the limitations of the current formulation. Therefore, to reiterate, one is naturally led to delve into the systematics of HS theories that, for the reasons mentioned so far, are the most natural candidates to describe in a fully background independent way Quantum Gravity, as well as the generalizations of classical \emph{geometry} to a full HS geometry underlying this completion.

In this respect the crucial role of quartic and higher-order interactions can be appreciated just looking at ST, where one can observe the presence of an infinite number of $\a^{\,\prime}$ corrections, in net contrast to the simpler cubic level containing, for given spins, a finite number of them. Hence, it is not a surprise from this viewpoint that starting from the quartic order the deformation program has encountered along the years severe difficulties, both in the metric-like formulation and in the frame-like one\footnote{There are indeed difficulties in constructing a consistent first-order Lagrangian description (see \cite{Boulanger:2011dd} for a proposal of a Lagrangian for the Vasiliev system that requires an enlargement of the usual Vasiliev's setting).}.

An infinite number of $\a^{\,\prime}$ corrections is actually tantamount to an intrinsic non-locality of the corresponding space-time description. Although, causality should be enforced only at the level of the observable quantities in the usual form of Einstenian locality \cite{Strocchi:2005yk}:
\be
\left[\cO(x),\cO(y)\right]\,=\,0\,,\qquad (x-y)^2>0\,,\label{Einstein locality}
\ee
for a space-like separation of the supports of the observables. This leaves open in principle the possibility that non-observable objects like gauge-fields interact in a seemingly unexpected non-local way. A further observation can be made in this respect and is related to the possible singular nature that goes in hand with non-localities. A very instructive example is given by a non-locality of the form
\be
\frac{1}{\square\,-\,m^{\,2}}\,,\label{massivebox}
\ee
that naively seems to be more tractable than one of the form\footnote{The reason why it is more tractable is that the corresponding convolution kernel has an exponential fall-off at infinity while the convolution kernel of $\tfrac{1}{\square}$ has in general a polynomial fall-off.}
\be
\frac{1}{\square}\,,\label{masslessbox}
\ee
because the operator in \eqref{massivebox} can be expanded as a formal series yielding
\be
\frac{1}{\square\,-\,m^{\,2}}\,=\,-\,\frac{1}{m^{2}}\,\sum_{n=0}^{\infty}\left(\frac{\square}{m^{\,2}}\right)^{\,n}\,,\label{expansion}
\ee
that displays a perturbatively local expansion in the number of derivatives. Let us stress, though, that the formal expansion of eq.~\eqref{expansion} embodies a degree of singularity that is comparable to the one of $\tfrac1\square$ because the above series has a finite radius of convergence. Hence, \emph{off-shell} non-localities behave in general in similar ways and distinguishing between perturbatively local and explicitly non-local ones should be supplemented by some further prescription\footnote{See e.g. \cite{Moeller:2002vx} for a discussion about the meaning of a non-local differential equation containing infinitely many time derivatives.}.
Let us conclude this brief detour stressing that non-localities like $\tfrac{1}{\square}$ are ubiquitous in field theory at the level of scattering amplitudes where they are contained within the propagators. Actually, together with the Feynmann $i\e$ prescription, they give rise to a general type of non-local structure that is compatible with causality and unitarity\footnote{Notice here that the above $\tfrac{1}{\square}$ terms are never of the ill-defined form $\tfrac{1}{0}$ because within the scattering amplitudes they are not acting on the external states. These kind of non-localities are instead related, in momentum space, to inverses of the Mandelstam variables so that their singular nature generates the pole contributions associated to resonances in accordance with unitarity, while the $i\e$ prescription implies Einstein locality \eqref{Einstein locality} at the level of the observables.}.

\section*{Noether Procedure}

It is instructive to introduce our arguments starting from the usual field theory perspective on the problem of HS interactions given by the Noether procedure. The latter played a key role in the construction of supergravity \cite{Freedman:1976xh}, and in its various incarnations has played a crucial role in order to solve for HS cubic couplings in explicit cases. From this point of view, the HS problem can be reformulated as equivalent to finding, order by order in the number of fields, a deformation of the free system of the form
\be
S[\phi]\,=\,\sum_s S^{\,(2)}[\phi_{\m_1\ldots\m_s}]\,+\,\e\, S^{\,(3)}[\phi_{\m_1\ldots\m_s}]\,+\,\e^{\,2}\,S^{\,(4)}[\phi_{\m_1\ldots\m_s}]\,+\,O(\e^{\,3})\,,\label{Deform}
\ee
including at least one field of spin $s\,>\,2$ and where the contribution $S^{\,(3)}$ is cubic, $S^{\,(4)}$ is quartic, and so on. Consistency of the deformation \eqref{Deform} translates into an equivalence class of deformations of the linearized gauge symmetries of the type
\be
\d_{\L}\,\phi_{\,\m_1\ldots\m_s}\,=\,\d^{\,(0)}_{\L}\,\phi_{\,\m_1\ldots\m_s}\,+\,\e\,\d^{\,(1)}_{\L}\,\phi_{\,\m_1\ldots\m_s}\,+\, \e^{\,2}\, \d^{\,(2)}_{\L}\,\phi_{\,\m_1\ldots\m_s}\,+\,O(\e^{\,3})\,,
\ee
leaving invariant $S[\phi]$ order by order, and defined modulo local redefinitions of fields and gauge parameters of the form
\be
\begin{split}
&\phi_{\,\m_1\ldots\m_s}\,\ra\,\phi_{\,\m_1\ldots\m_s}\,+\,\e\,f(\phi)_{\,\m_1\ldots\m_s}\,+\,O(\e^{\,2})\,,\\
&\L_{\,\m_1\ldots\m_{s-1}}\,\ra\,\L_{\,\m_1\ldots\m_{s-1}}\,+\,\e\,\zeta(\phi,\L)_{\,\m_1\ldots\m_{s-1}}\,+\,O(\e^{\,2})\,.
\end{split}
\ee
In its general form above the Noether procedure has presented along the years a number of difficulties that have appeared already at the quadratic and cubic levels. In the following we are going to reexpress it exploiting two key ingredients that have been recently recognized. These two ingredients are the ambient space approach and the restriction to the transverse and traceless (TT) part of the Lagrangian, before addressing the problem of completing it to its full version.

Before introducing these two ingredients in more detail, it can be interesting to make some more comments on the possible relations between the Noether program and the AdS/CFT correspondence. The key observation moves from the conjectured link between any ``fully consistent'' theory living in AdS, that could be thought of as a consistent outcome of the Noether procedure, and a corresponding CFT on its boundary. In this respect it is tempting to think that the latter correspondence may capture those consistency requirements encoded by the Noether procedure, embodying them in a different but equivalent fashion. More in detail, it would be interesting to understand the meaning of these links and in particular the dictionary between the Noether procedure requirements in the bulk on one side and the conformal symmetry at the boundary on the other side\footnote{We are referring here in particular to the bootstrap program at the CFT level.}. Most importantly this would possibly clarify the role played by Lagrangian locality, that is usually assumed in the bulk and seems to have no naive counterpart on the CFT side, in relation to more fundamental concepts like Causality and Unitarity \cite{Heemskerk:2009pn,Fitzpatrick:2011hu,Fitzpatrick:2011dm}. These observations strengthen the feeling that the Lagrangian locality constraint should be relaxed when solving the Noether procedure in full generality, while the AdS/CFT correspondence can give some hints on the possible alternative requirements that would rule out inconsistent options. Surely enough, it would be interesting to investigate the latter alternatives in order both to gain a better understanding of the Noether procedure itself and to extend the analysis from AdS to flat-space, that turns to be much more involved and elusive because of the lack of a guiding principle like AdS/CFT in this case\footnote{See e.g. \cite{Polchinski:1999ry,Susskind:1998vk,Okuda:2010ym,Penedones:2010ue} for the study of the flat limit of some AdS scattering amplitudes. Moreover, it is worth mentioning that the CFT results can be obtained \emph{a priori}
from the AdS results attaching the Boundary-to-Bulk propagators to the vertices.}. On the other hand this point of view may also shed some new light on the very nature of the AdS/CFT correspondence itself, whose generality, in a sense, goes far beyond ST or any other particular framework\footnote{See for instance \cite{Apolo:2012gg} for a discussion of an holographic model whose bulk dual does not contain the graviton but massive spin-2 fields.}.

In the following we are going to introduce the various ingredients that we shall exploit in order to address the Noether procedure scheme.

\section*{Ambient-space formalism}

Already in the flat case the full cubic vertices are highly non-trivial, and one can expect the structure of (A)dS cubic vertices to be even more complicated due to the non-commutativity of the covariant derivatives. The ambient space formalism puts in this respect constant curvature backgrounds on the same footing, representing them by proper embeddings into diverse signature flat space time. Actually, the ambient space formalism has a long history that goes back to the seminal paper of Dirac \cite{Dirac:1936fq} and has been used in the context of HS in \cite{Fronsdal:1978vb,Metsaev:1995re,Metsaev:1997nj, Biswas:2002nk}.
The key point of this formalism are the simplifications that arise when rewriting intrinsic (A)dS quantities in terms of simpler flat-space ones.\footnote{
The ambient-space formalism has been used for a large number of applications. See e.g.
\cite{Bars:1997bz,Bekaert:2003uc, Hallowell:2005np,Barnich:2006pc,Fotopoulos:2006ci,Francia:2008hd,Boulanger:2008up,Boulanger:2008kw,Alkalaev:2009vm}.}
Recently, it was also exploited in order to construct spin-$s$ gauge interactions with a scalar field \cite{Bekaert:2010hk}.

The key feature of the ambient-space formalism is to regard the (A)dS space as the codimension-one hyper-surface \mt{X^{2}=\s L^{2}}, with $\s=\pm$, in an ambient flat space-time parameterized by Cartesian coordinates $X^{\sst M}$ with ${\st M}=0, 1, \cdots, d$\,. In this formalism, the ambient-space HS fields $\Phi_{\sst M_{1}\cdots M_{s}}$ that are homogeneous in $X^{\sst M}$ and tangent to the hyper-surface, are in one-to-one correspondence to the (A)dS fields $\varphi_{\mu_{1}\cdots \mu_{s}}$\,. Moreover, the field equations and the gauge transformations, of (A)dS fields, can be derived from those of the ambient-space fields by a radial-dimensional reduction.

The only subtlety of this formalism arises from the formally diverging radial integral at the level of the action. This can be cured with a $\d$-function insertion of the form $\delta\big(\sqrt{\s X^{2}}-L\big)$\,. The presence of the $\delta$-function is the main difference between the purely flat-space constructions and the ambient-space ones and represents an original contribution of the author, which appeared in\footnote{See also \cite{Bars:2010xi} and references therein for similar ideas.} \cite{Joung:2011ww,Joung:2012rv,Joung:2012fv}. It requires particular care since it spoils the usual flat-space property that the integral of a total derivative vanishes.

\section*{TT part of the Lagrangian}

This second ingredient has to be considered as a strategy that is very useful in order to divide the initial problem in well defined and conceptually simpler ones. Indeed, one of the main lessons in the recent construction of flat-space cubic interactions \cite{Manvelyan:2010jr,Sagnotti:2010at} is that the complete expressions of the vertices are determined by their on-shell forms.
The latter may be regarded as the TT part of the corresponding Lagrangian, considered as an equivalence class modulo traces and divergences of the fields. In this approach the ambient-space representative of the kinetic term for a HS field $\Phi_{M_{1}\cdots M_{s}}$ becomes simply
\be
\int d^{d+1} X\ \d\left(\sqrt{\s X^{2}}-L\right)\ \Big[\Phi^{\,M_{1}\cdots M{s}}\,\square\,\Phi_{M_{1}\cdots M_{s}}+\ldots\Big]\,,\label{quad}
\ee
where the ellipsis refer to terms proportional to divergences and traces. Indeed, the above Lagrangian is invariant under \mt{\delta_E^{(0)}\, \Phi_{M_{1}\cdots M_{s}}= \partial_{(M_{1}} E_{M_{2}\cdots M_{s})}} when quotienting modulo the corresponding Fierz system for the gauge parameter.
The key observation is that the interaction problem can be addressed first at this level, while completing the Lagrangian to its full version requires a tedious but well defined procedure. In order to avoid any confusion, let us stress here that the TT part of the Lagrangian so far introduced is \emph{not} to be considered as a projection of the original Lagrangian. The ellipsis are there to recall this fact, while we are just concentrating on a particular portion of the full Lagrangian (the TT part). In this sense one can properly study the appearance of possible non-localities at this level. For instance the kinetic term in eq.~\eqref{quad} is clearly local involving only two derivatives while possible non-localities can come here only from the ellipsis in eq.~\eqref{quad}. Let us also mention that this splitting has a natural physical interpretation at the S-matrix level where the TT-part of the couplings is the only leftover contribution that couples on-shell propagating degrees of freedom while any piece that is proportional to divergences or traces would vanish identically.

\emph{Generating functions} are an additional ingredient that we shall exploit henceforth. One defines a \emph{master} field $\Phi(X\,,U)$ as the ambient generating function
\be
\Phi(X\,,U)\,=\,\sum_{s\,=\,0}^{\infty}\,\tfrac{1}{s!}\ \Phi_{\,\m_1\ldots\m_s}(X)\,U^{\m_1}\ldots\,U^{\m_s}\,,
\ee
whose components carry arbitrary bosonic representations of the ambient Lorentz group\footnote{In principle we can consider also generating functions of mixed-symmetry field but in this Thesis we concentrate our attention on the totally-symmetric representations.}. On-shell, the corresponding representations are described by the Fierz system
\be\label{on-shell}
\begin{split}
\square\,\Phi(X,U)\,&=\,0\,,\\
\partial_{U}\cdot\partial_X\,\Phi(X,U)\,&=\,0\,,\\
\partial_U\cdot\partial_U\,\Phi(X,U)\,&=\,0\,,
\end{split}
\ee
together with the on-shell gauge invariance
\be
\delta^{(0)}_E\Phi(X,U)\,=\,U\cdot\partial_{X}\,E(X,U)\,,
\ee
where the gauge parameter satisfies an analogous Fierz system
\be
\begin{split}
\square\,E(X,U)\,&=\,0\,,\\
\partial_{U}\cdot\partial_X\,E(X,U)\,&=\,0\,,\\
\partial_U\cdot\partial_U\,E(X,U)\,&=\,0\,.
\end{split}
\ee
As we have anticipated, the above equations are to be supplemented by homogeneity and tangentiality constraints on both the fields and the gauge parameters in order to keep an effective $d$-dimensional description. Hence, for massless fields one finally ends up with
\begin{align}
X\cdot\partial_U\,\Phi(X,U)\,&=\,0\,,& X\cdot\partial_{U}\,E(X,U)\,&=\,0\nn
(X\cdot\partial_X-U\cdot\partial_U+2)\,\Phi(X,U)\,&=\,0\,,& (X\cdot\partial_X-U\cdot\partial_U)\,E(X,U)\,&=\,0\,.
\end{align}
We are then able to recognize a simpler incarnation of the Noether procedure at the level of $n$-point functions that realize both the \emph{linearized} gauge symmetries and the \emph{global} symmetries of the free system above, to be contrasted with the $n$-point Lagrangian couplings, that nonetheless can be directly extracted from these data. Hence, defining by $\tilde{C}^{(n)\,\text{TT}}_{12\ldots n}(\partial_{X_i}\,,U_i\,,\ldots)$ the generating function of the TT part of the color-ordered HS $n$-point functions\footnote{In the following we shall avoid the superscript TT so that the generating function should be always considered as referred to the TT part of the corresponding couplings unless explicitly stated.}, one is led to the linear differential equations in the $U_i$'s
\be
\partial_{X_i}\cdot\partial_{U_i}\,\tilde{C}^{(n)\,\textbf{TT}}_{12\ldots n}(\partial_{X_i}\,,U_i\,,\ldots)\,\approx 0\ ,\qquad i\,=\,1\ ,\ldots\ ,n\ ,\label{onshellN}
\ee
in which the approximate equality means on-shell and where since the measure contains a $\d$-function insertion one has to take care of the non-vanishing contributions coming from the ambient total derivatives. As anticipated, the strategy is then to extend the above solutions adding traces and/or divergences in order to recover the \emph{same} condition
\be
\partial_{X_i}\cdot\partial_{U_i}\,\tilde{C}^{(n)\,\textbf{full}}_{12\ldots n}(\partial_{X_i}\,,U_i\,,\ldots)\,\approx 0\ ,\qquad i\,=\,1\ ,\ldots\ ,n\ ,
\ee
but where now the equality is modulo the full Lagrangian equations of motion (EoM), without quotienting modulo divergences and traces.
\begin{figure}[h]\label{fig:intro}
\begin{center}
\includegraphics[width=10cm]{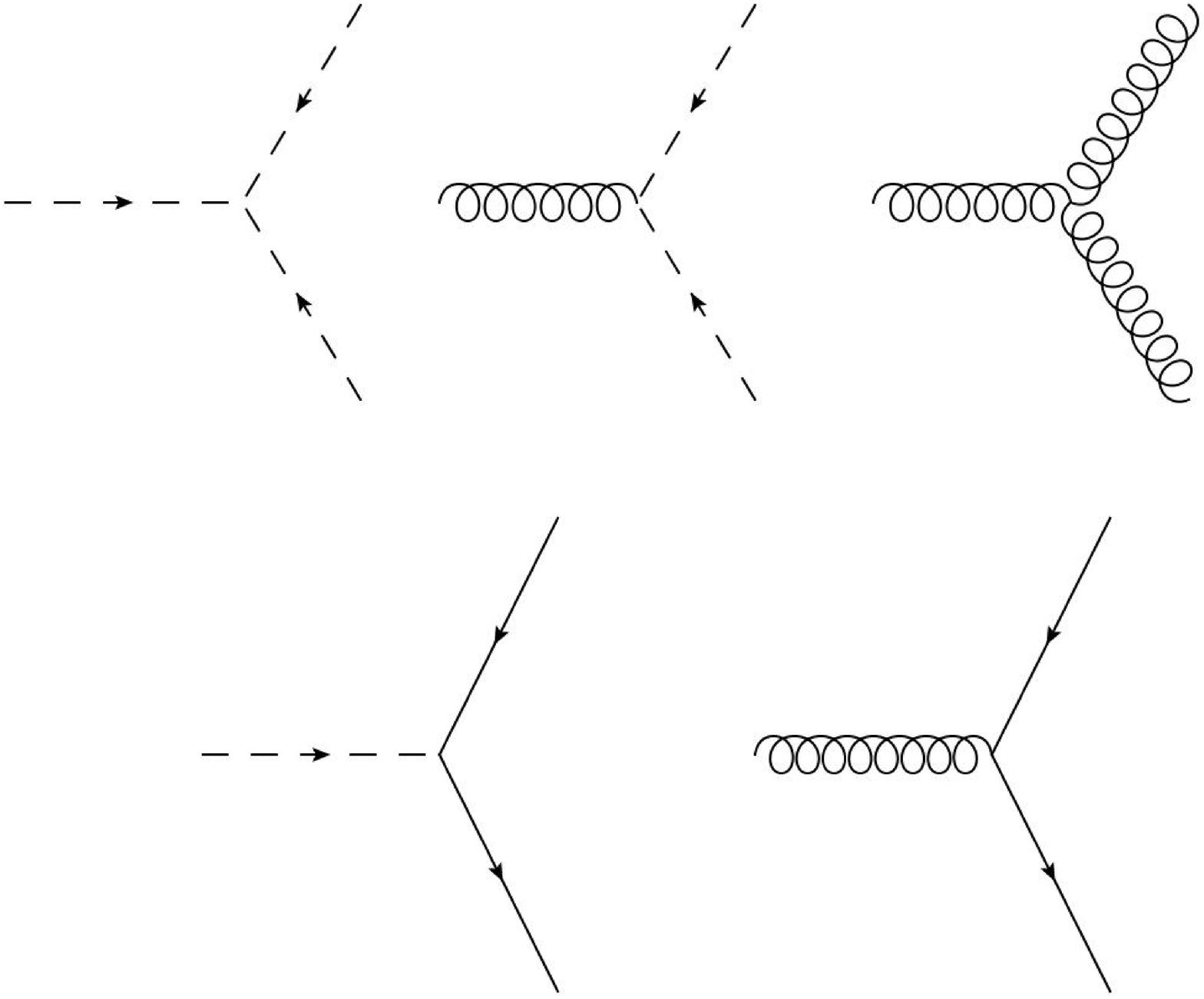}
\caption{Various cubic building blocks for the $G^{\,(i)}_{12\ldots n}$'s.}
\end{center}
\end{figure}
As shown in \cite{Taronna:2011kt}, the solution to this problem can be actually expressed in terms of powers of the standard color-ordered $n$-point functions $G_{12\ldots n}^{\,(i)}$ of the most general theory built from a gauge boson, a scalar field and a spin-$1/2$ fermion, that are uniquely specified by the cubic couplings in fig.~1\ref{fig:intro}. One should add, in principle, also the usual Yang-Mills quartic couplings, that however can be reconstructed here interpreting them as counterterms restoring the \emph{linearized} gauge invariance of the tree-level amplitudes, thus making clear the fundamental role of the latter with respect to the former. Notice here the role of the YM cubic couplings whose powers generate at the cubic order all HS cubic couplings in any constant curvature background providing a suggestive picture for the systematics of the HS interactions.
More in detail, at the cubic level the TT part of all consistent (A)dS cubic interactions reads
{\allowdisplaybreaks
\ba
\label{sol ads}
	&& S^{\sst (3)} = \frac{1}{3!} \sum_{s_{1},s_{2},s_{3}=0}^{\infty}
	\sum_{n=0}^{\text{min}\{s_{1},s_{2},s_{3}\}} g^{s_{1}s_{2}s_{3},n}_{a_{1}a_{2}a_{3}}\,
	\int {d^{d+1}X}\,\delta\Big(\sqrt{\s\,X^{2}}-L\Big)\,\times \\
	&&\times\,
	\big[\partial_{U_{1}}\!\!\cdot(\partial_{X_{23}}\!\!+\alpha\,\partial_{X})\big]^{s_{1}-n}\,
	\big[\!-2\,\partial_{U_{2}}\!\!\cdot(\partial_{X_{1}}\!\!
	-\tfrac{\alpha-1}{\alpha+1}\,\partial_{X})\big]^{s_{2}-n}\,
	\nn
    &&\hspace{200pt}\times\,\big[\,2\,\partial_{U_{3}}\!\!\cdot(\partial_{X_{1}}\!\!-\tfrac{\alpha+1}{\alpha-1}\,\partial_{X})\big]^{s_{3}-n}
	\nn
	&& \times \,
	\big[ \partial_{U_{2}}\!\!\cdot\partial_{U_{3}}\,
	\partial_{U_{1}}\!\!\cdot(\partial_{X_{23}}\!\!+\beta\,\partial_{X})
	-2\,\partial_{U_{3}}\!\!\cdot\partial_{U_{1}}\,
	\partial_{U_{2}}\!\!\cdot(\partial_{X_{1}}\!\!+\tfrac{\alpha-\beta}{\alpha+1}\,\partial_{X}) \nn
	&& \hspace{150pt}+\, 2\,\partial_{U_{1}}\!\!\cdot\partial_{U_{2}}\,
	\partial_{U_{3}}\!\!\cdot(\partial_{X_{1}}\!\!+\tfrac{\alpha-\beta}{\alpha-1}\,\partial_{X})
	\big]^{n}\nn
	&&\hspace{100pt}\times\,\Phi^{a_{1}}(X_{\sst 1},U_{\sst 1})\
	\Phi^{a_{2}}(X_{\sst 2},U_{\sst 2})\ \Phi^{a_{3}}(X_{\sst 3},U_{\sst 3})\,
	\Big|_{\overset{X_{i}=X}{\sst U_{i}0}}\,,
	\nonumber
\ea}\!\!
where the form of the vertices is encoded in a differential operator acting
on the generating functions of the ambient-space fields
\be
\Phi^{\sst a}(X,U)\,=\,\sum_{s\,=\,0}^\infty\tfrac{1}{s!}\
\Phi^{\sst a}_{\sst M_{1}\cdots M_s}(X)\,U^{\sst M_1}\cdots U^{\sst M_s}\,,
\ee
while
\be
\partial_{X^{\sst M}}=\partial_{X_{1}^{\sst M}}+\partial_{X_{2}^{\sst M}}+\partial_{X_{3}^{\sst M}}
\ee
is the ambient total derivative, and
\be
\partial_{X_{ij}^{\sst M}}:=\partial_{X_{i}^{\sst M}}-\partial_{X_{j}^{\sst M}}\,.
\ee
Let us stress that different choices of $\alpha$ and $\beta$ parameterize an ambiguity in writing the building blocks and hence their contribution vanishes identically upon integration. The number of \emph{ambient-space derivatives} in \eqref{sol ads} is
\be
\Delta=s_{1}+s_{2}+s_{3}-2n\,,
\ee
but, when radially reduced, different portions of the (A)dS vertices involve different number of covariant derivatives: \mt{\Delta,\ \Delta-2,\, \dots\,, 1} (or $0$), while whenever the number of derivatives decreases by two the corresponding mass-dimension is compensated by the cosmological constant $\Lambda:=1/L^{2}$\,.\footnote{The correct relation between the cosmological constant $\Lambda_{\sst\text{C.C.}}$ and the radius of (A)dS is \mt{\Lambda_{\sst\text{C.C.}}=(d-1)(d-2)/(2\,L^{2})=\Lambda\,(d-1)(d-2)/2}\,. However in this Thesis, for simplicity, we call also $\Lambda$ cosmological constant.}
This structure makes clear the relation between \eqref{sol ads} and the FV vertices, where the inverse-power expansion in $\Lambda$\, appears. For instance, concentrating on the gravitational couplings (\mt{s_{1}=s_{2}=s} and \mt{s_{3}=n=2}) in \eqref{sol ads},
the action can be recast in terms of an inverse-power series in $\Lambda$ as
\be
	S^{\sst (2)}+S^{\sst (3)} =  \frac{\lambda_{s}}{G_N}\ 	\sum_{r=2}^{s}\,
	\frac{1}{\Lambda^{r-2}}\,\int_{(A)dS_{d}}\cL_{r}\,.
	 \label{act ss2}
\ee
In order to get this expression, we made use of the redefinitions
\be
g^{ss{\sst 2,2}}=\Lambda^{2-s}\sqrt{G_N}\ \lambda_{s}\,,\qquad \varphi^{\sst (s)}=\phi^{\sst (s)}/\sqrt{G_N}\,,
\ee
where $G_N$ is the gravitational coupling constant. The $\cL_{r}$'s are cubic vertices which are separately gauge invariant under the spin 2 gauge transformations and can be written schematically as
\be
	\cL_{r}=D^{2(r-1)}\,h\ \phi^{\sst (s)} \phi^{\sst (s)}
	+\Lambda\,D^{2(r-2)}\,h\ \phi^{\sst (s)} \phi^{\sst (s)}\,,
	\label{FV ver}
\ee
where $\cL_{2}$ should involve the gravitational minimal coupling. Notice also that the inverse-power $\Lambda$-expansion draws its origin from the redefinition of the coupling constant $g^{ss{\sst 2,2}}$\,, which makes the two-derivative part of the vertex independent of $\Lambda$\,.

Coming back to our main discussion let us stress that eq.~\eqref{onshellN} fixes the tensor structure of the generating function $\tilde{C}^{(n)}_{12\ldots n}$, but leaves open the possibility of multiplying each of them with arbitrary coefficients and, starting from the quartic order, also with functions of Mandelstam-like invariants. Moreover, the general solution to eq.~\eqref{onshellN} for $n\geq 4$ can be also expressed as\footnote{Let me thank Euihun Joung for useful discussions on this point.}
\be
\tilde C^{(n)}\,=\,\cK^{(n)}\left(\partial_{X_i}\cdot\partial_{X_j}, H_{ij}^{(1)}, H_{ijk}^{(2)}\right)\,,
\ee
in terms of a generic function of the Mandelstam variables $\partial_{X_i}\cdot\partial_{X_j}$ and of the following building blocks:
\be
H_{ij}^{(1)}\,=\,\frac{{U_i}\cdot\partial_{X_j}\,{U_j}\cdot\partial_{X_i}}{\partial_{X_i}\cdot\partial_{X_j}}\, -\,{U_i}\cdot {U_j}\,,\quad H_{ijk}^{(2)}\,=\,\frac{{U_i}\cdot\partial_{X_j}}{\partial_{X_i}\cdot\partial_{X_j}}\,-\, \frac{{U_i}\cdot\partial_{X_k}}{\partial_{X_i}\cdot\partial_{X_k}}\,.
\ee
However, this representation hides the important relations with the current exchange amplitudes that should be used in order to constrain the consistent theories. Hence, in the following we shall keep the description in terms of the $G^{(i)}$'s even if the latter can be rewritten in terms of the above $H$-building blocks.

Moreover, let us underline a sort of \emph{correspondence} between tree-level $n$-point functions and $n$-point Lagrangian couplings. The former are associated to a linearized on-shell gauge symmetry related to the free system of eq.~\eqref{on-shell} together with the corresponding global symmetries. The latter encode somehow \emph{geometrical} principles, together with a fully non-linear deformation of the original gauge symmetries that could be captured, in principle, by a resummation of the full tower of non-linear couplings. For instance, in the case of spin-$2$ this resummation rebuilds the usual Riemaniann geometry, as observed by Deser in \cite{Deser:1969wk}, and references therein, and in the case of HS fields should encode, in a similar fashion, the HS geometry.
One should also gain a deeper understanding of the \emph{non-linear} symmetries and of the global symmetries of the system, that need to be related to some HS algebra putting more constraints in principle on the available choices at the cubic order.  We leave this important analysis for the future.

In this Thesis we take the aforementioned perspective in order to study how field theory can in principle overcome some difficulties encountered over the years and trying to understand, without requiring explicitly a local Lagrangian description, what can be the requirements leading to consistent interacting HS theories. The main discussion is done at the level of the leading term of the couplings that by the way coincides with the corresponding flat limit. The result can be roughly summarized by the generating function
\be
\tilde{C}^{(n)}_{1234}\left(\partial_{X_i},U_i\right)\,=\,-\,\frac{1}{su}\ \exp\left[-\,su\,\left(G^{(0)}_{1234}+G^{(1)}_{1234}+G^{(2)}_{1234}\right)\right]\,,\label{Cn}
\ee
of four-point amplitudes involving massless HS fields where the $G^{(i)}_{1234}$'s are the gauge boson four-point functions discussed in Section~\ref{sec:YMfour} while the generating function should be considered modulo functions of the Mandelstam variables that do not introduce poles of order greater than one, in analogy with the coefficients $g_{\sst a_1a_2a_3}$ in eq.~\eqref{sol ads}. Furthermore, the $G^{(i)}$'s have been multiplied by Mandelstam invariants in order to get only single poles. In this way, their current exchange parts reconstruct HS exchanges after combining properly the power expansion of eq.~\eqref{Cn}. This can be done choosing the relative functions of the Mandelstam invariants in order to match the corresponding contributions coming from the cubic level. In this way we are able to identify the current exchanges that cannot be pursued at higher orders in a local fashion while we are able to exhibit the local completion of the others.

This structure of the four-point functions arises as the solution of the Noether procedure equation for a general field theory with massless particles of any spin, but we expect only a few choices of the relative coefficient as well as of the spectrum of the theory to be consistent. Moreover, while embodying an infinite class of local quartic couplings, the above result contains also the seeds for the difficulties that have been faced along the years. However, the physical interpretation and the need for the corresponding Lagrangian non-localities is still a subtle issue and the peculiar form of the amplitudes one arrives at for HS fields in flat space clashes in general with commonly accepted ideas about the S-matrix structure that reflect some difficulties in defining an S-matrix for (infinitely many) massless particles (see e.g. the discussion in the introduction of \cite{Weinberg:1964ew} or \cite{Eden:1966}). The full amplitude generating function, containing also Chan-Paton factors \cite{Paton:1969je}, is finally recovered summing over all non-cyclic permutations of the external legs, as
\begin{multline}
\cA(\Phi_1,\Phi_2,\ldots,\Phi_n)\,=\,\sum_\s \tilde{C}^{(n)}_{1\s(2)\ldots\s(n)}(\partial_{X_i},U_{i})\\\star_{\sst 12\ldots n}\,\text{Tr}\Big[\Phi_1(X_1,U_1)\,\Phi_{\s(2)}(X_{\s(2)},U_{\s(2)})\, \ldots\, \Phi_{\s(n)}(X_{\s(n)},U_{\s(n)})\Big]\,,\label{IntroAmplitude}
\end{multline}
where the trace is over the color indices carried by the polarization generating functions $\Phi_{i}$, recovering in this way a kind of generalized \emph{open-string}-like form. Interestingly, from four points onwards, there are more possibilities, since some permutations with respect to the labels $\{1\,,\ldots\,,n\}$ of the $G_{12\ldots n}^{(i)}$'s are independent for $n\,\geq\,4$. This means that one can combine two or more cyclically independent $G^{(i)}$'s, eliminating the Chan-Paton factors and recovering in this way a kind of \emph{closed-string}-like amplitudes from which the usual gravitational four-point functions emerge, together with HS generalizations. In the four-point case, for instance, for each $G^{(i)}_{1234}$ there are two independent options and one can recover in this fashion the \emph{closed-string}-like generating function
\be
\tilde{C}^{(4)}(\partial_{X_i},U_i,U_i^{\,\prime})\,=\, \left(\sum_\s\ \tilde{C}^{(4)}_{1\s(2)\s(3)\s(4)}(\partial_{X_i},U_{i})\ \tilde{C}^{(4)}_{1\s(2)\s(4)\s(3)}(\partial_{X_i},U^{\,\prime}_{\,i})\right)\,,\label{Shs2}
\ee
together with analogous generalizations to higher orders. These results are analyzed in a number of examples pointing out some differences between the graviton and the colored spin-$2$ fields but leaving for the future a detailed analysis of the generalized \emph{closed-string}-like amplitudes together with possible generalizations of the Bern-Carrasco-Johansson (BCJ) construction of \cite{Bern:2010ue} to HS.

A general lesson to be drawn from the results that we have summarized is that HS $n$-point functions in flat space appear to go in hand with a peculiar feature: while they are still built from current exchanges and local terms, they generally factorize only on (infinite) \emph{subsets} of the current exchanges available at the cubic level, lacking finite numbers of lower-spin contributions. This unusual feature, when present, is equivalent to a \emph{non-local} nature of the corresponding HS Lagrangian couplings. This anyway results from conventional amplitudes built from Feynman propagators, and one can expect them to be still compatible with the notion of causality and with the cluster property, even though at the moment more effort needs to be done in order to clarify the situation. For instance, potential clashes with tree-level unitarity may be possibly related to the fact that infinitely many degrees of freedom have to contribute to the \emph{same} residue as soon as non-localities appear ending up with a breakdown of analyticity. In this respect we can only anticipate that within what we shall call \emph{minimal scheme}, even admitting non-localities, no pole arising in the amplitude can lack an interpretation as an external particle participating in the physical process. This can give some hope to arrive to an understanding of HS interactions, even though we are not able at present to give a definite answer about the consistency of the proposed scheme in flat space. We have also in mind to complete the analysis of the Lagrangian couplings, extending the above results to (A)dS along the same lines of the cubic level, and to exploit the AdS/CFT correspondence in order to see whether locality can be preserved in these more general cases, and if not what are the available alternatives.

Among other things, we shall discuss from the same perspective the role of the spin-$2$ excitation present in the Vasiliev system that admits in principle Chan-Paton factors, trying to give an answer to a puzzle pointed out in \cite{Francia:2007qt} together with a very interesting open question about its true nature. Indeed, at the massless level a mixing between the singlet part of colored spin-$2$ components and a combination that is strictly uncolored and plays the role of a graviton, anticipated in \cite{Francia:2007qt}, is here justified by the existence of two different kinds of amplitudes, the first of the \emph{open-string}-type and the second of the \emph{closed-string}-type. In the following, we shall see that a non-abelian colored spin-$2$ field brings about non-localities, and from this point of view it is naturally related to the massive excitations present in the Open String spectrum, while only the spin-$2$ components interacting with a \emph{closed-string}-like amplitude can be directly related to the usual graviton. This interesting feature can hopefully shed some new light on the non-local nature of the Vasiliev's system itself, where such colored spin-2 fields interact consistently, together with the possible links between Vasiliev's system and ST.

The non-local nature of HS Lagrangian couplings puts our discussion of QFT on more general grounds, that ought to be better understood, as we anticipated. At the same time this discussion, and in particular our generalization of \emph{open-string}-like and \emph{closed-string}-like amplitudes, reinforces the feeling that ST hides within its structure a number of potentially profound lessons for Field Theory.

\vspace{20pt}

The plan of the Thesis is the following. In Chapter~$1$, we introduce the ambient space formalism in full detail, discussing the Noether procedure from a general perspective. In Chapter~$2$ we apply the formalism developed in Chapter~$1$ to the quadratic level and present in detail the procedure that starting from the TT part generates the full Lagrangian. In Chapter~$3$ we describe the three-point couplings along the lines of \cite{Joung:2011ww,Joung:2012rv}. In Chapter~$4$ we extend the discussion to the quartic order, following \cite{Taronna:2011kt}. First we consider the simpler case of Yang-Mills theory and then the general setting of HS theories, pointing out the differences between \emph{open-string}-like amplitudes and \emph{closed-string}-like amplitudes and how the usual flat space problems arise. Some outlooks together with further discussions of the results described in this Thesis are summarized in the conclusions. Finally, the Appendices contain some details of the computations.

\renewcommand{\theequation}{\thesection.\arabic{equation}} \csname
@addtoreset\endcsname{equation}{section}

\clearpage{\pagestyle{empty}\cleardoublepage}

\chapter{Ambient Space Formulation}                

\rhead[\fancyplain{}{\bfseries
Ambient Space Formulation}]{\fancyplain{}{\bfseries\thepage}}
\lhead[\fancyplain{}{\bfseries\thepage}]{\fancyplain{}{\bfseries\rightmark}}

In this chapter we are going to introduce one of the main ingredients that we shall exploit in this Thesis. This is the Ambient space formalism whose main idea dates back to the works of Dirac \cite{Dirac:1936fq}, where the simplifications that arise making use of the isometric embedding of (A)dS spacetime as a codimension one hyperboloid inside a flat auxiliary space were first exploited. These ideas were then extended all over the years by many other people, most notably in the context of HS gauge theories by Fronsdal \cite{Fronsdal:1978vb}, Metsaev \cite{Metsaev:1995re,Metsaev:1997nj}, Biswas and Siegel \cite{Biswas:2002nk} and were also used for a large number of applications \cite{Bekaert:2003uc, Hallowell:2005np,Barnich:2006pc,Fotopoulos:2006ci,Francia:2008hd,Boulanger:2008up,Boulanger:2008kw,Alkalaev:2009vm,Bekaert:2010hk,Alkalaev:2011zv}.

\section{(A)dS Geometry}\label{(A)dS Geometry}

The key feature of the ambient-space formalism, as we have anticipated, is to regard the (A)dS space as the codimension-one hyper-surface
\be
X^{2}=\s L^{2}\,,
\ee
embedded into an ambient $(d+1)$-dimensional flat space-time parameterized by Cartesian coordinates $X^{\sst M}$ with ${\st M}=0, 1, \cdots, d$\,:
\be
i:\,{\text{(A)dS}_d} \hookrightarrow \{X\in\mathbb{R}^{d+1}\ \text{s.t.}\ \s X^2> 0\}:\quad x^{\m}\ra X^M(x^\m)\,.
\ee
Here, $\s$ is a sign positive for dS and negative for AdS. In order to avoid ambiguities, as well as to fix the notation, we shall discuss only the cases of Lorentzian dS or Euclidean AdS spaces. With these choices one can fix the signature of the ambient space to be Minkowskian:
\be
\eta_{MN}\,=\,(-1,+1,\ldots,+1)\,.
\ee
However, the Lorentzian AdS as well as the Euclidean dS cases are straightforward generalizations. Notice also that Lorentzian dS and Euclidean AdS are related by an analytic continuation of the (A)dS radius $L\ra iL$. The embedding formalism makes in this way manifest the isometry group of (A)dS that coincides with the isometry group of the ambient space $SO(d,1)$.

The ambient space metric can be conveniently written in radial coordinates $(R, x^{\m})$,
\be
X^M\,=\,R\hat{X}^M(x)\,,\qquad \hat{X}^M\,=\,\frac{X^M}{\sqrt{\s X^2}}\,,
\ee
foliated by (A)dS sections as
\be
ds_{d+1}^2\,=\,\s\,dR^2+\frac{R^2}{L^2}\,g_{\m\n}(x)dx^\m dx^\n\,,
\ee
where
\be
g_{\m\n}(x)\,=\,L^2\,\eta_{MN}\,\frac{\partial\hat{X}^M(x)}{\partial x^{\m}}\,\frac{\partial\hat{X}^N(x)}{\partial x^{\n}}\,,
\ee
is the (A)dS induced metric. For convenience one can also introduce the ambient space and the intrinsic vielbeins, respectively by
\be
ds_{d+1}^2\,=\,E^R E^R+\eta_{\a\b}\,E^\a E^\b\,,\qquad
ds_{\text{(A)dS}}^2\,=\,\eta_{\a\b}\,e^\a e^\b\,.
\ee
Let us recall that the vielbein is in general a local orthonormal frame
\be
e^\a(x)={e^\a}_\m(x)\, dx^\m\,,
\ee
defined by
\be
{e^\a}_\m(x)\,{e^\b}_\n(x)\,\eta_{\a\b}=\,g_{\m\n}(x)\,.
\ee
together with its dual vector fields
\be
e_\a(x)\,=\,{e_\a}^\m(x)\,\partial_\m\,,
\ee
where $e_\a^\m(x)$ is the inverse vielbein:
\be
{e^\a}_{\m}(x)\,{e_\b}^\m(x)\,=\,\d^\a_\b\,.
\ee
Their transformations under infinitesimal local Lorentz transformations are given by
\be
\d e^\a\,=\,{\e^{\,\a}}_\b(x) e^\b(x)\,,\qquad \d e_\a\,=\,-{\e^{\,\b}}_\a(x) \,e_\b(x)\,,\label{vieltrasf}
\ee
while the transformation of the spin connection reads
\be
\delta\o^{\a\b}(x)\,=\,{\e^{\,\a}}_\g(x)\o^{\g\b}(x)+{\e^{\,\b}}_\g(x)\o^{\a\g}(x)-d\e^{\,\a\b}\,,\label{omega transf}
\ee
so that the curvature two-form is
\be
R^{\a\b}\,=\,\tfrac{1}{2}\,{R_{\m\n}}^{\a\b}\,dx^\m\wedge dx^\n\,=\,d\o^{\a\b}+\o^{\a\g}\wedge{\o_\g}^\b\,.
\ee
A totally-symmetric rank-$s$ tensor field in the moving basis takes the form
\be
\varphi(x)\,=\,\tfrac{1}{s!}\,\varphi_{\a_1\ldots\a_s}(x)\,e^{\a_1}(x)\otimes\ldots\otimes e^{\a_s}(x)\,,
\ee
and in the following, introducing a fixed auxiliary vector $u^\a$, we are going to consider generating functions of the above tensor fields of the form
\be
	\varphi^{a}(x,u):=
	\sum_{s=0}^{\infty}\,\frac1{s!}\ \varphi^{a\,(s)}_{\mu_{1}\ldots\mu_{s}}(x)\
	u\cdot e^{\mu_{1}}(x)\,\cdots\, u\cdot e^{\mu_{s}}(x)\,,\label{intgenfunc}
\ee
where the contraction with the flat auxiliary variables $u^{\alpha}$ is via
the inverse vielbein $e^{\ \mu}_{\alpha}(x)$\,:
$u\cdot e^{\mu}(x)=u^{\alpha}\,e^{\ \mu}_{\alpha}(x)$\,,
and $\st a$ is a \emph{color} index associated with the Chan-Paton factors.
A local Lorentz transformation will also act on this functions since the inverse vielbein transforms. Hence, one can conveniently express the latter transformation properties as
\be
\delta\varphi(x,u)\,=\,-{\e^{\,\a}}_{\b}(x)\,u^\b\partial_{u^\a}\,\varphi(x,u)\,,\label{infinitesimal}
\ee
that reduces to \eqref{vieltrasf} if applied to $u\cdot e^{\m}(x)$.
One can also consider finite Lorentz transformations integrating \eqref{infinitesimal} to
\be
\varphi_e(x,u)\ra\varphi_{e^\prime}(x,u)\,=\,e^{-{\e^{\,\a}}_{\b}(x)\,u^\b\partial_{u^\a}}\varphi_e(x,u)\,.
\ee
Within this formalism it is natural to compute the transformed form of the derivative $\partial_{\m}$ from a constant frame ${e_\a}^{\m}(x)={\d_\a}^\m$ to a generic moving frame recovering, in accordance with eq.~\eqref{omega transf}
\begin{multline}
e^{-{\e^{\,\a}}_{\b}(x)\,u^\b\partial_{u^\a}}\partial_{\m}\varphi_\d(x,u)\,=\,\left[\partial_{\m}-(\partial_\m{\e^{\,\a}}_{\b}(x)+\text{O}(\e^2))\, u^{\sst\b}\partial_{u^\a}\right]\,\varphi_e(x,u)\\
=\,\left[\partial_{\m}-\left(e^{-\e(x)}\right)^\a_\g\,\partial_{\m}\,\left(e^{\e(x)}\vphantom{e^{-\e}}\right)^\g_\b\, u^{\sst\b}\partial_{u^\a}\right]\,\varphi_e(x,u)\,,\label{derivativetransf}
\end{multline}
and where by definition of local frame:
\be
\left(e^{-\e(x)}\vphantom{e^{-\e}}\right)^\a_\b\, \d^\b_{\m}\,=\,{e^\a}_{\m}(x)\,.\label{constant frame vielbein}
\ee
Hence, in the following it will prove convenient to define the covariant derivative as
\be
D_\m\,=\,\nabla_\m\,+\,\tfrac{1}{2}\,\o_\m^{\,\a\b}(x)\,u_{[\st\a}\partial_{u^{\b]}}\,,\label{covder}
\ee
where $\nabla_\m$ is the usual covariant derivative acting on tensor indices, while $\o_\m^{\,\a\b}$ is the spin connection.

The formalism so far described proves to be quite convenient in the ambient space approach introduced at the beginning of this section. Indeed, one can lift the intrinsic (A)dS generating functions of eq.~\eqref{intgenfunc} to ambient space flat ones
\be
\Phi^{\st a}(X,U)\,=\,\sum_{s\,=\,0}^\infty\tfrac{1}{s!}\
\Phi^{\st a}_{\sst M_{1}\cdots M_s}(X)\,U^{\sst M_1}\cdots U^{\sst M_s}\,,
\ee
where we have defined the ambient space auxiliary variables $U^{A}$ in analogy with eq.~\eqref{intgenfunc}, while we have used a constant vielbein ${E_A}^M={\d_A}^M$. The condition under which the lifting procedure turns to be well definite and one-to-one were explained by Fronsdal in \cite{Fronsdal:1978vb} and are:
\begin{itemize}
  \item homogeneity in $X^M$:
  \be
  (X\cdot\partial_X\,-\,U\cdot\partial_U+2+\m)\,\Phi(X,U)\,=\,0\,,\label{homo}
  \ee
  \item tangentiality to constant $R$ surfaces:
  \be
  X\cdot\partial_U \, \Phi(X,U)\,=\,0\,.\label{tangent}
  \ee
\end{itemize}
The above conditions translate at the level of the tensor fields respectively as
\begin{itemize}
  \item homogeneity in $X^M$:
  \be
  (X\cdot\partial_X\,-\,\D_s)\,\Phi_{M_1\ldots M_s}(X)\,=\,0\,,
  \ee
  \item tangentiality to constant $R$ surfaces:
  \be
  X^{M_1}\,\Phi_{M_1\ldots M_s}\,=\,0\,.
  \ee
\end{itemize}
More in detail, the latter conditions make the pull-back of a tensor field in the ambient space to a tensor field in (A)dS:
\begin{multline}
i^*:T_{\mathbb{R}^{d+1}}\ra T_{(A)dS_d}:\\ \Phi_{M_1\ldots M_s}(X)\ra \varphi_{\m_1\ldots \m_s}(x)\,=\,\frac{\partial X^{M_1}}{\partial x^{\m_1}}\ldots \frac{\partial X^{M_s}}{\partial x^{\m_s}}\ \Phi_{M_1\ldots M_s}(X(x))\,,
\end{multline}
or, in terms of generating functions:
\be
i^*: \Phi(X,U)\,\ra\,\varphi(x,u)\,=\,\exp\left(u^{\, \m}\,\tfrac{\partial X^M}{\partial x^\m}\,\partial_{\sst U^{\sst M}}\right)\,\Phi(X(x),U)\Big|_{U=0}\,,
\ee
one-to-one, recovering a well definite isomorphism of the associated tensor bundles. Indeed, homogeneity in the radial coordinates gives a well defined ambient space extensions of any function defined on (A)dS, while the tangentiality condition takes care of the kernel of the pull-back that coincides precisely with the radial components of tensors:
\be
X_M\,\frac{\partial X^{M}}{\partial x^{\m}}\,=\,\frac{1}{2}\frac{\partial}{\partial x^{\m}}\,X^2\,=\,0\,.
\ee
Hence, in this formalism, the ambient-space HS fields $\Phi_{\sst M_{1}\cdots M_{s}}$ that are homogeneous in $X^{\sst M}$ and tangent to the hyper-surface, are in one-to-one correspondence to the (A)dS fields $\varphi_{\mu_{1}\cdots \mu_{s}}$\,.
Moreover, also the various differential operators that one usually defines in terms of intrinsic coordinates can be lifted in the ambient space modulo the above prescriptions. In order to see explicitly how this works let us consider the ambient space coordinate transformation in the radial frame $X^M\ra(R,x^\m)$ together with the auxiliary variable splitting in radial and tangent components \mt{U^M\ra(v,u^\m)}. This corresponds to move from the constant ambient frame ${\d^A}_M$, with corresponding generating functions of tensor fields of the form
\be
\Phi_\d(X,U)\,=\,\sum_s \tfrac{1}{s!}\,\Phi^{(\d)}_{M_1\ldots M_s}(X)\, U\cdot{\d^{M_1}}\ldots U\cdot{\d^{M_s}}\,,\label{ambient gen}
\ee
to the radial frame ${E^A}_M$, where the analogous generating functions of tensor fields look like
\be
\Phi_E(X,U)\,=\,\sum_s \tfrac{1}{s!}\,\Phi^{(E)}_{M_1\ldots M_s}(X)\, U\cdot{E^{M_1}}\ldots U\cdot{E^{M_s}}\,.
\ee
Using eq.~\eqref{derivativetransf} one can then compute the covariant derivative in the radial frame as
\be
D_M\,=\,\nabla_M\,-\,{\O_M^{\,A}}_B(x)\,{ U^B}\partial_{U^A}\,,\qquad {\O_M^{\,A}}_B(x)\,=\,{E^A}_N(x)\,\partial_M\,{E_B}^N(x)\,.\label{connection}
\ee
It is now convenient to express the ambient space vielbein $E^A(X)$ in terms of the (A)dS vielbein $e^{\,\a}(x)$ as
{\small{
\be
	E^{\sst R}_{\ {\sst R}}(R,x)=\sqrt{\s}\,,\quad E^{\sst R}_{\ \mu}(R,x)=0\,,
	\qquad
	E^{\alpha}_{\ {\sst R}}(R,x)=0\,,\quad E^{\alpha}_{\ \mu}(R,x)
	= \tfrac {R}{L}\,e^{\alpha}_{\ \mu}(x)\,,
\ee}}
that is equivalent to
\be
E^R\,=\, \sqrt{\s}\, dR\,,\qquad E^\a\,=\,\tfrac{R}{L} \,e^\a\,.
\ee
The above relations can be also translated as
{\small{
\begin{gather}
{E^A}_M(x)\,=\,({E^R}_M,{E^\a}_M)\,=\,\left({\st\sqrt{\s}}\,\tfrac{\partial R}{\partial X^M},\tfrac{R}{L}\,{e^\a}_\m\,\tfrac{\partial x^\m}{\partial X^M}\right)\,=\,\left(\tfrac{1}{\sqrt{\s}}\,\hat{X}_M,\tfrac{R}{L}\, {e^\a}_\m\,\tfrac{\partial x^\m}{\partial X^M}\right)\,,\nonumber\\
{E_A}^M(x)\,=\,({E_R}^M,{E_\a}^M)\,=\,\left(\tfrac{1}{\sqrt{\s}}\,\hat{X}^M, \tfrac{L}{R}\,{e_\a}^\m\,\tfrac{\partial X^M}{\partial x^\m}\right)\,.
\end{gather}}}
\!\!Along the same lines, one can also compute the ambient space connection ${\O_M^{\,A}}_B(x)$ in the radial frame in terms of the intrinsic connection ${\o_{\m}^{\,\a}}_\b(x)$. Applying eq.~\eqref{connection} one ends up with
\be
{\O_M^{\,A}}_B(x)\,=\,\left({\O_M^{\,R}}_\b(x),{\O_M^{\,\a}}_\b(x)\right)\,=\,\tfrac{\partial x^\m}{\partial X^M}\,\left({E^R}_N\,\partial_\m \,{E^N}_\b,{E^\a}_N\,\partial_\m \,{E^N}_\b\right)\,,
\ee
so that one can check that \mt{{\o_\m^\a}_\b\,=\,{E^\a}_N\,\partial_\m \,{E^N}_\b} satisfies the compatibility equation
\be
de^{\,\a}+{\o^{\,\a}}_\g e^{\,\g}\,=\,0\,,
\ee
and hence can be identified with the (A)dS spin-connection. On the other hand,
\be
{E^R}_N\,\partial_\m \,{E^N}_\b\,=\,-{E^N}_\b\,\partial_\m\,{E^R}_N\,=\,-\tfrac{1}{L\sqrt{\s}} \, e_{\b\,\m}\,,
\ee
and so the ambient space spin-connection in the radial frame takes the form
\be
{\O_M^{\,A}}_B(x)\,=\,\left({\O_M^{\,R}}_\b(x),{\O_M^{\,\a}}_\b(x)\right)\,=\,\tfrac{\partial x^\m}{\partial X^M}\,\left(-\tfrac{1}{L\sqrt{\s}}\,e_{\b\,\m}(x),{\o_\m^\a}_\b(x)\right)\,.
\ee
Using the above formulas one can easily perform the radial reduction of the ambient space derivative in radial coordinates. More in details:
{\small{
\begin{multline}
e^{{\e^A}_B(x)\,U^B\partial_{U^A}}\,\partial_M\,\Phi_\d(X,U)\,=\,(\partial_M-{\O_M^A}_B \, U^B\partial_{U^A})\Phi_E(X,U)\\
=\left[\s\,\hat{X}_M\,\partial_R\,+\,\tfrac{\partial x^\m}{\partial X^M}\,\left(\partial_\m-{\o_\m^\a}_\b\,u^b\partial_{u^\a}\right.\right.\\ +\left.\left.\tfrac{1}{L\sqrt{\s}}\,e_{\b\,\m}\,u^\b\partial_v\,-\tfrac{1}{L\sqrt{\s}}\,{e^\b}_\m\,v\,\partial_{u^\b} \right)\right]\,\Phi_E(R,x;v,u)\,,
\end{multline}}}
\!\!from which, factoring the (A)dS vielbein, one recovers
\be
\partial_M\ra \s\,\hat{X}_M(x)\,\partial_R\,+\,\tfrac{\partial x^\m}{\partial X^M}(R,x)\,{e^\a}_\m(x)\,\left[D_\a+\tfrac{1}{L\sqrt{\s}}\,\left(u_\a\partial_v\,-\,v\,\partial_{u^\a}\right)\right]\,.\label{derivative}
\ee
Analogously, recalling eq.~\eqref{constant frame vielbein}, one can obtain similar relations also for the auxiliary variables and the corresponding derivatives. In details the following relations
\begin{gather}
e^{{\e^A}_B(x)\,U^B\partial_{U^A}}\,\partial_{U^M}\,\Phi_\d(X,U)\,=\,{\left(e^{-\e}\right)^A}_M\,\partial_{U^A}\,\Phi_E(X,U)\,,\\
e^{{\e^A}_B(x)\,U^B\partial_{U^A}}\,U^M\,\Phi_\d(X,U)\,=\,{\left(e^{+\e}\right)_A}^M\,U^A\,\Phi_E(X,U)\,,
\end{gather}
imply
\begin{gather}
\partial_{U^M}\ra \tfrac{1}{\sqrt{\s}}\,\hat{X}_M\,\partial_v+\tfrac{R}{L}\,{e^\a}_\m(x)\tfrac{\partial x^\m}{\partial X^M}\,\partial_{u^\a}\,,\label{auxderivative}\\
U^M\ra \tfrac{1}{\sqrt{\s}}\,\hat{X}_M\,v+\tfrac{L}{R}\,{e_\a}^\m(x)\tfrac{\partial X^M}{\partial x^\m}\,u^\a\label{aux}\,.
\end{gather}
For convenience, it is also useful to write explicit expressions for the intrinsic form of the various differential operators acting on the ambient space generating functions of eq.~\eqref{ambient gen}. The relevant operators are the ambient space gradient \mt{U\cdot\partial_X}, the divergence \mt{\partial_{U}\cdot\partial_X}, the Laplacian \mt{\partial_X^{\,2}} and the trace operator \mt{\partial_U\cdot \partial_U} and their intrinsic form can be easily recovered from eqs.~\eqref{derivative}, \eqref{auxderivative} and \eqref{aux}:
{\small
\begin{eqnarray}\label{ambientoperator0}
U\cdot \partial_X\!\!\!&=&\!\!\!\tfrac{1}{\sqrt{\s}}\, v\,\partial_R\,+\,\tfrac{L}{R}\,\left[u\cdot e^\m\,D_{\m}+\tfrac{1}{L\sqrt{\s}}\,\left(u^2\partial_v-v\,u\cdot\partial_u\right)\right]\,,\\
\partial_U\cdot \partial_X\!\!\!&=&\!\!\!\tfrac{1}{\sqrt{\s}}\,\partial_v\partial_R+\tfrac{L}{R}\left[\partial_u\cdot e^\m \,D_\m+\tfrac{1}{L\sqrt{\s}}\,\left(d\,\partial_v\,+\,u\cdot\partial_u\,\partial_v-v\,\partial_u\cdot\partial_u\right)\right]\,,\nonumber\\
\partial_U\cdot\partial_U\!\!\!&=&\!\!\!\partial_v^{\,2}+\partial_u\cdot\partial_u\,,\nonumber
\end{eqnarray}}
where we have used:
\be
\begin{split}
\hat{X}^{M}\,\tfrac{\partial x^\m}{\partial X^M}\,=\,\partial_R \,x^\m\,&=\,0\,,\qquad \hat{X}_M \tfrac{\partial X^M}{\partial x^\m}\,=\,\tfrac{\s}{2R}\,\partial_{x^\m} R^2\,=\,0\,,\\
\tfrac{\partial X^M}{\partial x^\m}\,\tfrac{\partial x^\n}{\partial X^M}\,&=\,\d^\n_\m\,,\qquad \tfrac{\partial x^\m}{\partial X_M}\,\tfrac{\partial x^\n}{\partial X^M}\,=\,\tfrac{L^2}{R^2}\,g^{\,\m\n}(x)\,.\label{ident1}
\end{split}
\ee
The radial frame expression for \mt{\partial_X^{\,2}} requires a little more work and is given by:
\begin{multline}
\partial_X^{\,2}\,=\,\left\{\s\,\hat{X}_M(x)\,\partial_R\,+\,\tfrac{\partial x^\m}{\partial X^M}(R,x)\,{e^\a}_\m(x)\,\left[D_\a+\tfrac{1}{L\sqrt{\s}}\,\left(u_\a\partial_v\,-\,v\,\partial_{u^\a}\right)\right]\right\}^2\\
=\s\,\partial_R^{\,2}\,+\,\s\,\tfrac{d}{R}\,\partial_R\,+\,\tfrac{L^2}{R^2}\,\left[D^2+\tfrac{2}{L\,\sqrt{\s}}\, \left(\partial_v\,u\cdot D\,-\,v\,\partial_u\cdot D\right)\right.\\\left.+\tfrac{1}{L^2\s}\left(u^2\partial_v^2-2v\,\partial_v\,u\cdot\partial_u-D\,v\,\partial_v+v^2\partial_u\cdot\partial_u- u\cdot\partial_u\right)\vphantom{D^2+\tfrac{2}{L\,\sqrt{\s}}\, \left(\partial_v\,u\cdot D\,-\,v\,\partial_u\cdot D\right)}\right]\,,\label{box completo}
\end{multline}
where we have used the identities in eq.~\eqref{ident1} together with
\be
\tfrac{\partial x^\m}{\partial X^M}\, D_{\m}\, \hat{X}^M\,=\,\tfrac{\partial x^\m}{\partial X^M}\,\partial_{x^\m}\,\hat{X}^M\,=\, \tfrac{d}{R}\,,
\ee
while \mt{D^2} is the (A)dS Laplacian expressed in terms of the covariant derivative \eqref{covder}.
The above expressions simplify a bit if restricted to tangent generating functions. Hence, setting $\partial_v\,=\,0$, one recovers:
\begin{eqnarray}\label{ambientoperator}
U\cdot \partial_X\!\!&=&\!\!\tfrac{1}{\sqrt{\s}}\, v\,\partial_R\,+\,\tfrac{L}{R}\,\left[u\cdot e^\m\,D_{\m}+\tfrac{1}{L\sqrt{\s}}\,v\,u\cdot\partial_u\right]\,,\\
\partial_U\cdot \partial_X\!\!&=&\!\!\tfrac{L}{R}\left[\partial_u\cdot e^\m \,D_\m+\tfrac{1}{L\sqrt{\s}}\,v\,\partial_u\cdot\partial_u\right]\,,\nonumber\\
\partial_U\cdot\partial_U\!\!&=&\!\!\partial_u\cdot\partial_u\,,\nonumber\\
\partial_X^{\,2}\!\!&=&\!\!\s\,\partial_R^{\,2}\,+\,\s\,\tfrac{d}{R}\,\partial_R\nn
&&\hspace{20pt}+\,\tfrac{L^2}{R^2}\,\left[D^2- \tfrac{2}{L\,\sqrt{\s}}\,v\,\partial_u\cdot D+\tfrac{1}{L^2\s}\left(v^2\partial_u\cdot\partial_u- u\cdot\partial_u\right)\vphantom{\square_{\text{(A)dS}}+\tfrac{2}{L\,\sqrt{\s}}\, \left(\partial_v\,u\cdot D\,-\,v\,\partial_u\cdot D\right)}\right]\,.\nonumber
\end{eqnarray}

In the next section we shall apply the formalism so far developed to the description of unitary spin-$s$ representations on constant curvature backgrounds.


\section{(A)dS Dynamics}\label{(A)dS Dynamics}


In this section we introduce the ambient space description of the standard totally-symmetric spin-$s$ representations of the isometry groups of any constant curvature background. It is well known from the old works of Fierz, Pauli, Bargmann, and Wigner \cite{Fierz:1939zz,Fierz:1939ix,Bargmann:1946me} that the various representations of the corresponding isometry groups of the maximally symmetric spaces can be specified by proper wave equations together with some irreducibility constraints.
In flat space these wave equations were discovered for bosonic fields by Fierz \cite{Fierz:1939zz} and are respectively
\be
(\square-m_s^2) \,\varphi_{\m_1\ldots \m_s}(x)\,=\,0\,,\qquad \partial^{\m_1}\varphi_{\m_1\ldots\m_s}(x)\,=\,0\,,\qquad {\varphi^{\m}}_{\m\m_3\ldots\m_s}(x)\,=\,0\,.
\ee
The first equation assigns the value of the Casimir operator $p^{\,2}$, the second equation takes care of the translational part of the Poincar\'e group, removing the negative-norm states, while the third equation is related to irreducibility of the representation.
The massless limit $m_s=0$ of this system presents some subtleties and indeed in order to keep a tensorial helicity representation of the Lorentz group one needs to introduce a gauge symmetry quotienting by the following equivalence relation
\be
\varphi_{\m_1\ldots \m_s}(x)\sim \varphi_{\m_1\ldots\m_s}(x)+\partial_{(\m_1}\varepsilon_{\m_2\ldots\m_s)}(x)\,,
\ee
where the gauge parameter $\varepsilon_{\m_2\ldots\m_s}(x)$ satisfies by consistency an analogous Fierz system:
\be
\square \,\varepsilon_{\m_1\ldots \m_{s-1}}(x)\,=\,0\,,\qquad \partial^{\m_1}\varepsilon_{\m_1\ldots\m_{s-1}}(x)\,=\,0\,,\qquad {\varepsilon^{\m}}_{\m\m_3\ldots\m_{s-1}}(x)\,=\,0\,.
\ee
In terms of generating functions the above equations become
\be
(\square-M^2) \,\varphi(x,u)\,=\,0\,,\qquad \partial_u\cdot\partial_x\,\varphi(x,u)\,=\,0\,,\qquad \partial_u\cdot \partial_u\,{\varphi}(x,u)\,=\,0\,,
\ee
where $M$ is here a mass operator, while in the massless case $M\,=\,0$ the equivalence relation can be written as
\be
\varphi(x,u)\sim\varphi(x,u)\,+\,u\cdot\partial_x\,\varepsilon(x,u)\,,
\ee
together with the gauge parameter Fierz system
\be
\square \,\varepsilon(x,u)\,=\,0\,,\qquad \partial_u\cdot\partial_x\,\varepsilon(x,u)\,=\,0\,,\qquad \partial_u\cdot \partial_u\,{\varepsilon}(x,u)\,=\,0\,.
\ee
These equations can be easily generalized on constant curvature backgrounds\footnote{One can perform a local Lorentz transformation on the Fierz system going to a generic moving frame and then extend the system to a generic curved background exploiting covariance.} where they look like
\be
(D^2-M^2) \,\varphi(x,u)\,=\,0\,,\qquad \partial_u\cdot e^\m D_\m\,\varphi(x,u)\,=\,0\,,\qquad \partial_u\cdot \partial_u\,{\varphi}(x,u)\,=\,0\,,\label{(A)dS Fierz}
\ee
while in the massless case the equivalence relation becomes
\be
\varphi(x,u)\sim\varphi(x,u)\,+\,u\cdot e^\m D_\m\,\varepsilon(x,u)\,,
\ee
again together with the corresponding Fierz system for the gauge parameters\footnote{Notice that in (A)dS the massless case does not correspond to $M=0$ but to a particular non-vanishing value of the mass.}
\be
(D^2-\tilde{M}^2)\,\varepsilon(x,u)\,=\,0\,,\qquad \partial_u\cdot e^\m D_\m\,\varepsilon(x,u)\,=\,0\,,\qquad \partial_u\cdot \partial_u\,{\varepsilon}(x,u)\,=\,0\,.
\ee
The key feature of the ambient space techniques that we have introduced in Section~\ref{(A)dS Geometry} is that thanks to the isomorphism between intrinsic tensors and homogeneous and tangent ambient space tensors one can rewrite the (A)dS Fierz system in terms of ambient space quantities making it possible to simplify all problems coming from the non commuting nature of covariant derivatives. Indeed, from eq.~\eqref{ambientoperator} one recovers that the ambient space Fierz system
\be
\square\,\Phi(X,U)\,=\,0\,,\qquad \partial_U\cdot\partial_X\,\Phi(X,U)\,=\,0\,,\qquad\partial_U\cdot\partial_U\,\Phi(X,U)\,=\,0\,,\label{ambient Fierz}
\ee
supplemented with the homogeneity and tangentiality constraints
\be
(X\cdot\partial_X-U\cdot\partial_U+2+\m)\,\Phi(X,U)\,=\,0\,,\qquad X\cdot\partial_U\,\Phi(X,U)\,=\,0\,,\label{tangen homo field}
\ee
reduces to the intrinsic Fierz system \eqref{(A)dS Fierz} with a mass squared operator
\be
M^2\,=\,-\,\tfrac{1}{L^2\,\s}\,\left[(\m-u\cdot\partial_u+2)(\m-u\cdot\partial_u-d+3)-u\cdot\partial_u\right]\,,\label{massAdS}
\ee
where we have used
\be
	\Phi(R,x;v,u)= \left(\tfrac RL\right)^{u\cdot\partial_{u}-2-\mu}\,\varphi(x,u)\,.
	\label{doh}
\ee
Notice that for dS, where $\s=1$\,, the parameter $\mu$ is in general a complex number, hence, in order for the fields to be
real one has to add the complex conjugate in eq.~\eqref{doh}. It can be interesting to notice that the trace constraint present within the Fierz system of eq.~\eqref{ambient Fierz} turns to be redundant whenever the tangential constraint is considered just because on transverse fields one gains the following identity:
\be
\partial_U\cdot\partial_U\,\Phi(X,U)\,=\,\partial_U\cdot\partial_X\,X\cdot\partial_U\,\Phi(X,U)\,=\,0\,.
\ee
Hence, the description of reducible fields in (A)dS turns to be more involved than its flat-space counterpart where the trace and the divergence of the field are completely independent.

In the following we are going to study in detail the gauge symmetries that can be preserved for the various unitary representations of the corresponding isometry group.

\subsection{Massless representations}

For that regards massless representations of the (A)dS isometry group one can find them working in the ambient-space formalism simply requiring compatibility of the Fierz system \eqref{ambient Fierz} and of the homogeneity and tangential constraints with gauge transformations of the form
\be
\delta_E^{(0)}\Phi(X,U)\,=\,U\cdot\partial_X\,E(X,U)\,.\label{masslesstransf}
\ee
Here, by consistency with the homogeneity constraint on the fields, the gauge parameter satisfies the following homogeneity constraint
\be
(X\cdot\partial_X-U\cdot\partial_U+\m)E(X,U)\,=\,0\,.\label{homo gauge}
\ee
Even if the ambient Fierz system is compatible with such gauge symmetries provided the gauge parameter satisfies the same Fierz system
\be
\square\,E(X,U)\,=\,0\,,\qquad \partial_U\cdot\partial_X\,E(X,U)\,=\,0\,,\qquad\partial_U\cdot\partial_U\,E(X,U)\,=\,0\,,
\ee
regardless the value of $\m$, arbitrary values of $\m$ are in general incompatible with the tangentiality constraint, as can be easily seen from
\be
X\cdot\partial_U\,U\cdot\partial_X\,E(X,U)\,=\,\left(U\cdot\partial_X\,X\cdot\partial_U-\m\right)\,E(X,U)\,.\label{compcond}
\ee
From this respect it is easy to notice that the choice $\m=0$ if combined with the algebraic constraint
\be
X\cdot\partial_U\,E(X,U)\,=\,0\,,\label{tanggauge}
\ee
makes the gauge transformations compatible with the tangentiality of the gauge field recovering in this way the standard massless representations. Indeed, with this choice the ambient gauge symmetry \eqref{masslesstransf} reduces to the usual massless intrinsic gauge symmetry
\be
\delta\varphi(x,u)\,=\,u\cdot e^\m D_\m\,\varepsilon(x,u)\,.
\ee
Plugging $\m=0$ into eq.~\eqref{massAdS} one then finds with ambient space techniques the value of the (A)dS mass corresponding to massless representations of the isometry group that in terms of the spin reads
\be
M^2\,=\,-\,\tfrac{1}{L^2\,\s}\,\left[(s-2)(s-3+d)-s\right]\,,
\ee
and agrees with standard results.
Actually, in the massless case, the ambient space construction so far consider is nothing but equivalent to ask for the intrinsic expression of $U\cdot\partial_X$ in eq.~\eqref{ambientoperator0} to reduce to $u\cdot e^\m\, D_\m$ when acting on gauge parameters of the form
\be
E(R,x;v,u)\,=\,\left(\tfrac{R}{L}\right)^{u\cdot\partial_u-\m}\,\varepsilon(x;v,u)\,,
\ee
that solves the corresponding homogeneity constraints \eqref{homo gauge}. Notice however that the ambient space construction does not require the intrinsic form of the various differential operators like $U\cdot\partial_X$, while all standard properties of the spin-$s$ representations follow from their ambient realization.

\subsection{Massive representations}

On the contrary in the cases $\m\neq 0$ the same compatibility condition \eqref{compcond} translates into the differential like constraint
\be
\left(U\cdot\partial_X\,X\cdot\partial_U-\m\right)\,E(X,U)\,=\,0\,,
\ee
for the gauge parameter that in turn is compatible with tangentiality provided
\be
\big[\,(U\cdot\partial_X)^2\, (X\cdot\partial_U)^2-2\,\mu\,(\mu-1)\,\big]\,E(X,U)=0\,.
\ee
Again if $\m\neq 1$ one can go on and in general if
\be
[\m]_s\,:=\,\m(\m-1)\ldots(\m-s+1)\neq0\,,
\ee
one can iterate $s$ times this procedure ending up with
\be
\left\{(U\cdot\partial_X)^s(X\cdot\partial_U)^s-s! \,[\m]_s\right\}\,E(X,U)\,=\,0\,.\label{itercompat}
\ee
Hence, since the spin-$(s-1)$ component $E^{(s-1)}(X,U)$ of the gauge parameter generating function $E(X,U)$ trivially satisfies
\be
(X\cdot\partial_U)^{s}\,E^{(s-1)}(X,U)\,=\,0\,,
\ee
whenever $[\m]_s\neq 0$ the spin-$(s-1)$ component of eq.~\eqref{itercompat} implies that the gauge parameter $E^{(s-1)}(X,U)$ has to vanish identically and hence no gauge symmetry can be preserved compatibly with the tangentiality constraint. In this way, even though the ambient space Fierz system is exactly a massless system in ambient space, for generic values of $\m$ no gauge symmetry can be preserved compatibly with the tangentiality constraint.

\subsection{Partially-massless representations}

Continuing the analysis of the previous section, it is interesting to notice that the integers values of $\m$ have a special role since in those cases the iteration that led to eq.~\eqref{itercompat} stops before implying the vanishing of the gauge parameter. Indeed, if $\m=r\in \mathbb{N}$, a non-vanishing solution for the gauge parameter exists for $s > r$ such that
\be
(X\cdot\partial_U)^r\,E(X,U)\,=\,0\,.
\ee
In this case it is convenient to introduce an auxiliary tangent gauge parameter $\O(X,U)$ defined by
\be
\O(X,U)\,=\,(X\cdot\partial_U)^r E(X,U)\,,
\ee
in terms of which the gauge transformations read
\be
\delta_\O^{(0)}\Phi(X,U)\,=\,(U\cdot\partial_X)^{r+1}\O(X,U)\,.
\ee

To summarize, since for massless fields one would like to preserve a gauge transformation of the form \eqref{masslesstransf} with a gauge parameter at most subjected to algebraic constraints, the corresponding massless representations are recovered for $\m\,=\,0$ and for tangent and homogenous gauge parameters.
On the other hand, more general solutions to the tangential compatibility are given by partially-massless representations \cite{Deser:1983mm} corresponding to non-zero integer values of $\m$. These are however unitary only in dS, while for other values of $\m$ one recovers a standard massive representation \footnote{Notice also that for $\m\neq 0$ one can preserve the usual form of gauge transformations dropping the tangentiality constraint (see e.g. \cite{Joung:2012rv} for more details). This is equivalent to perform a Stueckelberg shift introducing additional gauge symmetries together with the corresponding Stueckelberg fields playing the role of auxiliary fields. This is actually a different definition of the Fronsdal isomorphism built in terms of the equivalence classes $\Phi(X,U)\sim\Phi(X,U)+X\cdot U\,\tilde{E}(X,U)$. In this section however we were interested in the unitary gauge in which we keep only the minimum gauge invariance sufficient to maintain covariance on-shell. Let us mention that for $\m\neq 0$ in order to write a local quadratic Lagrangian it is needed to introduce a number of Stueckelberg fields playing the role of auxiliary fields in (A)dS, dropping in general the tangentiality constraint.}. In the following we shall concentrate on massless fields for which these subtleties are absent referring to \cite{Joung:2012rv} for more details.

Before concluding this section let us study the flat limit from the ambient space perspective. This construction can be interesting because puts on very similar grounds all constant curvature backgrounds and also because gives the possibility to recover the flat vertices starting from the (A)dS ones. From our perspective the flat-space limit $L\ra \infty$ can be still realized at the ambient space level. In order to achieve this, first one needs to place the origin of the ambient space in a point on the hypersurface $X^2\,=\,\s L^2$. This can be done translating the coordinate system as
\be
X^M\ \ra \ X^M+L \hat{N}^M\,,\label{flat limit}
\ee
where $N^M$ is a constant vector in the ambient space that satisfies
\be
\hat{N}^2\,=\,\s\,.
\ee
Then, in the flat limit the homogeneity and tangent constraints become
\be
\left(\hat{N}\cdot\partial_X-\sqrt{-\s}\,M\right)\,\Phi(X,U)\,=\,0\qquad \hat{N}\cdot\partial_U\,\Phi(X,U)\,=\,0\,,
\ee
respectively, where we have considered $\m\ra\infty$ and $L\ra \infty$, keeping $\tfrac{\m}{L}=M$ finite, while the Fierz system stays the same as in eq.~\eqref{ambient Fierz}. It is worth mentioning that, has we have anticipated, in the flat limit the traceless and transverse constraints are no more redundant. Hence, one has now the possibility to describe massless reducible representations just dropping the traceless constraint at the level of the Fierz system. This feature is quite interesting and resonates with the subtleties that arise when describing reducible representations in (A)dS space in relation to their flat space counterparts \cite{Metsaev:1995re}. From our perspective these subtleties acquire a natural interpretation related to the fact that in order to describe a massless representation one has to choose a certain degree of homogeneity for the field in relation to its spin. Therefore, a reducible representation, containing different spin components, would require different degrees of homogeneity that are not mutually compatible each other. We leave for future work a more detailed analysis of these features in our approach.

\section{Ambient space measure}\label{Ambient measure}

So far we have described the various spin-$s$ representations of the constant curvature backgrounds from the ambient space perspective. What is left in order to complete the discussion is to define the concept of ambient space Action principle. The simplest possibility is to consider a similar setting to that of the standard dimensional reduction framework in which one of the dimensions is compactified on a circle. More in detail, this attempt would correspond to a tentative Action of the form
\be
S\,=\,\int d^{d+1}X\ \cL[\Phi]\,,
\ee
Unfortunately, this kind of ansatz turns to work well at the quadratic level but has some problems as soon as one considers interactions. Indeed the action in this way is not truly $d$-dimensional because is written in terms of a $(d+1)$-dimensional measure. More precisely the radial integration can be a problem at the interacting level due to the appearance of infinities related to the fixed degree of homogeneities of the fields. Indeed, the ambient-space integral splits into the (A)dS one together with an additional radial
integral as
\be
	\int \frac{d^{d+1}X} L
	=\int_{0}^{\infty} \tfrac{d R}L \left(\tfrac{R}{L}\right)^{d}\,\int_{{\text (A)dS}} d^{d}x\sqrt{-g}\,.
\ee
In order to overcome this difficulties we have to keep in mind that ambient space techniques are just a translation of intrinsic (A)dS quantities in terms of ambient space ones, while the tangentiality and homogeneity constraints that one introduces at the level of the ambient space have the role of keeping the description effectively $d$-dimensional. This suggest to simply rewrite the $d$-dimensional intrinsic measure of (A)dS space in terms of ambient space quantities so that one is naturally led to introduce a $\d$-function uplifting the intrinsic (A)dS measure as\footnote{Let us stress at this point that since we are working in the truncated ambient space $\s X^{2}>0$ we could have chosen as well the measure
\be
\d^{d+1} X\ \d\left(\s X^{2}-L^{2}\right)\,,
\ee
simplifying some of the following formulas. We have decided in order to avoid some confusion to keep the same convention of \cite{Joung:2011ww,Joung:2012rv}.}
\be
d^{d}x\,\sqrt{-\s g}\,=\,d^{d+1}X\,\delta\left(\sqrt{\s X^{2}}-L\right)\,,
\ee
Moreover, this formalism makes it possible to extend the discussion about the flat limit considered above also at the level of the ambient space measure so that performing the change of coordinates of eq.~\eqref{flat limit} and taking the $L\ra \infty$ limit one ends up with the ambient space measure
\be
d^{d+1}X\,\s\ \delta\left(\hat{N}\cdot X\right)\,,
\ee
that is just a $d$-dimensional measure on the flat hyperplane orthogonal to the unit vector $\hat{N}$.

Having defined the ambient space (A)dS measure we can now start constructing Lagrangians. Let us mention that the effect of the $\d$-function insertion will take care of all subtleties that arise in in constant curvature backgrounds as for instance the loss of translation invariance at the ambient space level, together with its implications as, for instance, the fact that total total derivatives are not any more vanishing.

This is actually the counterpart of a well known fact when lifting to curved backgrounds flat space quantities. Indeed the presence of the $\d$-function makes clear that the (A)dS deformation of a total derivative is not any more in general a total derivative in (A)dS. Indeed, even if
\be
\int d^{d+1}X\,\partial_{X^M}\left(\ldots\right)\,=\,0\,,
\ee
this is not true after inserting the $\d$-function so that
\begin{multline}
\int d^{d+1}X\,\delta\left(\sqrt{\s X^{2}}-L\right) \partial_{X^M}\left(\ldots\right)\\=\,-\int d^{d+1}X\,\delta^{\prime}\left(\sqrt{\s X^{2}}-L\right) \tfrac{\s}{L}\,{X^M}\left(\ldots\right)\,\neq \,0\,,
\end{multline}
where by definition
\be
\delta^{\,\prime}(R-L)\,:=\,\tfrac{L}{R}\,\partial_{R}\, \delta(R-L)\,.
\ee
This is actually the same feature that presents itself from the intrinsic point of view when substituting ordinary derivatives with covariant ones so that the deformation of a flat total derivative will not give rise in general to a pure total derivative in (A)dS but will give rise to a total derivative plus lower derivative contributions that are not total derivatives and that in the ambient space formalism come from the action of $\partial_{X}$ on the $\d$-function\footnote{Consider for instance the very simple intrinsic flat total derivative
\be
(\partial_{\n}\,\partial_{\m}\varphi)\,\varphi\,+\,\partial_\n\varphi\,\partial_\m\varphi\,.
\ee
This can be lifted as an example as
\be
(D_\n D_\m\varphi)\,\varphi\,+\,D_\n\varphi\,D_\m\varphi\,,
\ee
whose $2$ derivative piece is still a total derivative. However, the (A)dS deformation introduces a piece with no derivative that hence is not a total derivative.}. (A)dS total derivatives can be instead easily written in our approach as
\be
\int d^{d+1}X\ \partial_{X^M}\left[\delta\left(\sqrt{\s X^{2}}-L\right) \left(\ldots\right)\right]\,,
\ee
avoiding many of the difficulties that can arise when working with the non-commuting covariant derivatives.
In order to simplify the formalism it is convenient to introduce a further auxiliary variable $\hat{\d}$ in terms of which one can encode all derivatives of the $\delta$-function as
\be
\delta(R-L)\,(\s\,\hat{\d})^n\,=\,\delta^{(n)}(R-L)\,,
\ee
where by definition
\be
\delta^{(n)}(R-L)\,=\,[\tfrac{L}{R}\,{\partial_{R}}]^n\,\delta(R-L)\,.
\ee
With this prescription one can encode any distribution that is a sum of derivatives of the delta function into a function of the auxiliary variable $\hat{\d}$ as
\be
\sum_{n=0}^\infty a_n\,\delta^{(n)}(R-L):=\,\delta(R-L)\,a(\s\,\hat{\d})\,,\qquad a(x)\,=\,\sum_{n=0}^\infty a_n\, x^n\,.
\ee
Let us mention that the $\hat{\d}$ prescription is also a formal device in order to encode in a single function different portions scaling with different powers of $\tfrac{1}{L}$. This feature will actually prove to be useful in (A)dS where in general any Lagrangian contains a tail of lower derivative contributions scaling differently in terms of the cosmological constant. Indeed, the observation is very simple and is based on the identity
\begin{multline}
\delta^{(n)}(R-L)\,R^\l\,=\,
\tfrac{1}{2}\,(2 L)^n\,\partial_{R^\prime}^n\,\delta(\sqrt{R^\prime}-L)\,(R^\prime)^{(\l-1)/2}\\=\,\tfrac{1}{2}\,(-2 L)^n\delta(\sqrt{R^\prime}-L)\partial_{R^\prime}^n(R^\prime)^{(\l-1)/2}\,=\,\tfrac{(-2)^n\,[(\l-1)/2]_n}{L^{n}}\,\delta(R-L)\, R^\l\,,\label{hat delta}
\end{multline}
where we have used the change of variable $R^\prime\,=\,R^2$ so that negative powers of $L$ can be absorbed into derivatives of the $\d$-function and hence into $\hat{\d}$.
Similar considerations applies also to the flat limit where the auxiliary variable $\hat{\d}$ can be simply replaced by the differential operator
\be
\hat{\d}\,\ra\,\hat{N}\cdot \partial_X\,.
\ee
In the following for simplicity we shall use the short-hand notation
\be
d^{d+1}X\,\delta\left(\sqrt{\s X^{2}}-L\right)\,:=\,d^{d+1}X\,\delta\,,
\ee
without explicitly writing the argument of the $\d$-function when no ambiguity is present while, in order to avoid any confusion, we shall use the notation $\d_E$, with the subscript referring to the gauge parameter, when considering the gauge transformations.
Having settled the formalism we can start applying it in order to find order by order Lagrangians for HS totally symmetric fields. In the following sections we shall first describe the Noether procedure from a general perspective in order to address the problem in the subsequent chapters.

\section{Noether Procedure}\label{Noether Procedure}

Noether procedure can be considered as one of the key techniques in order to construct consistent Lagrangians for gauge theories and in its various incarnations has played a crucial role in order to solve for cubic HS couplings \cite{Berends:1979kg,Berends:1984rq,Berends:1985xx,Boulanger:2006gr,Boulanger:2008tg,Boulanger:2011se}.
From this point of view the HS problem can be reformulated as equivalent to finding, order by order in the number of fields, a deformation of the free system.
Actually this approach is quite general and can be applied even at the quadratic level promoting the on-shell gauge symmetry introduced in Section~\ref{(A)dS Dynamics} to genuine off-shell ones. More in detail one starts expanding the fully non-linear ambient-space action order by order in the number of fields as
\be
S\,=\,\int d^{d+1}X\,\d\,\left[\cL^{(2)}\,+\,\cL^{(3)}\,+\,\cL^{(4)}\,+\,\ldots\right]\,,
\ee
where $\cL^{(2)}$ is quadratic in the fields, $\cL^{(3)}$ is cubic and so on. The same expansion is considered at the level of the gauge algebra transformations as
\be
\delta_E\Phi(X,U)\,=\,\delta_E^{(0)}\Phi(X,U)\,+\,\delta_E^{(1)}\Phi(X,U)\,+\,\delta_E^{(2)}\Phi(X,U)\,+\,\ldots
\ee
where $\delta_E^{(0)}$ has the same form of the on-shell gauge symmetry recovered\footnote{Notice that, as we have mentioned, dropping the tangentiality constraint one can end up with a gauge invariant formulation of massive fields that is usually called Stueckelberg formulation. In this way it is possible in principle to apply the Noether procedure also in those more general cases.} in Section~\eqref{(A)dS Dynamics}, while both the Lagrangian and the gauge transformations are defined modulo local redefinitions of the fields and of the gauge parameters of the form
\be
\begin{split}
\Phi(X,U)\,&\ra\,\Phi(X,U)\,+\,f^{(1)}(\Phi)\,+\,f^{(2)}(\Phi)\,+\,\ldots\,,\\
E(X,U)\,&\ra\,E(X,U)\,+\,g^{(1)}(\Phi,E)\,+\,g^{(2)}(\Phi,E)\,+\,\ldots\,.
\end{split}
\ee
Let us recall for completeness that at the level of the gauge algebra it is sufficient to limit the attention to the linearized order in the gauge parameter. After this perturbative expansion also the condition that the action be gauge invariant splits analogously ending up with the following infinite set of consistency conditions:
\be
\begin{split}\label{Noether equations}
  \d^{(0)}_E\,\cL^{(2)}\,&=\,0\,,\\
  \d^{(1)}_E\,\cL^{(2)}\,+\,\d^{(0)}_E\,\cL^{(3)}\,&=\,0\,,\\
  \d^{(2)}_E\,\cL^{(2)}\,+\,\d^{(1)}_E\,\cL^{(2)}\,+\,\d^{(0)}_E\,\cL^{(4)}\,&=\,0\,,\\
  \ldots
\end{split}
\ee
where for brevity we have dropped the integral sign and the measure. The above equations can be solved order by order iteratively and under some further assumptions give as solutions consistent gauge theories at each order. The general strategy goes as follows:
\begin{enumerate}
  \item Solve for the quadratic part of the Lagrangian\footnote{One has to fix what kind of wave equations have to be reproduced by the free action. In other words one has to decide what kind of representations of the isometry group to consider.} (Free Theory),
  \item Having fixed the quadratic part any $\d^{(1)}_E\cL^{(2)}$ is just proportional to the free EoMs regardless the precise form of $\d^{(1)}$ and hence one can first solve the simplest equation
      \be
      \delta^{(0)}_E\,\cL^{(3)}\,\approx\,0\,,\label{cubic}
      \ee
      where henceforth $\approx$ means modulo the free EoMs.
  \item Having the solution to eq.~\eqref{cubic} one can now solve for $\d^{(1)}$ and go to the next equation in the list.
  \item Let us now suppose that we have solved the first $n$ equations completely. This means that we know the form of $\cL^{(i)}$ and of $\d^{(i-2)}$ for any $i\leq n+1$
  \item We can now solve the $(n+1)$-th equation first solving for $\cL^{(n+2)}$ modulo the free EoMs and then extracting the corresponding deformation of the gauge transformation $\d^{(n)}$.
\end{enumerate}
Before going on with our discussion it is worth stressing that the set of equations so far considered is kind of trivial if no other assumption is made on the structure of the interactions, as was first observed in \cite{Barnich:1993vg} when discussing the role of locality in relation to the Noether procedure. The key point is from this respect the fact that one would like to recover unitary theories and so it would be important to have a criterium to check the possible violations of Unitarity. This is not a mystery at the quadratic level, where unitarity of the theory is related to the representation theory of the corresponding isometry group and manifests itself in terms of the sign of the kinetic terms. On the other hand, at the interacting level and in particular from the quartic order on, as it will be clearer in the following, the usual criteria that has been used is \emph{Lagrangian locality}. However, the necessity of the latter is debatable and it should be considered just as a \emph{sufficient} condition in order to ensure tree-level unitarity, and hence, at least the classical consistency of the theory. From this point of view it is worth mentioning that for long time ST and also Vasiliev's system or more recently the AdS/CFT correspondence have put in question the \emph{necessity} of requiring a local Lagrangian description. The only local commutation relations have to hold at the level of the observable quantities that, as more gauge symmetries one adds to the system, tend to lose completely their local nature as for instance happens already in General Relativity. In the following, having in mind this discussion we are going to relax the standard locality hypothesis at the Lagrangian level with the aim of understanding its implications in relation to the systematics of HS interactions. Further constraints can come for instance from a detailed study of the global symmetries of the system, whose structure is already very constraining at the cubic level, or even from the AdS/CFT correspondence, from which one can hopefully extract the right consistency criteria. We leave this important analysis for the future.

Coming back to the Noether procedure, the structure of the above equations can be further simplified exploiting generating function techniques in the ambient space formalism. First of all let us notice from this respect that by considering all possible forms of ambient-space Lagrangian vertices one does not loose any generality since any (A)dS Lagrangian vertex can be written as an ambient-space one making use of the projector\footnote{See e.g. \cite{Bekaert:2010hk} for more details}:
\be
P^{MN}\,=\,\frac{X^M X^N-\eta^{MN}\,X^2}{X^2}\,.
\ee
However, as suggested by the projector itself the possible ambient-space Lagrangian couplings can be more general than the flat-space ones allowing a non trivial $X$-dependence. This can be also understood as a result of the break down of translational symmetry at the level of the ambient space induced by the $\d$-function insertion.
All that being said, any Lagrangian coupling involving a given number $n\geq 2$ of fields can be rewritten in terms of the corresponding ambient-space generating function as
\begin{multline}
\cL^{(n)}\,=\,\tfrac{1}{n!}\ C^{(n)}_{a_1 \,\ldots\,a_n}\left(\hat{\d};X ; \partial_{X_1},\ldots,\partial_{X_n};U_1,\ldots,U_n\right)\\\star_{1\ldots n} \Phi^{a_1}(X_1,U_1)\ldots \Phi^{a_n}(X_n,U_n)\Big|_{X_i\,=\,X}\,,\label{coupling genfunc}
\end{multline}
where for completeness we have also explicitly introduced the color indices $a_i$ associated to the Chan-Paton factors, while the subscript on the various partial derivatives is meant to specify on which generating function of the fields the derivatives act upon. Here we have also introduced the inner-product $\star_i$ that we shall refer to as $\star$-contraction between generating functions
\be
\star_i: \left(\Phi(U_i),\Psi(U_i)\right)\ra \Phi\star_i\Psi\,=\,e^{\partial_{U_i}\cdot\partial_{U_i^{\prime}}}\,\Phi(U_i)\,\Psi(U_i^\prime)\Big|_{U_i=0}\,,\label{star}
\ee
where the subscript $i$ is meant to specify what sets of auxiliary variables are contracted together especially whenever there is more then one choice\footnote{In our case for instance one of the generating functions depends on more then one auxiliary variable.}. Let us mentioned that one can explicitly perform the $\star$-contraction with the formal substitution
\be
U_i\ra \partial_{U_i}\,,
\ee
and depending on the situation in the following it will prove convenient to work with the operatorial version of eq.~\eqref{coupling genfunc} that can be presented as
\begin{multline}
\cL^{(n)}\,=\,\tfrac{1}{n!}\ C^{(n)}_{a_1 \,\ldots\,a_n}\left(\hat{\d};X ;\partial_{X_1},\ldots,\partial_{X_n};\partial_{U_1},\ldots,\partial_{U_n}\right)\\\times \Phi^{a_1}(X_1,U_1)\ldots \Phi^{a_n}(X_n,U_n)\Big|_{_{X_i=X}^{U_i=0}}\,,\label{coupling genfunc operator}
\end{multline}
Furthermore, if one restricts the attention to parity-even Lorentz-invariant couplings, one can consider functions of the following Lorentz-invariant building blocks
\be
X^2\,,\qquad X\cdot\partial_{X_i}\,,\qquad \partial_{X_i}\cdot\partial_{X_j}\,,\qquad X\cdot{U_i}\,,\qquad U_i\cdot\partial_{X_j}\,,\qquad U_i\cdot U_j\,,
\ee
modulo integration by parts, where by convention we have chosen the ordering prescription in which all $X$ dependent quantities are placed on the left with respect to the derivatives. However, even though we can have Lagrangian generating functions with an explicit $X$ dependence this is indeed redundant and can be in fact neglected:
\begin{itemize}
\item
First, $X^{2}$ simply becomes $L^{2}$ after the radial integration,
and can be absorbed into the definition of $C_{a_{1}a_{2}a_{3}}$\,.
\item
Second, $X\cdot \partial_{X_{i}}$ is equivalent to $X_{i}\cdot\partial_{X_{i}}$ which essentially counts the number of $X_{i}$'s and so can be absorbed into $C_{a_{1}a_{2}a_{3}}$ as well.
\item
Finally, $X\cdot {U_{i}}$ is equivalent to $X_{i}\cdot{U_{i}}$ which is nothing but the tangent condition \eqref{tangent} after the $\star$-contraction. Hence, when it acts directly on the fields it vanishes, while
acting on the derivatives produces
\ba
	&& (X\cdot \partial_{U})\ \partial_{X^{\sst M_{1}}}\,\cdots\,\partial_{X^{\sst M_{n}}}\,\Phi \\
	&& \quad =-
	\sum_{m=1}^{n}\,\partial_{X^{\sst M_{1}}}\,\cdots\,\partial_{X^{\sst M_{m-1}}}\,
	\partial_{U^{\sst M_{m}}}\,
	\partial_{X^{\sst M_{m+1}}}\,\cdots\,\partial_{X^{\sst M_{m-1}}}\,\Phi\,,\nonumber
\ea
so that $X\cdot {U_{i}}$ is equivalent to a linear combination of the other Lorentz invariants.
\end{itemize}
At the end one can write without loss of generality the Lagrangian generating function of eq.~\eqref{coupling genfunc} as
\begin{multline}
\cL^{(n)}\,=\,\tfrac{1}{n!}\ C^{(n)}_{a_1 \,\ldots\,a_n}\left(\hat{\d};\,\partial_{X_i}\cdot\partial_{X_j};\,{U_i}\cdot\partial_{X_j};\,{U_i}\cdot{U_j}\right)\\\star_{1\ldots n} \Phi^{a_1}(X_1,U_1)\ldots \Phi^{a_n}(X_n,U_n)\Big|_{X_i\,=\,X}\,.\label{coupling genfunc 2}
\end{multline}
For the later convenience let us also introduce the color ordered part of the interactions that will play a key role in Chapter~4. Indeed one can explicitly introduce Chan-Paton factors $T^{a_i}$ as in ST carrying the color indices $a_i$ so that one can rewrite the Lagrangian couplings in the form
\begin{multline}
\cL^{(n)}\,=\,\tfrac{1}{n!}\ \sum_{\s_{n-1}}\ C^{(n)}_{0\s(1)\ldots\s(n)}\left(\hat{\d};\,\partial_{X_i}\cdot\partial_{X_j};\,{U_i}\cdot\partial_{X_j};\,{U_i}\cdot{U_j}\right)\\\star_{01\ldots n} \text{Tr} \left[\Phi_0(X_0,U_0)\,\Phi_{\s(1)}(X_{\s(1)},U_{\s(1)})\ldots \Phi_{\s(n)}(X_{\s(n)},U_{\s(n)})\right]\Big|_{X_i\,=\,X}\,,\label{coupling genfunc 3}
\end{multline}
where the fields have been redefined as
\be
\Phi\,=\,\Phi^{a_1}\,T^{a_1}\,,
\ee
the trace is over the Chan-Paton factors, the sum is over all permutations of $n-1$ elements and we have defined the cyclic color ordered portions of the generating functions $C^{(n)}_{01\ldots n}$.
Let us stress that we have defined this \emph{color-ordered} representation of the generating functions because it is intimately related to what one can call \emph{open-string}-like couplings having a natural planar structure. In Chapter~\ref{sec:four-point} we will also discuss other types of couplings that one can call similarly of the \emph{closed-string}-type in order to underline how string results actually reflect interesting field theory properties that are still to be completely understood.

All Noether procedure equations \eqref{Noether equations} translate into differential conditions for the generating functions $C^{(n)}_{a_1\ldots a_n}$.
For instance the equations
\be
\delta_{E_i}^{(0)}\,\cL^{(3)}\,\approx 0\,,
\ee
where the gauge variation is taken with respect to the field $\Phi^{a_i}(X_i,U_i)$, imply in the massless case the linear homogeneous first order differential equation
\be
\delta\ \left[\partial_{U_i}\cdot\partial_{X_i}\,C^{(3)}_{a_1 \,a_2\,a_3}\left(\hat{\d};\,\partial_{X_j}\cdot\partial_{X_k};\,{U_j}\cdot\partial_{X_k};\,{U_j}\cdot{U_k}\right)\right]\,\approx\,0\,,\label{diff eq}
\ee
or equivalently the commutator equation
\be
\delta\ \left[C^{(3)}_{a_1 \,a_2\,a_3}\left(\hat{\d};\,\partial_{X_j}\cdot\partial_{X_k};\,\partial_{U_j}\cdot\partial_{X_k};\,\partial_{U_j}\cdot\partial_{U_k} \right),\,U_i\cdot\partial_{X_i}\right]_{_{X_i=X}^{U_i=0}}\,\approx\,0\,,\label{comm eq}
\ee
that have to be satisfied modulo the corresponding homogeneity and tangentiality constraints on the fields and gauge parameters.
Analogously, in the partially-massless cases for each partially-massless field $\Phi^{a_i}(X_i,U_i)$ at the partially-massless point $r_i$, the corresponding coupling should be a solution of the higher-order differential equation
\be
\delta\ \left[(\partial_{U_i}\cdot\partial_{X_i})^{r_i+1}\,C^{(3)}_{a_1 \,a_2\,a_3}\left(\hat{\d};\,\partial_{X_j}\cdot\partial_{X_k};\,{U_j}\cdot\partial_{X_k};\,{U_j}\cdot{U_k}\right)\right]\,\approx\,0\,,
\ee
or of the commutator equation
\be
\delta\ \left[C^{(3)}_{a_1 \,a_2\,a_3}\left(\hat{\d};\,\partial_{X_j}\cdot\partial_{X_k};\,\partial_{U_j}\cdot\partial_{X_k};\,\partial_{U_j}\cdot\partial_{U_k} \right),\,(U_i\cdot\partial_{X_i})^{r_i+1}\right]_{_{X_i=X}^{U_i=0}}\,\approx\,0\,.
\ee
Obviously the differential constraints on the vertices have to be imposed only in relation to the fields that require a gauge symmetry as compatibility conditions with the gauge symmetry itself while, for instance, massive fields in the unitary gauge do not impose additional constraints on the vertices apart of course from the requirement of unitarity that as we have remarked should be imposed on top of the Noether procedure. From this respect the gauge invariant formulation \emph{\`a la} Stueckelberg of massive fields can be of help (see e.g. \cite{Joung:2012rv} for more details).
In the following, starting from the quadratic order we are going to systematically study the solution of the Noether procedure equations in the massless case. For massive and partially-massless couplings the logic is very similar to that of the massless case and we refer to \cite{Joung:2012rv} for more details.

\clearpage{\pagestyle{empty}\cleardoublepage}
\rhead[\fancyplain{}{\bfseries
HS Free Theory}]{\fancyplain{}{\bfseries\thepage}}
\lhead[\fancyplain{}{\bfseries\thepage}]{\fancyplain{}{\bfseries\rightmark}}
\chapter{HS Free Theory on constant curvature backgrounds}                

In this chapter we are going to apply the formalism introduced in the previous chapter in order to construct free Lagrangians for totally symmetric HS fields on constant curvature backgrounds. See \cite{Francia:2002aa,Francia:2005bu,Francia:2007qt,Francia:2007ee,Francia:2008hd,Campoleoni:2012th} for previous constructions in flat and (A)dS backgrounds in the intrinsic formulation and also \cite{Alkalaev:2009vm} for a discussion of the radial reduction method in combination with the BRST formalism. This chapter is based on the appendices of \cite{Taronna:2011kt,Joung:2011ww}. While addressing here as a warm up exercise the problem at the quadratic level, we shall discuss the interactions at the cubic order and beyond in the next Chapters. Among the other things we shall also define the concepts of transverse and traceless (TT) part of the lagrangian that we shall heavily use in the following.

\section{Free Ambient Lagrangians}

In this section, as a warm up exercise, we are going to present the solution to the Noether procedure at the quadratic level. Our point of view is to take as starting point the Fierz system introduced in Section~\ref{(A)dS Dynamics} finding a Lagrangian that upon gauge fixing reduces to it on-shell. Since the dynamical equation at the free level is
\be
\square \, \Phi(X,U)\,=\,0\,,\label{box}
\ee
being of the second order, while the other equations are at most of first order, the natural starting point for a quadratic action describing those representations is
\begin{multline}
S^{(2)}\,=\,\tfrac{1}{2}\int d^{d+1} X\,\d\left(\sqrt{\s X^2}-L\right)\\\times\left[\vphantom{\Big|}\delta_{a_1 a_2}\,e^{\partial_{U_1}\cdot \partial_{U_2}}\,\Phi^{a_1}(X_1, U_1)\,\square_2\,\Phi^{a_2}(X_2,U_2)\,+\,\ldots\right]_{_{X_i=X}^{U_i=0}}\,,\label{quadratic TT}
\end{multline}
where the ellipsis denote, henceforth, terms proportional to divergences and traces of the fields as well as possible auxiliary fields.
The corresponding generating function
\be
C^{(2)}_{a_1 a_2}\,=\,\delta_{a_1 a_2}\,e^{U_1\cdot U_2}\,\square_2\,,
\ee
is in turn a solution of the differential equation
\be
\partial_{U_i}\cdot\partial_{X_i}\,C^{(2)}_{a_1 a_2}\,=\,0\,,
\ee
modulo the tangentiality and homogeneity constraints, if both the field and the gauge parameter are also transverse and traceless.
Of course keeping constraints that are not purely algebraic at the Lagrangian level is most of the time problematic. Moreover since we would like to recover those constraints after on-shell gauge fixing one can exploit the Noether procedure in order to complete the Lagrangian to its full version. More in detail we have to compensate the non vanishing gauge variation of \eqref{quadratic TT} with terms that vanish if the transversality and traceless (TT) constraints are enforced. Computing the gauge variation of the action \eqref{quadratic TT} and leaving implicit for a moment the Chan-Paton indices one ends up with
\be
\delta^{(0)}_ES^{(2)}\,=\,-\int d^{d+1} X \,\d\ e^{\partial_{U_1}\cdot \partial_{U_2}}\left(\partial_{U_1}\cdot\partial_{X_1}\,\Phi(X_1,U_1)\,\square_2\, E(X_2,U_2)\vphantom{\Big|}\right)_{_{X_i+X}^{U_i=0}}\,,
\ee
where we have used
\be
\square_2\,=\,\partial_X\cdot\left(\partial_{X_2}-\partial_{X_1}\right)\,+\,\square_1\,,\qquad \partial_X:=\partial_{X_1}+\partial_{X_2}\,,\label{box by parts}
\ee
together with the tangentiality and homogeneity constraints of the fields and gauge parameters.
Then, proceeding as in the Appendix C of \cite{Taronna:2011kt} one can then define a so called generalized de Donder tensor by the condition
\be
\begin{split}\label{deDonder}
\cD(X,U)\,&=\,\partial_U\cdot\partial_X\,\Phi(X,U)\,+\,U\cdot\partial_X\,\cA(X,U)\,,\\ \delta^{(0)}_{E}\,\cA(X,U)\,&=\,-\partial_U\cdot\partial_X\,E(X,U)\,,\qquad(X\cdot\partial_X-U\cdot\partial_U)\cA(X,U)\,=\,0\,.
\end{split}
\ee
whose basic property is
\be
\delta^{(0)}_{E}\,\cD(X,U)\,=\,\square\,E(X,U)\,,
\ee
while $\cA(X,U)$ contains only tensor structures that vanish on the TT constraints\footnote{Notice that if we discard the trace constraint from the Fierz system then the tensor $\cA$ cannot be built from traces.} and does not satisfy any tangentiality constraint.
In terms of $\cD(X,U)$ and $\cA(X,U)$ it is then possible to write a gauge invariant ambient-space Lagrangian as
\begin{multline}
S^{(2)}\,=\,\frac{1}{2}\int d^{d+1} X \, \d\left[\Phi(X_1,U_1)\,\star\,\square_2\, \Phi(X_2,U_2)\vphantom{\Big|}\right.\\ \left.\vphantom{\Big|}+\,\cD(X_1,U_1)\,\star\,\cD(X_2,U_2)\,-\,\cA(X_1,U_1)\,\star\,\square_2\,\cA(X_2,U_2)\right]\,.\label{quadratic action}
\end{multline}
This general form for the quadratic Lagrangian is independent from the particular form of the operator $\cA$ and what is left is to present the solutions to the eqs.~\eqref{deDonder}. If we restrict the attention to local solutions without auxiliary fields and giving rise to a two derivative action, the available options are quite limited and at the end the only possible choice is given by
\be
\cA(X,U)\,=\,-\,\tfrac{1}{2}\,\partial_U\cdot\partial_U\,\Phi(X,U)\,,\label{trace}
\ee
that actually is compatible with eqs.~\eqref{deDonder} only for traceless gauge parameters:
\be
\partial_U\cdot\partial_U\,E(X,U)\,=\,0\,.
\ee
In more details, the presence of this leftover constraint means that in order to increase the number of independent gauge symmetries either auxiliary fields or higher derivatives should be added while, keeping this minimal formulation, there is a maximal number of gauge symmetries that is compatible with the original Fierz system we started from. This phenomenon is actually quite similar to what happened at the level of the Fierz system itself when we have solved for the compatible gauge symmetries. Indeed there we have asked for the minimum number of gauge symmetries that was compatible with the Fierz system without adding auxiliary fields. On the other hand, we could have preserved a completely unconstrained gauge symmetry just introducing an auxiliary field for any broken gauge symmetry. This kind of result is usually achieved with the help of the Stueckelberg shift so that one is forced to add auxiliary fields that can be settled to zero upon a partial gauge fixing, recovering finally the original minimal form of the gauge symmetry. In particular, in the case of the trace constraint we can address the problem already at the level of the Fierz system considering the following Stueckelberg shift:
\be
\partial_U\cdot\partial_U\,\Phi(X,U)\,\ra\,\partial_U\cdot\partial_U\,\Phi(X,U)\,-\,U\cdot\partial_X\,\a(X,U)\,,
\ee
where by construction
\be
\delta^{(0)}\a(X,U)\,=\,\partial_U\cdot\partial_U\,E(X,U)\,,
\ee
so that one lose any trace constraint on the gauge parameter. The extension of the solution for the tensor $\cA(X,U)$ is then straightforward:
\be
\cA(X,U)\,=\,-\tfrac{1}{2}\,\left[\vphantom{\Big|}\partial_U\cdot\partial_U\,\Phi(X,U)\,-\,U\cdot\partial_X\,\a(X,U)\right]\,.\label{trace compensator}
\ee
Having recovered the ambient space Lagrangian one has to check if the amount of gauge symmetry is anyway sufficient in order to gauge fix it back to the original Fierz system. This presents in general some subtleties and for instance in our case is related to the fact that there exist a combination of the double trace of the field that is identically gauge invariant:
\be
\d^{(0)}_E\,\partial_U\cdot\partial_U\left[\vphantom{\Big|}\partial_U\cdot\partial_U\,\Phi(X,U)\,-\,U\cdot\partial_X\,\a(X,U)\right]\,=\,0\,.
\ee
Hence, one needs to set it to zero
\be
\partial_U\cdot\partial_U\left[\vphantom{\Big|}\partial_U\cdot\partial_U\,\Phi(X,U)\,-\,U\cdot\partial_X\,\a(X,U)\right]\,=\,0\,,\label{doubly traceless}
\ee
in order to avoid gauge fixing problems. This can be done introducing a further Lagrange multiplier $\b(X,U)$ as
\be
S^{(2)}_\b\,=\,\int d^{d+1} X \, \d\ \left\{\b(X,U)\star \partial_U\cdot\partial_U\left[\vphantom{\Big|}\partial_U\cdot\partial_U\,\Phi(X,U)\,-\,U\cdot\partial_X\,\a(X,U)\right]\right\},\label{Lagrange multiplier}
\ee
along the lines of\footnote{Notice that in our approach the Lagrange multiplier does not transform under gauge transformations:
\be
\delta^{(0)}_E\,\b(X,U)\,=\,0\,.
\ee} \cite{Francia:2005bu} while eq.~\eqref{doubly traceless} reduces to the double trace constraint on $\Phi(X,U)$ whenever the compensator $\a$ is removed. Let us mention in fact that the same problem was present even before, since for traceless gauge parameters the double trace of the field is identically gauge invariant\footnote{Setting the double-trace to zero is strictly required whenever the double trace has spin greater than zero because any massless helicity representation cannot be consistent without the corresponding gauge symmetry.}. Finally, the above Lagrangian \eqref{quadratic action}, given the tensors $\cA$ in \eqref{trace} and \eqref{trace compensator} respectively, reduces in the flat limit to the Fronsdal Lagrangian \cite{Fronsdal:1978rb} and, supplemented with a Lagrange multiplier piece of eq.~\eqref{Lagrange multiplier}, to the minimal unconstrained Lagrangian of \cite{Francia:2005bu}, respectively. Hence, they are the corresponding (A)dS deformations written in the ambient space language and reduce exactly to the corresponding intrinsic form \cite{Francia:2007qt} as one can check explicitly. Before going on with the discussion let us also mention another possible solution to eqs.~\eqref{deDonder}. Indeed, by choosing a transverse gauge parameter
\be
\partial_U\cdot\partial_X\,E(X,U)\,=\,0\,,
\ee
one can find the solution
\be
\cA(X,U)\,=\,0\,,
\ee
for which the Lagrangian \eqref{quadratic action} reduces simply to
\be
S^{(2)}\,=\,\int d^{d+1} X \, \d\ \left\{\Phi(X,U)\star\left[\vphantom{\Big|}\square\,-\,U\cdot\partial_X\,\partial_{U}\cdot\partial_X\right]\Phi(X,U)\right\}\,,\label{transverse Lagrangian}
\ee
recovering the transverse Lagrangians studied in \cite{Francia:2010qp,Francia:2011qa} in a ambient space representation as a particular solution for the tensor $\cD$ in eq.~\eqref{deDonder}. This Lagrangian is compatible with the ambient space gauge symmetry only for tangent and transverse gauge parameters. However, these two conditions together imply also the trace constraint
\be
\partial_{U}\cdot\partial_U\,E(X,U)\,=\,\partial_U\cdot\partial_X \, X\cdot\partial_U\,E(X,U)\,=\,0\,\label{traceless gauge}
\ee
so that the trace of the field turns to be identically gauge invariant and has to be set to zero
\be
\partial_{U}\cdot\partial_U\,\Phi(X,U)\,=\,0\,,\label{trace const}
\ee
by consistency, along the same lines as above when the double trace constraint has been introduced.
In this way the above Lagrangian \eqref{transverse Lagrangian} describes on-shell massless \emph{irreducible} representations of the corresponding isometry group. Let us mention that eq.~\eqref{traceless gauge} holds only for finite $L$ while in the flat limit $L\ra \infty$ this condition is lost so that one is not forced to require a traceless gauge parameter. In this case in the flat limit one recovers the so called triplet system \cite{Heanault:book,Bonelli:2003zu,Sagnotti:2003qa} describing a reducible representation containing spin $s$, $s-2$ down to $1$ or $0$ for odd and even $s$ respectively. The fact that traceless and transversality conditions are linked in (A)dS suggests that the corresponding description of reducible trace-full representations in (A)dS is more involved. This can be actually appreciated already at the level of the Fierz system and resonates, as we have already mentioned, with the fact that it is in general more difficult to describe massless reducible representations in (A)dS. In our formalism the difficulty is related to the fixed degree of homogeneity that one assigns to a given field. Indeed, in this case all traces will have the same degree of homogeneity independently from their spin violating eq.~\eqref{homo gauge}. Hence, their degree of homogeneity is in general different from the value for which one gets a massless representation. This pattern is actually the same that one encounters when dealing with reducible mixed-symmetry representations. For instance, allowing non-vanishing traces in the above case one would in general end up with a massless spin-$s$ field plus a bunch of massive lower-spin fields (traces) recovering a bigger representation with respect to the shorter flat space counterpart.
From this respect, more efforts are needed in order to extend the discussion presented so far to more general reducible representations and also to partially-massless and massive fields of all symmetry type and we leave this for future work.

Before closing this section it is worth discussing the role of locality at this level as we have remarked in Section~\ref{Noether Procedure}. As we have anticipated, the key concept is rather unitarity and at this order it is entirely related to the representation theory and hence to the general form of eq.~\eqref{quadratic TT} regardless of the precise structure of the ellipsis. From this respect one can in principle allow any non-local solution for the tensor structure $\cA$ that upon on-shell gauge fixing implies the TT constraints and hence the correct propagator compatibly with the representation theory. This is actually the first and simplest instance in which locality proves to be too strong and manifests its role as a sufficient condition implying unitarity, while allowing non-localities there are still a number of consistent Unitary options that otherwise one would have lost\footnote{See e.g. \cite{Francia:2002aa,Francia:2007ee,Francia:2007qt,Francia:2010ap,Francia:2010qp} for the discussion of free non-local Lagrangians.}.

\section{TT part of the action}\label{TT part}

Before closing this section let us reiterate on the logic that we have taken in order to find solutions to the Noether procedure. This logic will be indeed our key approach at higher orders. The starting point has been the Fierz system translating into the general form of the kinetic term in eq.~\eqref{quadratic TT}, while the Noether procedure has produced its completion containing also terms proportional to divergences and traces. One can indeed observe how the part of the Lagrangian proportional to $\square$ is gauge invariant modulo divergences and traces of the fields while, modulo the EoMs, all traces and divergences have a gauge variation that closes on the TT constraints. This suggest to solve the Noether procedure first modulo the TT constraints and only after, finding the completion of the TT part. This perspective is actually in the spirit of the important work of Metsaev \cite{Metsaev:2005ar,Metsaev:2007rn}. There, all consistent cubic interactions
involving massive and massless HS fields in flat space-time were constructed in the light-cone formalism restricting the attention to the physical DoF and avoiding in this way many of the problems that have been observed in the previous literature related to the appearance of unphysical degrees of freedom. See e.g. \cite{Porrati:1993in, Doria:1994cz,
Cucchieri:1994tx,Giannakis:1998wi,Klishevich:1998yt,
Deser:2000dz, Deser:2001dt,
Porrati:2008gv,Porrati:2008rm,Porrati:2008ha,Zinoviev:2008ck,Zinoviev:2009hu,
Porrati:2009bs,Zinoviev:2010av, Zinoviev:2011fv,Rahman:2011ik}
for some recent works on the consistency of the
electromagnetic (EM) and gravitational couplings to massive HS fields.\footnote{See also \cite{Deser:2006zx} for the study of EM interactions of partially-massless spin 2 fields.}
It is worth noticing that,
as shown for spin 2 in \cite{Argyres:1989cu,Porrati:2011uu} and for arbitrary spins in \cite{Porrati:2010hm},
ST provides a solution for the case of constant  EM background.
See also \cite{Taronna:2010qq,Sagnotti:2010at} for an analysis of HS interactions in the open bosonic string and
\cite{Bianchi:2010es,Polyakov:2010sk,Schlotterer:2010kk,Bianchi:2011se,Lee:2012ku} for studies on scattering amplitudes of HS states in superstring and heterotic string theories.
Other works on cubic interactions of massive HS fields in (A)dS
can be found in \cite{Zinoviev:2006im,Zinoviev:2008jz}.

The aforementioned difficulties in finding consistent interactions
manifest themselves only at the full \emph{off-shell} level,\footnote{
By ``full \emph{off-shell} level'' we mean the entire Lagrangian including traces and divergences of fields,
as opposed to its TT part.} while they can be circumvented restricting the attention to the physical DoF.
In this approach, what is left is to find the \emph{complete} expressions associated with those vertices.
Starting from the TT parts of the interactions, that can be viewed as the covariant versions of Metsaev's lightcone vertices,
the corresponding complete forms within the Fronsdal setting were obtained recently in \cite{Manvelyan:2010jr,Sagnotti:2010at}.
Moreover, the computation of (tree-level) correlation functions does not require the full vertices but only their TT parts.\footnote{See \emph{e.g} \cite{Sagnotti:2010at,Taronna:2011kt} for the analysis of higher-order interactions of massless particles in flat space.}
Therefore, although they ought to be completed, the TT parts of the vertices are also interesting in their own right.
Motivated by this observation, recently the TT parts of the cubic interactions
of massive and (partially-)massless HS fields in (A)dS were identified in \cite{Joung:2011ww,Joung:2012rv}.\footnote{See \cite{Vasiliev:2011xf} for a discussion of the same problem in the frame-like approach.}
\begin{figure}[h]\label{fig:TT}
\begin{center}
\includegraphics[width=11cm]{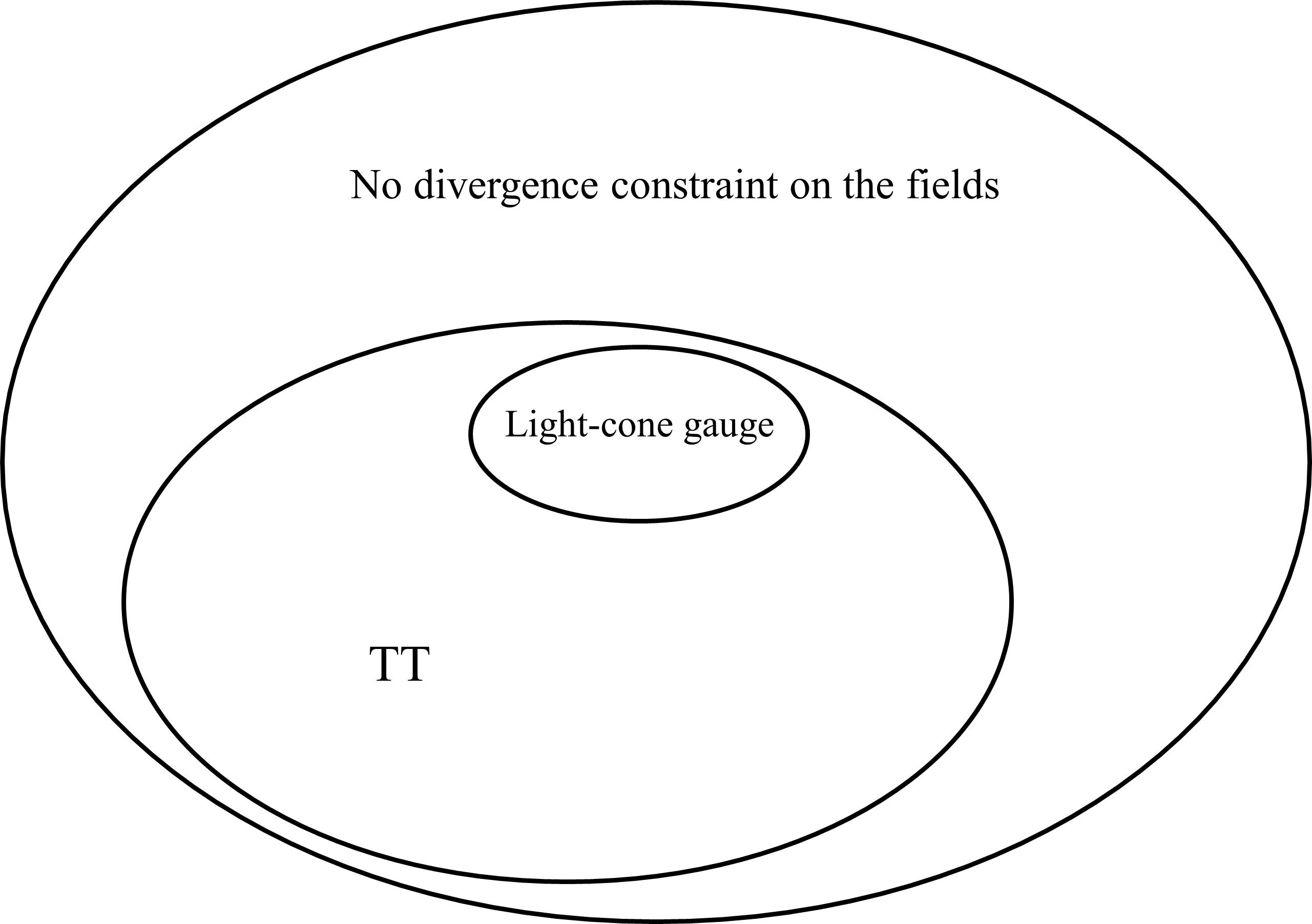}
\caption{Various domain where to apply the Noether procedure. The TT domain is the smallest one for which one can keep a covariant description but it is well defined only on-shell or off-shell as an equivalence class.}
\end{center}
\end{figure}
More in detail in our approach we shall define the TT part of the Action as the corresponding equivalence class modulo traces and divergences of the fields:
\be
S_{\,TT}\,=\,\{S/\sim\}\,,\qquad S_1\sim S_2\ \ra\ S_1\,=\,S_2 + f(\Phi,\,\partial_X\cdot\partial_U\,\Phi,\,\partial_{U}\cdot\partial_{U}\,\Phi)\,,
\ee
with $f$ any (possibly non-local) functional involving the field and proportional to its divergences or traces. Let us reiterate here that in the above definition there is no non-local projection while, in order to arrive to the full Lagrangian, it is required to fix the portion proportional to divergences and traces as we have shown at the quadratic level. The concept of Lagrangian locality translates as usual in the form of polynomial dependence of the Lagrangian on the ambient space derivatives and extends also to the TT part of the Lagrangian, if there exists a local representative within the corresponding equivalence class.
At the quadratic level for instance a representative of the corresponding equivalence class is local and coincides with \eqref{quadratic TT}. The corresponding EoMs then read
\be
\square\,\Phi(X,U)\,+\,\ldots\,\approx\,0\,,\label{free EoMs}
\ee
under the same equivalence relation, together with possible equations for the auxiliary fields. Moreover, the whole Noether procedure has a well definite lifting to the corresponding equivalence classes. Indeed, for any action representative
\be
S\,=\,S_{TT}\,+\,\ldots\,
\ee
the equation
\be
\delta_E S\,\approx\,0\,,
\ee
implies
\be
\delta_E S_{TT}\,\approx\,\ldots\,,
\ee
since
\be
\delta_E(\ldots)\,\approx\,\ldots\,.
\ee
modulo the free EoMs \eqref{free EoMs}. Hence, the TT part of the action is well defined, while after having solved for it one should find its completion as we have done at the quadratic level above (See fig.~$2.1$ for a graphic representation of the setting described so far). Let us also mention that from this perspective the TT part of the action is completely independent on the number of auxiliary fields and in general from the explicit form of the ellipsis in eq.~\eqref{quadratic TT}. This is similar to what happens at the quadratic level where the $\square$ part of the action \eqref{quadratic TT} is fixed by the Fierz system or equivalently by the representation theory. Moreover, the classification of the TT part can be thought of as the classification of all possible on-shell interactions and hence can be considered on very similar grounds underlying an interesting link between the following, say, \emph{bulk} construction and the analogous problem from the CFT side of constructing the most general correlation function in a boundary CFT containing currents of arbitrary rank\footnote{See also \cite{Giombi:2011rz} for the corresponding analysis in three dimensions.} \cite{Stanev:2012nq,Zhiboedov:2012bm,Alkalaev:2012rg}. Moreover let us reiterate that the computation of (tree-level) correlation functions does not require the full vertices but only their TT parts. Hence, the TT part of the vertices acquire some interest on its own right not only as a starting point in order to construct a full HS Lagrangian. All that being said, in the following we shall restrict our attention to the solution of the Noether procedure modulo divergences and traces that we shall refer as TT part of the vertices. We shall review the completion of the flat vertices along the lines of \cite{Manvelyan:2010jr,Sagnotti:2010at,Taronna:2011kt} in Appendix~\ref{Completion}.

\clearpage{\pagestyle{empty}\cleardoublepage}
\rhead[\fancyplain{}{\bfseries
HS Interaction on constant curvature backgrounds}]{\fancyplain{}{\bfseries\thepage}}
\lhead[\fancyplain{}{\bfseries\thepage}]{\fancyplain{}{\bfseries\rightmark}}
\chapter{HS Interaction on constant curvature backgrounds}                

In this chapter we shall present the solution to the Noether procedure at the cubic level for arbitrary massless totally-symmetric fields. This Chapter is based on \cite{Taronna:2011kt,Joung:2011ww,Joung:2012rv}. As explained previously we restrict our attention to the TT part of the vertices while we shall describe the explicit completion in the flat-space case in Appendix~\ref{Completion}. In this Chapter we shall work mostly with coupling generating function of the form \eqref{coupling genfunc operator} where for convenience the $\star$-contraction has been explicitly performed.

\section{Cubic interactions: general setting}
\label{sec: cubic unitary}

In this section we construct the consistent parity-invariant cubic interactions
of massless totally-symmetric HS fields in (A)dS.
More precisely, we focus on those pieces which do not contain divergences and traces of the fields (TT parts). We begin with the most general expression for the cubic vertices introduced in Section~\ref{Noether Procedure}
\begin{multline}
\label{cubicact1}
S^{\sst {(3)}} \,=\, \frac1{3!}\int d^{d+1}X\ \delta\Big(\sqrt{\s\,X^2}-L\Big)\
C_{\sst {a_{1}a_{2}a_{3}}}(L^{-1}\,;\,\partial_{X_1},\partial_{X_2},\partial_{X_3}\,;\,
\partial_{U_1},\partial_{U_2},\partial_{U_3}) \\\times\ \Phi^{\sst a_{1}}(X_1,U_1)\ \Phi^{\sst a_{2}}(X_2,U_2)\ \Phi^{\sst a_{3}}(X_3,U_3)\ \Big|_{^{X_i=X}_{U_i=0}}+\,\ldots\,,
\end{multline}
where hereafter $C_{\sst a_{1}a_{2}a_{3}}$ denotes the TT part of the vertices, as defined in Section~\ref{TT part}, while the ellipsis represent its completion.
The cubic interactions in (A)dS
are in general inhomogeneous in the number of derivatives,
the lower-derivative parts being dressed by negative powers of $L$
compared to the highest-derivative one.
Hence, the TT parts of the vertices can be expanded as
\be
\label{seriesEx}
C_{\sst {a_{1}a_{2}a_{3}}}(L^{-1}\,;\,\partial_{X_i}\,,\,\partial_{U_i})
=\sum_{n=0}^\infty\, L^{-n}\,
C^{\sst [n]}_{\sst {a_{1}a_{2}a_{3}}}(Y,Z)\,,
\ee
where we have introduced the parity-preserving Lorentz invariants:
\be
\label{Y and Z}
Y_i=\partial_{U_i}\!\cdot\partial_{X_{i+1}}\,, \qquad
Z_i=\partial_{U_{i+1}}\!\!\cdot\partial_{U_{i-1}}\,,
\qquad [i\simeq i+3]\,,
\ee
in the operatorial convention, or
\be
Y_i={U_i}\!\cdot\partial_{X_{i+1}}\,, \qquad
Z_i={U_{i+1}}\!\!\cdot{U_{i-1}}\,,
\qquad [i\simeq i+3]\,,
\ee
in the other one where the $\star$-contraction has not been performed yet. In this way, choosing a particular set of $Y_{i}$'s, we have fixed any ambiguity related to the (A)dS deformation of total derivatives mentioned in Section~\ref{Ambient measure}, while in order to properly analyze the role of the total derivatives the latter are denoted henceforth by
\be
\partial_X\,:=\,\partial_{X_1}\,+\,\partial_{X_2}\,+\,\partial_{X_3}\,.
\ee
Moreover, it is worth noticing that at the cubic level and restricting the attention to the TT part of the interactions one can further simplify the general expressions of eq.~\eqref{coupling genfunc 2} and indeed we have dropped divergences, $\partial_{U_{i}}\!\cdot\partial_{X_{i}}$\,, traces, $\partial_{U_{i}}^{\,2}$ as well as
terms proportional to $\partial_{X_i}\!\cdot\partial_{X_j}$'s that being proportional to the field equations
up to total derivatives:
\be
\partial_{X_1}\cdot\partial_{X_2}\,=\,\tfrac{1}{2}\,\partial_X\cdot(\partial_{X_1}+\partial_{X_2}-\partial_{X_3})\,+ \,\tfrac{1}{2}(\partial_{X_3}^2-\partial_{X_1}^2-\partial_{X_2}^2)\,,
\ee
can be removed by proper field redefinitions. Notice on this respect that for that regards the TT part of the cubic vertex there is no possible room for non-local terms that are not singular on-shell just because either
\be
\frac{1}{\partial_{X_i}^2}\,\approx\,\frac{1}{0}\,\,
\ee
or
\be
(\partial_{X_1}\cdot\partial_{X_2})^{-1}\,\approx\,2\,[\partial_X\cdot(\partial_{X_1}+\partial_{X_2}-\partial_{X_3})]^{-1}\,,
\ee
that modulo integrations by parts is proportional to a combination of the degrees of homogeneities of the fields and hence is either of the singular form $\tfrac{1}{0}$ or can be reabsorbed by a redefinition of the coupling function, being just a constant overall factor. We are assuming here that observables are well-defined on-shell\footnote{Let us reiterate that we have defined locality of the level of the TT part as equivalent to the existence of a local representative in the equivalence class \eqref{TT part}. Above we have proved that a non-local cubic TT-part would inevitably led to ill-defined observables and hence no room for them is leftover.}.

For the above reasons one does not need to assume locality so far while one has indeed to check Unitarity as we have remarked in Section~\ref{Noether Procedure}. From this respect, one can easily remove from the vertex any temporal derivative of order higher than one with proper field redefinitions just exploiting the equations of motion
\be
\partial_t^2\,\approx\,\partial_x^2\,,
\ee
ending up with a coupling that contains at most one temporal derivative and for this reason does not propagate ghosts at this order. Even though non-localities do not enter explicitly at the level of the TT part of the coupling they can play a role in combination with traces and divergences just because in those cases they can still be well defined on-shell being of the generic form $\tfrac{0}{0}$. Their possible presence is related from this respect to the tensor structure $\cA$ that has been chosen at the quadratic level and that enters the EoMs. Their consistency rely on the consistency of the complete quadratic Lagrangian so that non-localities can still play a role, compatibly with unitarity also at the cubic level, even though the TT part of the vertex turns out to be of higher-derivative nature but local for any fixed spin of the fields.

Coming back to our discussion, in order to simplify the analysis, it is convenient to
recast the expansion \eqref{seriesEx} as in Section~\ref{Ambient measure} so that
each coefficient of \eqref{seriesEx} can be redefined as
\be
	L^{-n}\,C^{\sst [n]}_{\sst a_{1}a_{2}a_{3}}(Y,Z) =
	\hat\delta^{n}\,C^{\sst (n)}_{\sst a_{1}a_{2}a_{3}}(Y,Z)\,.
\ee
Notice that $C_{\sst a_{1}a_{2}a_{3}}^{\sst [n]}$ and
$C_{\sst a_{1}a_{2}a_{3}}^{\sst (n)}$
are different functions for $n\ge1$\,.
The entire couplings can be finally resummed as
\be
	C_{\sst a_{1}a_{2}a_{3}}(\hat\delta;Y,Z)
	=\sum_{n=0}^{\infty}\,\hat\delta^{n}\,
	C_{\sst a_{1}a_{2}a_{3}}^{\sst (n)}(Y,Z)\,,
	\label{exp lambda}
\ee
where we have used the same notation
for both {\small$C(L^{-1};\ldots)$}
and {\small$C(\hat\delta;\ldots)$}
although they are different functions. Moreover, an equivalent presentation of the same generating function that will be very useful in the following can be given reabsorbing all $\hat{\d}$ dependence within total derivatives. Indeed exploiting backwards the identity
\be
\delta^{(n)}\left(\sqrt{\s X^2}-L\right)\partial_{X^M}\,=\,-\,\s\,\delta^{(n+1)}\left(\sqrt{\s X^2}-L\right)\,\frac{X^M}{L}\,,
\ee
one can rewrite the vertex generating function as
\be
	\delta\Big(\sqrt{X^{2}}-L\Big)\
	C_{a_{1}a_{2}a_{3}}(\,\partial_{X}\, ;\,
	Y_i,Z_i)
	=\delta\Big(\sqrt{\s X^{2}}-L\Big)\sum_{n=0}^{\infty}\,\hat\d^n\,\,
	C_{a_{1}a_{2}a_{3}}^{{\sst [n]}}(\,Y_i,Z_i)\,,
	\label{delta exp}
\ee
where all the $L$ dependence of the vertex has been reabsorbed into the total derivative dependece of the function $C_{a_{1}a_{2}a_{3}}(\,\partial_{X}\,;\,Y_i,Z_i)$, that is again a different function with respect to the above ones.

Before going on let us make contact with the standard tensor notation providing an explicit example. For instance, a dS vertex of the form
\be
C(\hat\delta;Y,Z)=(Y_1^2\,Y_2\,Y_3\,Z_1+\text{cycl.})-\tfrac{\hat{\delta}}{L}\,(Y_1\,Y_2\,Z_1\,Z_2+\text{cycl.})+\tfrac{3}{4}\,\left(\tfrac{\hat{\delta}}{L}\right)^2\,Z_1\,Z_2\,Z_3\,,
\ee
in our notation translates in the usual tensor notation as
\begin{multline}
S^{\sst (3)}=\frac{1}{2}\,\int d^{d+1}X\ \delta\Big(\sqrt{X^2}-L\Big)\,\Big[\partial_{\sst P}\,\Phi^{\sst MN}\,\partial_{\sst M}\,\partial_{\sst N}\,\Phi_{\sst LQ}\,\partial^{\sst L}\,\Phi^{\sst PQ}\\
+\frac{d-5}{L^2}\,\Phi^{\sst M}_{\phantom{\sst M}\sst N}\,\partial_{\sst M}\,\Phi_{\sst LP}\,\partial^{\sst L}\,\Phi^{\sst NP}
+\frac{(d-3)(d-4)}{4\,L^4}\,\Phi^{\sst M}_{\phantom{\sst M}\sst N}\,\Phi^{\sst N}_{\phantom{\sst N}\sst P}\,\Phi^{\sst P}_{\phantom{\sst P}\sst M}\Big]\,.
\end{multline}

\section{Consistent cubic interactions of massless HS fields}
\label{sec:Consistent cubic}

So far, we have just improved our notation at the cubic level exploiting the simplifications that arise in this case. As we have discussed in Section~\eqref{Noether Procedure} gauge consistency can be studied order by order
(in the number of fields), and
at the cubic level gives
\be
\label{noetherproc}
\delta^{\sst (1)}_{i}S^{\sst (2)}+\delta^{\sst (0)}_{i}
S^{\sst (3)}=0 \quad \Rightarrow \quad
\delta^{\sst (0)}_{i}S^{\sst (3)} \approx 0\,,
\ee
where $\approx$ means equivalence modulo the free field equations
\be
\square\,\Phi(X,U)\,+\,\ldots\,=\,0\,,
\ee
and $\delta_{i}^{\sst (0)}$ is
the linearized gauge transformation \eqref{masslesstransf}
associated with the massless field $\Phi^{a_{i}}$\,.
Let us reiterate that the key point of our approach
is that the TT parts of the vertices
can be determined from the Noether procedure \eqref{noetherproc}
independently from the ellipses in \eqref{quadratic TT}.
This amounts to quotient the Noether equation \eqref{noetherproc}
by the Fierz systems of the fields $\Phi^{\sst a_{i}}$ and of the gauge parameters $E^{a_{i}}$\,.
In our notation, this is equivalent to impose,  for $i=1$\,,
\be
\label{gaugeconscond1}
\left[\,C_{\sst a_{1}a_{2}a_{3}}(\hat\delta;Y,Z)\,,\, U_1\cdot\partial_{X_1}\,\right]\Big|_{U_1=0}\approx 0\,,
\ee
modulo all the $\partial_{X_{i}}^{\,2}$'s\,, $\partial_{U_{i}}\!\cdot\partial_{X_{i}}$'s
and $\partial_{U_{i}}^{\,2}$'s\,.
Due to the presence of the delta function, the total derivative terms generated by
the gauge variation do not simply vanish, but contribute as
\be
\delta\Big(\sqrt{\s\,X^2}-L\Big)\,\partial_{X^M}=
-\,\delta\Big(\sqrt{\s\,X^2}-L\Big)\,\hat\delta\, \frac{X_{\sst M}}{L}\,.
\ee
Using the commutation relations \eqref{commrelations} together with the identity \eqref{fBcomm},
eq.~\eqref{gaugeconscond1} turns to be equivalent to
the following differential equation:
\be
\label{recurrrelmass}
\Big[\,Y_2\partial_{Z_3}-Y_3\partial_{Z_2}
+\tfrac{\hat\delta}{L}\left(Y_2\partial_{Y_2}-Y_3\partial_{Y_3}
-\tfrac{\mu_{2}-\mu_3}2
\right)\partial_{Y_{1}}\Big]\,C_{\sst a_{1}a_{2}a_{3}}(\hat\delta;Y,Z)=0\,,
\ee
where $\m_2$ and $\m_3$ are the possibly non-zero homogeneities of the other two fields taking part to the interactions. In the following we shall restrict the attention to the case $\m_i=0$ so that one should solve the system of three equations
\ba\label{massless system}
&&\!\!\!\!\!\!\!\!\!\Big[\,Y_{i+1}\partial_{Z_{i-1}}-Y_{i-1}\partial_{Z_{i+1}}
+\tfrac{\hat\delta}{L}\left(Y_{i+1}\partial_{Y_{i+1}}-Y_{i-1}\partial_{Y_{i-1}}
\right)\partial_{Y_{i}}\Big]\,C_{\sst a_{1}a_{2}a_{3}}(\hat\delta;Y,Z)\,=\,0\,,\nn &&i=1,2,3\,,
\ea
whose solutions are the consistent parity-invariant cubic interactions
involving massless HS fields in (A)dS.
Since $C_{\sst a_{1}a_{2}a_{3}}$ is a polynomial in
$\hat\delta$\,,
one can solve \eqref{recurrrelmass} iteratively starting from the lowest order in $\hat \d$. More explicitly the auxiliary variable $\hat{\delta}$ is nothing but a formal device in order to turn a differential recurrence relation into a single partial differential equation. Indeed, eliminating $\hat{\d}$, the simple commutator equation \eqref{gaugeconscond1} becomes
\be
	\delta\Big(\sqrt{\s\,X^{2}}-L\Big)\,\sum_{n=0}^{\infty}\,\hat\d^n\,
	\big[\, C_{a_{1}a_{2}a_{3}}^{{\sst (n)}}(\,Y_{i}\,,\,Z_{i}\,)\,,
	\,U_{\sst j}\!\cdot\partial_{X_{\sst j}}\,\big] \approx 0\,,
	\label{GC dS}
\ee
that taking into account the total derivatives and integrating them by parts implies the differential recurrence relation
\be
	\left(Y_{2}\,\partial_{Z_{\sst 3}}-Y_{3}\,\partial_{Z_{\sst 2}}\right) C^{\sst (n)}_{a_{1}a_{2}a_{3}} \,+\,\tfrac1{L} \left(Y_2\partial_{Y_{\sst 2}}-Y_3\partial_{Y_{\sst 3}}\right)\partial_{Y_{\sst 1}}
	 C^{\sst (n-1)}_{a_{1}a_{2}a_{3}}\,=\,0\,,
	\label{dfe dS}
\ee
where the $C_{\sst a_1a_2a_3}^{(n)}$ vanish for negative $n$. Before discussing the solutions of the system \eqref{massless system} let us notice that in the flat limit $L\ra\infty$ the differential equations simplify becoming the corresponding Noether procedure equations in flat space time\footnote{Some subtleties arise whenever one considers massive fields in the flat limit because taking the limit one is generically not allowed to rescale the homogeneities $\m_i$ in order to end up after the substitution
\be
	\lim_{L\to\infty}\, \tfrac1L\,\mu=-M\,.
\ee
with the flat massive Noether equation
\be
\Big[\, Y_2\partial_{Z_3}-Y_3\partial_{Z_2}
+\tfrac{\hat\delta}2\,(M_{2}-M_3)\,\partial_{Y_1}\Big]\,
C_{\sst a_{1}A_{2}A_{3}}(\hat\delta;Y,Z)=0\,,
\label{recur flat}
\ee
In these cases one needs to check that the value of the homogeneities is not at a particular point at which some enhancement of the number of solution can arise. In general particular care should be taken whenever $\m_{2}-\m_3\in 2\,\mathbb{Z}$ so that the limiting equation would still result to be \eqref{massless system flat}.
}:
\be\label{massless system flat}
\Big(\,Y_{i+1}\partial_{Z_{i-1}}-Y_{i-1}\partial_{Z_{i+1}}\Big)\,C_{\sst a_{1}a_{2}a_{3}}(\hat\delta;Y,Z)\,=\,0\,.
\ee
This means that any solution of the (A)dS Noether equation \eqref{massless system} in the flat limit gives rise to a consistent solution of the flat Noether procedure. Obviously, the opposite is not generically true and one should explicitly check if for each flat solution there exists a corresponding deformation starting with the same leading term and supplemented with $\tfrac{1}{L}$ corrections such that the former flat solution can be promoted to a full (A)dS solution.
In the following we shall solve this problem in two ways:
\begin{itemize}
\item First, we shall follow the derivation of \cite{Joung:2011ww,Joung:2012rv}, where the solution to the problem has been found asking for the possibility to lift the order zero solution to a full solution of \eqref{gaugeconscond1} deforming the former with total derivative contributions that as we have discussed can be used in order to encode the lower derivative tail of the vertex.
\item Second, we shall present the explicit polynomial solutions of the system of partial differential equations \eqref{massless system} solving directly the differential equation and building the most general deformation of the order zero solutions.
\end{itemize}
Finally, we shall link the two different ways of presenting the solutions showing how they are related by integrations by parts. Hence, let us discuss the solution for the order zero part of the equations, namely the flat limit \eqref{massless system flat}.
\subsection{Flat solution}
The general solution of the latter differential equations \eqref{massless system flat} is given by
\be
	C_{a_{1}a_{2}a_{3}}(\,Y_{1}\,,\,Y_{2}\,,\,Y_{3}\,,\,Z_{1}\,,\,Z_{2}\,,\,Z_{3}\,)
	= \cK_{a_{1}a_{2}a_{3}}(\,Y_{1}\,,\,Y_{2}\,,\,Y_{3}\,,\,G\,)\,,
	\label{sol flat}
\ee
where $G$ is defined as
\ba
	G &:=& Y_{1}\,Z_{1}+Y_{2}\,Z_{2}+Y_{3}\,Z_{3} \\
	&\ =&
	\partial_{U_{2}}\!\!\cdot\partial_{U_{3}}\,\partial_{U_{1}}\!\!\cdot\partial_{X_{2}}
	+\partial_{U_{3}}\!\!\cdot\partial_{U_{1}}\, \partial_{U_{2}}\!\!\cdot\partial_{X_{3}}
	+\partial_{U_{1}}\!\!\cdot\partial_{U_{2}}\,\partial_{U_{3}}\!\!\cdot\partial_{X_{1}}\,.\nonumber
\ea
The consistent cubic interactions are encoded in an arbitrary function $\cK_{a_{1}a_{2}a_{3}}$ \eqref{sol flat}, and when expanded for different spins, they are expressed as
\be
	\cK_{a_{1}a_{2}a_{3}}=
	\sum_{s_{1},s_{2},s_{3}} \!\!\!\!
	\sum_{n=0}^{{\text{min}}\{s_{1},s_{2},s_{3}\}}
	g^{s_{1}s_{2}s_{3},n}_{a_{1}a_{2}a_{3}}\ \,
	G^{n}\
	Y_{1}^{s_{1}-n}\,
	Y_{2}^{s_{2}-n}\,
	Y_{3}^{s_{3}-n}\,,
	\label{V b}
\ee
where the $g^{s_{1}s_{2}s_{3},n}_{a_{1}a_{2}a_{3}}$'s are coupling constants
that might be fixed by the consistency of higher-order interactions.
The number of derivatives of each vertex can be identified from the above expansion as
\be
	s_{1}+s_{2}+s_{3}-2n\,,
\ee
so that the mass-dimensions of the coupling constants are
\be
	\big[g^{s_{1}s_{2}s_{3},n}_{a_{1}a_{2}a_{3}}\big]
	= \tfrac{6-d}2-s_{1}-s_{2}-s_{3}+2n\,.
	\label{cc md}
\ee
Moreover, from the symmetry properties of the cubic action \eqref{cubicact1}, the coupling constants inherit the symmetries
\be
	g^{s_{2}s_{1}s_{3},n}_{a_{2}a_{1}a_{3}}=
	g^{s_{1}s_{3}s_{2},n}_{a_{1}a_{3}a_{2}}
	=(-1)^{s_{1}+s_{2}+s_{3}}\,g^{s_{1}s_{2}s_{3},n}_{a_{1}a_{2}a_{3}}\,.
	\label{cc cyc}
\ee
As a result, the uncolored case is consistent only when the
total spin $s_{1}+s_{2}+s_{3}$ is even.

Having classified the corresponding flat solution what is left, as we have anticipated, is to find their lifting. Before discussing explicitly this issue let us add some comments on the higher order solutions $C^{\sst (n)}_{\sst a_{1}a_{2}a_{3}}$. Indeed one can now solve iteratively all subsequent differential equations that for each \mt{n\ge1} are now inhomogeneous differential equations
whose solutions are fixed up to a solution of the corresponding homogeneous equation.
Each of these homogeneous solutions $\cK^\prime_{a_{1}a_{2}a_{3}}(Y_{i},G)$ induces a tail of particular solutions for higher orders, and provides
additional solutions of eqs.~\eqref{dfe dS}:
\be
	C_{a_{1}a_{2}a_{3}}=\,\delta\Big(\sqrt{\s\,X^{2}}-L\Big)\,\hat\delta^n\,\left[\cK'_{a_{1}a_{2}a_{3}}
	+\sum_{m=1}^{\infty} \,
	\hat\delta^{\,m}\,
	C^{\sst (m)}_{a_{1}a_{2}a_{3}}\right]\,.
	\label{C ser}
\ee
This observation seems to imply the presence of ambiguities in the (A)dS interactions,
but these additional solutions are in fact redundancies.
This is true because, after the radial integration,
different $\hat\delta^{\,n}$'s give just different spin-dependent constant factors as one can see from eq.~\eqref{hat delta}.
Therefore, any solution of the type \eqref{C ser} can be re-expressed in the form:
\be
	C_{a_{1}a_{2}a_{3}} =
	\delta\Big(\sqrt{\s\,X^{2}}-L\Big)\,\left[\tilde\cK'_{a_{1}a_{2}a_{3}}
	+\sum_{n=1}^{\infty}\,\hat\d^{\,n}\,
	\tilde C^{\sst (n)}_{a_{1}a_{2}a_{3}}\right]\,,
\ee
where $\tilde \cK'_{a_{1}a_{2}a_{3}}$ and $\tilde C^{\sst (n)}_{a_{1}a_{2}a_{3}}$
are proportional to $\cK'_{a_{1}a_{2}a_{3}}$ and $C^{\sst (n)}_{a_{1}a_{2}a_{3}}$
with some spin-dependent factors.
To reiterate, the aforementioned ambiguity can be recast into a redefinition of the original
\mt{C^{\sst (0)}_{a_{1}a_{2}a_{3}}=\cK_{a_{1}a_{2}a_{3}}}\,.
Hence, a consistent cubic interaction is univocally determined from the choice of $C^{\sst (0)}_{a_{1}a_{2}a_{3}}$\,.

\section{General solutions by total derivative deformation}

So far, we have shown that the consistent cubic interactions in (A)dS
can be obtained solving the differential equations \eqref{dfe dS}.
The $\delta^{\sst(0)}$-order solution was already obtained in terms of an arbitrary function
$\cK_{a_{1}a_{2}a_{3}}$\,, and what is left is to determine a particular solution for
the higher order parts $C^{\sst (n\ge 1)}_{a_{1}a_{2}a_{3}}$\,,
keeping in mind that the ambiguities in the latter are redundancies.

In the following, we construct at once the full cubic vertex
comprising the full higher order tail of the $C^{\sst (n)}_{a_{1}a_{2}a_{3}}$'s,
by making use of the following ansatz
\be
	C_{a_{1}a_{2}a_{3}}(\,\partial_{X}\, ;\,
	\partial_{X_{i}},\partial_{U_{i}})=
	\cK_{a_{1}a_{2}a_{3}}\big(\, \tilde{Y}_{i}\,,\,\tilde{G}\,\big)\,,
	\label{amb K}
\ee
where we have considered a total derivative deformation of the $\tilde{Y}_{i}$'s and $\tilde{G}$ of the form
\ba
	\tilde{Y}_{i} \eq Y_{i}+\alpha_{i}\,\partial_{U_{i}}\!\!\cdot\partial_{X}\,, \\
 	\tilde{G} \eq
	(Y_{1} + \beta_{1}\,\partial_{U_{1}}\!\!\cdot\partial_{X})\,Z_{1} +
	(Y_{2} + \beta_{2}\,\partial_{U_{2}}\!\!\cdot\partial_{X})\,Z_{2} +
	(Y_{3} + \beta_{3}\,\partial_{U_{3}}\!\!\cdot\partial_{X})\,Z_{3}\,.\nonumber
	\label{amb G}
\ea
Notice first that this ansatz is a highly restricted one, with only six constants,
compared to the general setting with an arbitrary number of $C^{\sst (n)}_{a_{1}a_{2}a_{3}}$'s.
Nonetheless, the motivation is straightforward: we attempt to keep
the form of the generating function $\cK_{a_{1}a_{2}a_{3}}$ fixed in terms of the same type of building blocks as in the flat case. Indeed if all order zero solutions really admit a proper lifting to (A)dS, as expected in the massless case, this would be equivalent to the existence of a proper deformation at the level of the building blocks themselves, otherwise one would most likely lose some particular couplings depending on the spin of the external particles.

Notice as well that, although \eqref{amb G} contains explicitly
total derivatives, the highest-derivative part of the vertices built from
\eqref{amb G} do not, ensuring its non-triviality. Moreover, let us mention that in general in order to avoid triviality problems it is always sufficient to start from the corresponding solutions at the zero-th order asking for a proper lower derivative deformation. Indeed, for any solution to the (A)dS Noether procedure the corresponding leading term\footnote{By leading term we mean the highest derivative piece of the (A)dS coupling that is not a total derivative. Notice on this respect that the flat limit is well defined only at the action level and not at the Lagrangian level due to the identical vanishing of total derivatives under the integral sign.} is always a proper solution of the corresponding flat Noether procedure equation. This means that the number of solutions to the flat Noether procedure is always bigger than the corresponding number of solutions to the (A)dS Noether procedure. Hence, inequivalent flat solution will always admit inequivalent lower derivative liftings whenever existing.

In order to examine the ansatz \eqref{amb K},
we compute the gauge variation of the latter
exercising some care in treating total derivatives.
We first provide our solutions, leaving
the detailed computation for the last part of this subsection.
Requiring that the gauge variation vanishes modulo $\partial_{X_{i}}^{2}\approx 0$\,, one ends up with the conditions %
\ba
	&& (\alpha_{1}+1)\,\alpha_{2}+1=0\,, \label{B1} \\
	&& (\alpha_{1}+1)(\beta_{2}+1)+\alpha_{1}\,\beta_{3}=0\,,\label{B2} \\
	&& (\beta_{1}+1)(\beta_{2}+1)+\beta_{3}(\beta_{1}+\beta_{2}+1)=0\,, \label{B3}
\ea
on the constants $\alpha_{i}$ and $\beta_{i}$ appearing in the ansatz \eqref{amb K},
together with their cyclic permutations on the subscripts $i$ of $\alpha_{i}$ and $\beta_{i}$\,. If a solution for these equations exists,
eq.~\eqref{amb K} provides the consistent cubic interactions. Actually, as expected, eqs.~(\ref{B1}\,-\,\ref{B3}) admit solutions parameterized by two constants $\alpha$ and $\beta$\,:
\ba
	 \alpha_{1}=\alpha\,,\qquad
	&\alpha_{2}=-\tfrac{1}{\alpha+1}\,,\qquad
	&\alpha_{3}=-\tfrac{\alpha+1}{\alpha}\,, \nonumber\\
	 \beta_{1}=\beta\,,\qquad
	&\beta_{2}=-\,\tfrac{\beta+1}{\alpha+1}\,,\qquad
	&\beta_{3}=-\,\tfrac{\alpha-\beta}{\alpha}\,,
	\label{sol cnst}
\ea
so that we can definitely conclude that the (A)dS deformation exists at the level of the building blocks themselves as we have argued before.
Regarding possible redundancies related to different choices of $\alpha$ and $\beta$ they can be in principle reabsorbed into the definition
of the functions $\cK_{a_{1}a_{2}a_{3}}$ so that one can work with a particular choice of $\alpha$ and $\beta$ without loss of generality. More precisely, as we shall see in the following, one can even prove that different choices of $\a$ and $\b$ are parameterizing just total derivative contributions that vanish identically.

Finally, the general solution for the (A)dS cubic-interaction problem is given by an arbitrary function
$\cK_{a_{1}a_{2}a_{3}}$ \eqref{amb K} together with eq.~\eqref{sol cnst}\,.
One can easily verify that when \mt{s_{1}=s_{2}=0} our results coincide
with the interaction vertices constructed in \cite{Bekaert:2010hk}. The coupling constants have the same mass-dimensions \eqref{cc md}
and the same permutation symmetries \eqref{cc cyc} as the flat-space ones, while each vertex is now not homogeneous in the number of the (A)dS-covariant derivatives since the ambient-space derivative \eqref{covder} is not.
However, the maximum number of derivatives of the vertex associated with $g^{s_{1}s_{2}s_{3},n}_{a_{1}a_{2}a_{3}}$ can be easily identified as \mt{s_{1}+s_{2}+s_{3}-2n}\,. In the next section we shall see in more detail how this non-homogeneity is related to
the inverse power expansion in the cosmological constant exhibited by the FV vertices.

\subsection{Proof at  the $\delta^{\sst (1)}$ level}

Here we show that  the gauge invariance of the ansatz \eqref{amb K} is equivalent to the conditions of eqs.~(\ref{B1}\,-\,\ref{B3})\,. Since all vertices of different spins are independent, we consider without loss of generality the case where $\cK_{a_{1}a_{2}a_{3}}$ is given by an exponential function:
\be
	\cK_{a_{1}a_{2}a_{3}}=k_{a_{1}a_{2}a_{3}}\,e^{L\,\cV}\,,
\ee
where $L$ is again the radius of (A)dS, $\cV$ is the sum of the arguments of $\cK_{a_{1}a_{2}a_{3}}$ in
eq.~\eqref{amb K} modulo a factor of two:
\ba
	\cV \edf \partial_{U_{1}}\!\!\cdot(\partial_{X_{23}}\!+\tilde{\a}_1\,\partial_{X})+
	\partial_{U_{2}}\!\!\cdot(\partial_{X_{31}}\!+\tilde{\alpha}_{2}\,\partial_{X})+
	\partial_{U_{3}}\!\!\cdot(\partial_{X_{12}}\!+\tilde{\alpha_{3}}\,\partial_{X}) \nonumber\\
	&&+\, \partial_{U_{2}}\!\!\cdot\partial_{U_{3}}\,
	\partial_{U_{1}}\!\!\cdot(\partial_{X_{23}}\!+\tilde{\beta}_{1}\,\partial_{X})
	+\partial_{U_{3}}\!\!\cdot\partial_{U_{1}}\,\partial_{U_{2}}\!\!\cdot
	(\partial_{X_{31}}\!+\tilde{\beta}_{2}\,\partial_{X})\nonumber\\
	&&+\,\partial_{U_{1}}\!\!\cdot\partial_{U_{2}}\,\partial_{U_{3}}\!\!\cdot
	(\partial_{X_{12}}\!+\tilde{\beta}_{3}\,\partial_{X})\,.
\ea
and where, for convenience, we have redefined the $\a_i$'s and $\b_i$'s as
\be
\a_i\ra \tilde{\a}_i\,=\,\tfrac{\a_i-1}2\,,\qquad \b_i\ra \tilde{\b}_i\,=\,\tfrac{\b_i-1}2\,.
\ee
It is now convenient to use the following compact notation for the cubic action:
\be
	S^{\sst (3)}=\frac1{3!}\int \ \delta\
	k\ e^{L\,\cV}\ \Phi_{1}\,\Phi_{2}\,\Phi_{3}\,\big|\,,
\ee
where we use as a shorthand notation $k$ and $\Phi_{i}$ in place of
$k_{a_{1}a_{2}a_{3}}$ and $\Phi^{a_{i}}(X_{i}, U_{i})$\,, while $|$ at the end of equation denotes
the evaluation $X_{i}=X$ and $U_{i}=0$\,. Performing the gauge variation with respect to $\Phi_{1}$\,, one then gets
\be
	\delta^{\sst (0)}_{E_{1}}\,S^{\sst (3)} = \frac1{3!}\int \ \delta\
	k \big[\,\cV\,,\,U_{1}\!\cdot\partial_{X_{1}}\big]\ e^{L\,\cV}
	\ E_{1}\,\Phi_{2}\,\Phi_{3}\,\big|\,,
\ee	
where the commutator $[\,\cV\,,\,U_{1}\!\cdot\partial_{X_{1}} ]$ is given by the total derivative terms:
\ba
\big[\,\cV\,,\,U_{1}\!\cdot\partial_{X_{1}}\big] &\approx&
	\partial_{X}\!\cdot\partial_{X_{23}}\,(1+\partial_{U_{2}}\!\!\cdot\partial_{U_{3}})
	+\partial_{X}\!\cdot\partial_{X_{1}}\,(\tilde\alpha_{1}+\tilde\beta_{1}\,\partial_{U_{2}}\!\!\cdot\partial_{U_{3}})\nonumber\\
	&&+\,\partial_{X}\!\cdot\partial_{U_{2}}\,\big(\tfrac12\,\partial_{X_{12}}\!\!\cdot\partial_{U_{3}}
	+\tilde\beta_{2}\,\partial_{X_{1}}\!\!\cdot\partial_{U_{3}}\big)\nn
	&&+\,\partial_{X}\!\cdot\partial_{U_{3}}\,\big(\tfrac12\,\partial_{X_{31}}\!\!\cdot\partial_{U_{2}}
	+\tilde\beta_{3}\,\partial_{X_{1}}\!\!\cdot\partial_{U_{2}}\big)\,. \qquad
\ea
After integrations by parts, one ends up with terms proportional to $X_{i}\!\cdot\partial_{X_{i}}$ and $X_{i}\!\cdot\partial_{U_{i}}$ which are exactly the operators appearing in the homogeneous and tangent conditions \eqref{tangen homo field}. To make these operators act directly on the fields, one must move them to the right of $e^{L\,\cV}$ using the commutators $[\,X_{i}\!\cdot\partial_{X_{i}}\,,\cV\,]$ and $[\,X_{i}\!\cdot\partial_{U_{i}}\,,\cV\,]$ computed in Appendix~\ref{sec: identities}. As a result one gets
\ba
	\delta^{\sst (0)}_{E_{1}}\,S^{\sst (3)} \approx -\frac1{3!}\int \ \delta^{\sst (1)}\
	k\ e^{L\,\cV}\ \cA\ E_{1}\,\Phi_{2}\,\Phi_{3}\,\big|\,,
\ea
with
\ba
	&& \cA =
	\big(1+\partial_{U_{2}}\!\!\cdot\partial_{U_{3}}\big)\\
	&&\qquad\times\,\big[\,\tfrac1L(X_{2}\!\cdot\partial_{X_{2}}-X_{3}\!\cdot\partial_{X_{3}})
	-(\tilde\alpha_{1}\,\partial_{X_{23}}+\partial_{X})\!\cdot\partial_{U_{1}}
	+(\tilde\alpha_{2}+1)\,\partial_{X_{3}}\!\!\cdot\partial_{U_{2}} \nonumber\\
	&& \hspace{50pt} -\,(\tilde\alpha_{3}-1)\,\partial_{X_{2}}\!\!\cdot\partial_{U_{3}}
	-(\tilde\beta_{1}\,\partial_{X_{23}}+\partial_{X})\!\cdot\partial_{U_{1}}\,\partial_{U_{2}}\!\!\cdot\partial_{U_{3}} \nonumber\\
	&&\hspace{50pt} +\,
	(\tilde\beta_{2}+1)\,\partial_{X_{3}}\!\!\cdot\partial_{U_{2}} \,\partial_{U_{3}}\!\!\cdot\partial_{U_{1}}
	-(\tilde\beta_{3}-1)\,\partial_{X_{2}}\!\!\cdot\partial_{U_{3}}\,\partial_{U_{1}}\!\!\cdot\partial_{U_{2}}\, \big]\nonumber\\
	&&\quad +\,\big(\tilde\alpha_{1}+\tilde\beta_{1}\,\partial_{U_{2}}\!\!\cdot\partial_{U_{3}}\big)\nonumber\\
    &&\qquad\times\,\big[\,\tfrac1L\,X_{1}\!\cdot\partial_{X_{1}}-(\tilde\alpha_{2}-1)\,\partial_{X_{1}}\!\!\cdot\partial_{U_{2}}
	-(\tilde\alpha_{3}+1)\,\partial_{X_{1}}\!\!\cdot\partial_{U_{3}}
	\nonumber\\
	&& \hspace{50pt}
	-\,(\tilde\beta_{2}-1)\,\partial_{X_{1}}\!\!\cdot\partial_{U_{2}} \,\partial_{U_{3}}\!\!\cdot\partial_{U_{1}}
	-(\tilde\beta_{3}+1)\,\partial_{X_{1}}\!\!\cdot\partial_{U_{3}}\,\partial_{U_{1}}\!\!\cdot\partial_{U_{2}}\, \big]\nonumber \\
	&&\quad +\,\big(\tfrac12\,\partial_{X_{12}}\!\!\cdot\partial_{U_{3}}
	+\tilde\beta_{2}\,\partial_{X_{1}}\!\!\cdot\partial_{U_{3}}\big)\nonumber\\
	&&\qquad\times\,\big[\,\tfrac1L\,X_{2}\!\cdot\partial_{U_{2}}
	-(\tilde\alpha_{3}-1)\,\partial_{U_{2}}\!\!\cdot\partial_{U_{3}}
	-(\tilde\alpha_{1}+1)\,\partial_{U_{1}}\!\!\cdot\partial_{U_{2}}
	\nonumber\\
	&& \hspace{50pt}
	-\,(\tilde\beta_{3}-1)\,\partial_{U_{2}}\!\!\cdot\partial_{U_{3}} \,\partial_{U_{1}}\!\!\cdot\partial_{U_{2}}
	-(\tilde\beta_{1}+1)\,\partial_{U_{1}}\!\!\cdot\partial_{U_{2}}\,\partial_{U_{2}}\!\!\cdot\partial_{U_{3}}\, \big]
	 \nn
	&&\quad +\,\big(\tfrac12\,\partial_{X_{31}}\!\!\cdot\partial_{U_{2}}
	+\tilde\beta_{3}\,\partial_{X_{1}}\!\!\cdot\partial_{U_{2}}\big)\nonumber\\
	&&\qquad\times\,\big[\,\tfrac1L\,X_{3}\!\cdot\partial_{U_{3}}
	-(\tilde\alpha_{1}-1)\,\partial_{U_{3}}\!\!\cdot\partial_{U_{1}}
	-(\tilde\alpha_{2}+1)\,\partial_{U_{2}}\!\!\cdot\partial_{U_{3}}
	\nn
	&& \hspace{50pt}
	-\,(\tilde\beta_{1}-1)\,\partial_{U_{3}}\!\!\cdot\partial_{U_{1}} \,\partial_{U_{2}}\!\!\cdot\partial_{U_{3}}
	-(\tilde\beta_{2}+1)\,\partial_{U_{2}}\!\!\cdot\partial_{U_{3}}\,\partial_{U_{3}}\!\!\cdot\partial_{U_{1}}\, \big]\,.
	\nonumber
\ea
The resulting term $\cA$ contains $X_{i}\!\cdot \partial_{X_{i}}$ and $X_{i}\!\cdot \partial_{U_{i}}$ as well as other terms coming from
the commutation relations. Here $X_{i}\!\cdot \partial_{U_{i}}$ vanishes by the tangent condition, and the homogenous condition replaces
\mt{X_{1}\!\cdot\partial_{X_{1}}} and \mt{X_{2}\!\cdot\partial_{X_{2}}-X_{3}\!\cdot\partial_{X_{3}}}
respectively, with \mt{U_{1}\!\cdot\partial_{U_{1}}} and \mt{U_{2}\!\cdot\partial_{U_{2}}-U_{3}\!\cdot\partial_{U_{3}}}\,. Since the last two depend on $U_{i}$\,, pushing them to the left of $e^{L\,\cV}$\,, they vanish when evaluated at $U_{i}=0$ and the only remaining
contributions are the commutators. Collecting all the resulting terms one finally ends up with
\ba
	\delta^{\sst (0)}_{E_{1}}\,S^{\sst (3)} \approx -\frac1{3!}\int\ \delta^{\sst (1)}\
	k\ (\,\cB+\cC\,)\ e^{L\,\cV}\ E_{1}\,\Phi_{2}\,\Phi_{3}\,\big|\,,
	\label{gv end}
\ea
where we have separated terms into the non-total-derivative part $\cB$ (which involves only $\partial_{X_{ij}^{\sst M}}$ but not $\partial_{X^{\sst M}}$) and the total-derivative part $\cC$ (which contains $\partial_{X^{\sst M}}$). If the gauge variation $\delta_{1}\,S^{\sst (3)}$ vanishes, $\cB$ should vanish as well since there is no way to compensate it. In order to simplify the discussion one can split $\cB$ as $\cB=\cB_{1}+\cB_{2}+\cB_{3}$, where the $\cB_{n}$'s are of order $n$ in the Lorentz invariants and are given respectively by
\ba
	\cB_{1}\!\!\eq\!\!\! \tfrac12\,\big[ (\tilde\alpha_{1}+1)(\tilde\alpha_{2}-1)+4\big]\,\partial_{X_{31}}\!\!\cdot\partial_{U_{2}}\nn
    &&\hspace{100pt}-\tfrac12\,\big[ (\tilde\alpha_{3}+1)(\tilde\alpha_{1}-1)+4\big]\,\partial_{X_{12}}\!\!\cdot\partial_{U_{3}}\,,
	\label {cB1}\\
	\cB_{2} \!\!\!\eq\!\!\!\tfrac12\,\big[(\tilde\alpha_{2}-1)(\tilde\beta_{1}+1)+(\tilde\alpha_{2}+1)(\tilde\beta_{3}-1)+4\big]\,
	\partial_{X_{31}}\!\!\cdot\partial_{U_{2}}\,\partial_{U_{2}}\!\!\cdot\partial_{U_{3}} \nn
	&&-\,\tfrac12\,\big[(\tilde\alpha_{3}-1)(\tilde\beta_{2}+1)+(\tilde\alpha_{3}+1)(\tilde\beta_{1}-1)+4\big]\,
	\partial_{X_{12}}\!\!\cdot\partial_{U_{3}}\,\partial_{U_{1}}\!\!\cdot\partial_{U_{2}}\,,\nn
    &&+\tfrac12\,\big[(\tilde\alpha_{1}+1)(\tilde\beta_{2}+1)+(\tilde\alpha_{1}-1)(\tilde\beta_{3}-1)\big]\nn
	&&\qquad\times\,
	\big(\partial_{X_{31}}\!\!\cdot\partial_{U_{2}}\,\partial_{U_{3}}\!\!\cdot\partial_{U_{1}}
	-\partial_{X_{12}}\!\!\cdot\partial_{U_{3}}\,\partial_{U_{1}}\!\!\cdot\partial_{U_{2}}\big) \label{cB2}\\
	\cB_{3}\!\!\!\eq\!\!\! \partial_{U_{2}}\!\!\cdot\partial_{U_{3}}\,\Big\{\,
	\tfrac12\,\big[(\tilde\beta_{1}+1)(\tilde\beta_{2}+1)+(\tilde\beta_{3}-1)(\tilde\beta_{1}+\tilde\beta_{2})\big]\,
	\partial_{X_{31}}\!\!\cdot\partial_{U_{2}}\,\partial_{U_{2}}\!\!\cdot\partial_{U_{3}} \nn
	&&\!\!\!\!\!\!-\tfrac12\,\big[(\tilde\beta_{3}+1)(\tilde\beta_{1}+1)+(\tilde\beta_{2}-1)(\tilde\beta_{3}+\tilde\beta_{1})\big]
    \partial_{X_{12}}\!\!\cdot\partial_{U_{3}}\,\partial_{U_{1}}\!\!\cdot\partial_{U_{2}}\,\Big\}\,.
	\label{cB3}
\ea
Since $\cB_{n}$'s are independent, each $\cB_{n}$ should vanish separately. Moreover, since we have considered so far the gauge consistency only with respect to $\delta^{\sst (0)}\,\Phi^{a_{1}}(X_{1},U_{1})$\, we have still to take into account the gauge invariance with respect to $\delta^{\sst (0)}\,\Phi^{a_{2}}(X_{2},U_{2})$ and $\delta^{\sst (0)}\,\Phi^{a_{3}}(X_{3},U_{3})$\,. These give the same conditions just recovered here for $\delta^{\sst (0)}\,\Phi^{a_{1}}(X_{1},U_{1})$ but with cyclic permutations on the subscripts $i$ of $\tilde\alpha_{i}$ and $\tilde\beta_{i}$\,. Finally, the equations $\cB_{1}=0$\,,
$\cB_{2}=0$  and $\cB_{3}=0$ give respectively
the conditions
\ba
	&& (\tilde\alpha_{1}+1)(\tilde\alpha_{2}-1)+4=0\,, \label{B11} \\
	&& (\tilde\alpha_{1}+1)(\tilde\beta_{2}+1)+(\tilde\alpha_{1}-1)(\tilde\beta_{3}-1)=0\,,\label{B22} \\
	&& (\tilde\beta_{1}+1)(\tilde\beta_{2}+1)+(\tilde\beta_{3}-1)(\tilde\beta_{1}+\tilde\beta_{2})=0\,, \label{B33}
\ea
that in terms of $\a_i$ and $\beta_i$ reduce exactly to eqs.~\eqref{B1}, \eqref{B2} and \eqref{B3}.

To complete the proof, one should also compute the total-derivative part $\cC$ in \eqref{gv end} and verify whether it imposes additional constraints on the
$\tilde\alpha_{i}$'s and $\tilde\beta_{i}$'s. Actually, $\cC$ is vanishing with the conditions (\ref{B1}\,-\,\ref{B3}), and hence
the latter equations are sufficient. However, this cannot be seen simply at the present level $\delta^{\sst (1)}$\,, but needs to be carefully analyzed at the next level $\delta^{\sst (2)}$\,. The details of the proof
can be found in the Appendix \ref{sec:C=0}\,.

\section{Cubic interactions from the differential equation}

In the previous section we have constructed the (A)dS lifting of all massless flat solutions considering a total derivative deformation of the corresponding flat building blocks and than enforcing eq.~\eqref{gaugeconscond1}. The existence of the deformed building blocks can be considered as a consequence of the fact that all flat vertices can be promoted to consistent (A)dS vertices. However, since the Noether equation \eqref{gaugeconscond1} is equivalent to the partial differential equations \eqref{massless system}, it can be of interest to recover the same solutions solving directly the differential equation and obtaining the result explicitly in terms of $\hat\delta$. In order to recover the solutions in this way one has to solve the differential equations \eqref{massless system} obtaining all of their polynomial solutions. This can be done, as we have anticipated, solving iteratively order by order in $\hat\delta$ the corresponding differential recurrence relations starting from the zero-th order solutions
\be
C^{(0)}_{\sst a_1a_2a_3}(Y_i,Z_i)\,=\,\cK^{(0)}_{\sst a_1a_2a_3}(Y_1,Y_2,Y_3,G)\,.
\ee
On the other hand, in order to solve the full differential equation it is convenient to consider the following ansatz
\be
\cK_{\sst a_1a_2a_3}(\hat\d,Y_i,Z_i)\,=\,\hat\cO\left(\tfrac{\hat\d}{L}\,Z_i\right)\,\cK^{(0)}_{\sst a_1a_2a_3}(Y_1,Y_2,Y_3,G)\,,
\ee
where the operator $\hat\cO$ acts on the zero-th order solution generating its lower derivative tail and contains the full explicit $Z_i$ dependence of the coupling. Having divided the explicit $Z_i$ dependence in the operator $\hat\cO$ and the $G$ dependence inside $\cK^{(0)}$ it is natural to rewrite the differential equations \eqref{massless system} in terms of partial derivatives making the following substitutions
\be
\partial_{Y_{\sst i}}\,=\,\bar{\partial}_{Y_{\sst i}}\,+\,Z_{i}\,\bar{\partial_{G}}\,,\qquad \partial_{Z_{\sst i}}\,=\,\bar{\partial}_{Z_{\sst i}}\,+\,Y_{i}\,\bar{\partial_{G}}\,.
\ee
The differential equations then becomes
\begin{multline}\label{massless system partial}
\Big[\,Y_{i+1}\bar\partial_{Z_{i-1}}-Y_{i-1}\bar\partial_{Z_{i+1}}+\tfrac{\hat\delta}{L}\left(Y_{i+1}\bar\partial_{Y_{i+1}}\, +\,Y_{i+1}Z_{i+1}\,\bar\partial_G-Y_{i-1}\bar\partial_{Y_{i-1}}\right.\\\left. -\,Y_{i-1}Z_{i-1}\,\bar\partial_G
\right)(\bar\partial_{Y_{i}}+Z_i\bar\partial_{G})\Big]\,\hat\cO\left(\tfrac{\hat\d}{L}\,Z_i\right)\,\cK^{(0)}_{\sst a_1a_2a_3}(Y_1,Y_2,Y_3,G)\,=\,0\,,
\end{multline}
and is divided into two pieces:
\begin{itemize}
\item The first piece $Y_{i-1}\bar\partial_{Z_{i+1}}-Y_{i+1}\bar\partial_{Z_{i-1}}$ whose partial derivatives acts only on $\hat\cO$,
\item The second piece of order $\tfrac{\hat\d}{L}$ whose derivatives act only on the order zero solution although it does not commute with the operator $\hat\cO$.
\end{itemize}
This structure of the differential equation supplemented by the fact that no constraint should be imposed on the order zero solution $\cK^{(0)}_{\sst a_1a_2a_3}(Y_1,Y_2,Y_3,G)$ implies the following operatorial differential equation for the operator $\hat{\cO}$:
\begin{multline}
(Y_{i+1}\bar\partial_{Z_{i-1}}-Y_{i-1}\bar\partial_{Z_{i+1}})\,\hat\cO\left(\tfrac{\hat\d}{L}\,Z_i\right)\,=\,-\,\tfrac{\hat\delta}{L} \left(Y_{i+1}\bar\partial_{Y_{i+1}}\,+\,Y_{i+1}Z_{i+1}\,\bar\partial_G\right.\\\left. -\,Y_{i-1}\bar\partial_{Y_{i-1}} -\,Y_{i-1}Z_{i-1}\,\bar\partial_G
\right)(\bar\partial_{Y_{i}}+Z_i\bar\partial_{G})\,\hat\cO\left(\tfrac{\hat\d}{L}\,Z_i\right)\,.
\end{multline}
The above equation can be easily integrated to a cyclic solution for the operator $\hat\cO$:\
\begin{multline}
\hat\cO\,=\,\exp\left[-\tfrac{\hat\d}{L}\big(Z_1\,\bar\partial_{Y_{\sst 2}}\,\bar\partial_{Y_{\sst 3}}\,+\,Z_2\,\bar\partial_{Y_{\sst 3}}\,\bar\partial_{Y_{\sst 1}}\,+\,Z_3\,\bar\partial_{Y_{\sst 1}}\,\bar\partial_{Y_{\sst 2}}\right.\\\left.+\,Z_{2}\,Z_3\,\bar\partial_{Y_{\sst1}}\bar\partial_G \,+\,Z_{3}\,Z_1\,\bar\partial_{Y_{\sst2}}\bar\partial_G\,+\,Z_{1}\,Z_2\,\bar\partial_{Y_{\sst3}}\bar\partial_G\right.\\\left. +\,Z_1\,Z_2\,Z_3\,\bar\partial_G^2\big)\right]\,,
\end{multline}
so that at the end we have recovered the most general polynomial solution of eq.~\eqref{massless system partial} and hence of the massless Noether procedure in (A)dS:
\begin{multline}
\cK_{\sst a_1a_2a_3}(\hat\d,Y_i,Z_i)\,=\,\exp\left[-\tfrac{\hat\d}{L}\big(Z_1\,\bar\partial_{Y_{\sst 2}}\,\bar\partial_{Y_{\sst 3}}\,+\,Z_2\,\bar\partial_{Y_{\sst 3}}\,\bar\partial_{Y_{\sst 1}}\,+\,Z_3\,\bar\partial_{Y_{\sst 1}}\,\bar\partial_{Y_{\sst 2}}\right.\\\left.+\,Z_{2}\,Z_3\,\bar\partial_{Y_{\sst1}}\bar\partial_G \,+\,Z_{3}\,Z_1\,\bar\partial_{Y_{\sst2}}\bar\partial_G\,+\,Z_{1}\,Z_2\,\bar\partial_{Y_{\sst3}}\bar\partial_G\right.\\\left. +\,Z_1\,Z_2\,Z_3\,\bar\partial_G^2\big)\right]\,\cK^{(0)}_{\sst a_1a_2a_3}(Y_i,G)\,,\label{explicit soluz}
\end{multline}
explicitly in terms of $\hat\d$. It is interesting to observe how the operator $\hat\cO$ generates the lower derivative tail corresponding to a given leading term in $\cK^{(0)}_{\sst a_1a_2a_3}$.
The simplest choices for the coupling function $\cK^{(0)}_{\sst a_1a_2a_3}$ is obviously the exponential so that the corresponding solution reads
\begin{multline}
\cK_{\sst a_1a_2a_3}(\hat\d,Y_i,Z_i)\,=\,\exp\left[L(Y_1+Y_2+Y_3+G)\right.\\\left.-\,{\hat\d}\big(Z_1\,+\,Z_2\,+\,Z_3\, +\,Z_{2}\,Z_3 \,+\,Z_{3}\,Z_1\,+\,Z_{1}\,Z_2\,+\,Z_1\,Z_2\,Z_3\big)\right]\,,\label{exponential}
\end{multline}
that further simplifies if one restricts the attention only to the highest derivative zero-th order solution for $\cK^{(0)}_{\sst a_1a_2a_3}$. In this case one recovers the suggestive result
\be
\cK^{\text{h.d.}}_{\sst a_1a_2a_3}(\hat\d,Y_i,Z_i)\,=\,\exp\left[L\,(Y_1+Y_2+Y_3)-{\hat\d}\big(Z_1\,+\,Z_2\,+\,Z_3\big)\right]\,,\label{highest derivative}
\ee
that has the same form of the string cubic amplitude recovered in \cite{Taronna:2010qq,Sagnotti:2010at} modulo the formal replacement
\be
\frac{\hat\d}{L}\ra\frac{2}{\a^{\prime}}\,.
\ee
Of course the above replacement is just formal but the latter analogy is pointing out how the tensor structure of the various couplings present in String Theory are the same as those arising from this very simple class of (A)dS couplings apart for the relative coefficient between pieces with a different number of derivatives. While we are planning to further investigate these analogies also in more general cases putting forward this analysis in the future, let us explicitly eliminate the $\hat\d$ auxiliary variable in this simple example exploiting eq.~\eqref{hat delta}.

Considering a generic monomial in the expansion of eq.~\eqref{highest derivative} and combining it with the radial part of the integration measure one gets:
\begin{multline}
dR \,R^{d}\,\d(R-L) \left(-\tfrac{\hat\d}{L}\right)^{m_1+m_2+m_3}Y_1^{s_1-m_2-m_3}Y_2^{s_2-m_3-m_1} Y_3^{s_3-m_1-m_2}\\Z_1^{m_1}Z_2^{m_2}Z_3^{m_3}\,R^{s_1+s_2+s_3-6}\,,
\end{multline}
where we have used the degree of homogeneity of massless fields $\D_i=s_i-2$. From the above expression one can compute the radial degree of homogeneity of each monomial that turns out to depend only on the number of derivative or on the total number $m$ of $Z_i$'s. Hence one can consider the following expansion
\be
\sum_{m=0}^\infty \tfrac{1}{m!}\ \delta(R-L)\,\left(-\tfrac{\hat\d}{L}\right)^m\ R^{d+2m-6}\ (Z_1+Z_2+Z_3)^m\,,
\ee
from which, applying eq.~\eqref{hat delta} one gets
\be
\sum_{m=0}^\infty \tfrac{1}{m!}\ \delta(R-L)\,\frac{2^m[\tfrac{d-7}{2}+m]_m}{(\s L^2)^m}\ R^{d+2m-6}\ (Z_1+Z_2+Z_3)^m\,,
\ee
that further simplifies in terms of the ascending pochhammer as
\be
\sum_{m=0}^\infty \tfrac{1}{m!}\ \delta(R-L)\,\frac{2^m(\tfrac{d-5}{2})_m}{(\s L^2)^m}\ R^{d+2m-6}\ (Z_1+Z_2+Z_3)^m\,.\label{resum}
\ee
Finally one can rewrite the full vertex as
\be
\cK^{\text{h.d.}}_{\sst a_1a_2a_3}(\hat{\d},Y_i,Z_i)\,=\,e^{\sqrt{\tfrac{\s\,L^2}{2}}\,\left(Y_1+Y_2+Y_3\right)}\,F(Z_1+Z_2+Z_3)\,,
\ee
where the function $F(z)$ can be explicitly computed resumming eq.~\eqref{resum} as
\be
F(z)\,=\,\sum_{m=0}^\infty \tfrac{1}{m!}\,(\tfrac{d-5}{2})_m\,z^m\,=\,(1-z)^{-\tfrac{d-5}{2}}\,.
\ee
A similar but slightly more complicated result can be recovered in the general case starting from \eqref{exponential} and obtaining
\begin{multline}
\cK_{\sst a_1a_2a_3}(\hat{\d},Y_i,Z_i)\,=\,e^{\sqrt{\tfrac{\s\,L^2}{2}}\,\left(Y_1+Y_2+Y_3\right)}\\\times\, F\left(\sqrt{\tfrac{\s\,L^2}{2}}\,G,Z_1+Z_2+Z_3,Z_1Z_2+Z_2Z_3+Z_3Z_1,Z_1Z_2Z_3\right)\,,
\end{multline}
in terms of a function $F$ of four variables given by
\begin{multline}
F(z_1,z_2,z_3,z_4)\,=\,\sum_{n_i=0}^\infty \tfrac{1}{n_1! n_2! n_3! n_4!}\,\left({\tiny{\tfrac{d-5}{2}+m_1+m_3+2m_4}}\right)_{m_2+m_3+m_4}\ \\\times \, z_1^{m_1}z_2^{m_2}z_3^{m_3}z_4^{m_4}\,.
\end{multline}

\section{Total derivatives or $\hat\d$?}

In the previous two sections we have seen two different presentation of the solution to the Noether procedure at the cubic level. The first is explicitly given in terms of deformed building blocks where the deformation is given by total derivatives that encode the lower derivative pieces of the vertex, while the second has been recovered directly as solution to the differential equation implied by the Noether procedure in terms of the variables $Y_i$ and $Z_i$. Both of them have their pros and cons and in this section we shall link these two different presentations carrying explicitly the integration by parts of the total derivative terms present in the building blocks. This can be of interest in order to relate different choices of the auxiliary variables $Y_i$ that are related by integrations by parts. In the following we shall also discuss the appearance of a new type of building blocks that trivialize the Noether procedure being identically tangent and gauge invariant. The latter provide in a simple way non-trivial solutions of the Noether differential equations whose precise form in terms of $Y_i$ and $Z_i$ is although very difficult to obtain because of the complicated integrations by parts involved in this case.

In order to make manifest the links between eqs.~\eqref{amb K} and \eqref{explicit soluz} in the following we shall explicitly integrate by parts all total derivative terms in \eqref{amb K}. In order to simplify the integrations by parts we can start from the simple exponential form of the coupling generating function
\begin{multline}
\d\left(\sqrt{\s X^2}-L\right)\,\cK_{\sst a_1a_2a_3}\left(\tilde{Y}_i,\tilde{G}\right)\\=\,\d\left(\sqrt{\s X^2}-L\right)\,\exp\big[Y_1\,\partial_{\l_1}+Y_2\partial_{\l_2}+Y_3\partial_{\l_3}+G\partial_\t\\+\partial_{U_1}\cdot\partial_X\, (\a_1\partial_{\l_1}+\b_1 \partial_\t Z_1)+\partial_{U_2}\cdot\partial_X\,(\a_2\partial_{\l_1}+\b_2 \partial_\t Z_2)\\+\partial_{U_3}\cdot\partial_X\,(\a_3\partial_{\l_1}+\b_3 \partial_\t Z_3)\big]\,\cK_{\sst a_1a_2a_3}\left(\l_i,\t\right)\Big|_{^{\l_i=0}_{\t=0}}\,,
\end{multline}
where we have introduced the auxiliary variables $\l_i$'s and $\t$ in order to put all relevant dependence at the exponent while we have also divided the $Y_i$ and $G$ parts from the corresponding total derivative deformation. The above expression is of the general form
\begin{multline}
\d\,\cK_{\sst a_1a_2a_3}\,=\,\d\,\exp\big[A_1\cdot\partial_X\,+\,A_2\cdot\partial_X\,+\,A_3\cdot\partial_X\\+B_1\cdot\partial_{X_1}\, +\,B_2\cdot\partial_{X_2}\,+\,B_3\cdot\partial_{X_3}\big]\,,
\end{multline}
where
\ba
A_i&=&(\a_i\partial_{\l_i}+\b_i\partial_\t\,Z_i)\,\partial_{U_i}\,,\\
B_i&=&(\partial_{\l_{i-1}}+\partial_\t Z_{i-1})\,\partial_{U_{i-1}}\,,
\ea
and can be integrated by parts term by term as
\ba
\d\,\cK_{\sst a_1a_2a_3}&=&\d\,\exp\big[A_1\cdot\partial_X\Big]\nn
&&\quad\times\,\exp\big[A_2\cdot\partial_X+A_3\cdot\partial_X+B_1\cdot\partial_{X_1} +B_2\cdot\partial_{X_2}+B_3\cdot\partial_{X_3}\big]\nn
&=&\d\,\exp\big[-\tfrac{\hat\d}{L}\,A_1\cdot X_1\Big]\nn
&&\quad\times\,\exp\big[A_2\cdot\partial_X+A_3\cdot\partial_X+B_1\cdot\partial_{X_1} +B_2\cdot\partial_{X_2}+B_3\cdot\partial_{X_3}\big]\nn
&=&\d\,\exp\big[A_2\cdot\partial_X+A_3\cdot\partial_X+B_1\cdot\partial_{X_1} +B_2\cdot\partial_{X_2}+B_3\cdot\partial_{X_3}\big]\nn
&&\quad\times\,\exp\big[\tfrac{\hat\d}{L}\,\left(A_1\cdot A_2+A_1\cdot A_3+A_1\cdot B_1\right)\big]\,.
\ea
where we have used that tangentiality implies
\be
\exp\big[-\tfrac{\hat\d}{L}\,A_1\cdot X_1\Big]\,\Phi(X_1,U_1)\,=\,\Phi(X_1,U_1)\,.
\ee
Iterating other two times the above procedure one finally gets
\begin{multline}
\d\,\cK_{\sst a_1a_2a_3}\,=\,\d\,\exp\big[B_1\cdot\partial_{X_1} +B_2\cdot\partial_{X_2}+B_3\cdot\partial_{X_3}\big]\\\times\,\exp\big[\tfrac{\hat\d}{L}\,\left(A_1\cdot A_2+A_2\cdot A_3+A_3\cdot A_1+A_1\cdot B_1+A_2\cdot B_2+A_3\cdot B_3\right)\big]\,,
\end{multline}
so that using the above expressions for $A_i$ and $B_i$ one recovers
\be
B_1\cdot\partial_{X_1} +B_2\cdot\partial_{X_2}+B_3\cdot\partial_{X_3}\,=\,Y_1\partial_{\l_1}+Y_2\partial_{\l_2}+Y_3\partial_{\l_3}+G\partial_{\t}\,,
\ee
together with
\begin{multline}
A_1\cdot A_2+A_2\cdot A_3+A_3\cdot A_1+A_1\cdot B_1+A_2\cdot B_2+A_3\cdot B_3\\=\,[\a_3(\a_2+1)]\,Z_1\,\partial_{\l_2}\partial_{\l_3}\,+\,[\a_1(\b_1+\b_3+1)+\b_2]\, Z_2Z_3\partial_{\l_1}\partial_\t\\+\,[\b_1\b_2+\b_2\b_3+\b_3\b_1+\b_1+\b_2+\b_3]\,Z_1Z_2Z_3\,\partial_\t^2\,+\,\text{cyclic}\,.
\end{multline}
Hence, exploiting eqs.~\eqref{B1}, \eqref{B2}, \eqref{B3} one finally gets
\begin{multline}
A_1\cdot A_2+A_2\cdot A_3+A_3\cdot A_1+A_1\cdot B_1+A_2\cdot B_2+A_3\cdot B_3\\=\,-\,\big[Z_1\,\partial_{\l_2}\partial_{\l_3}\,+\, Z_2Z_3\partial_{\l_1}\partial_\t\,+\,Z_1Z_2Z_3\,\partial_\t^2\,+\,\text{cyclic}\big]\,.
\end{multline}
The fact that no leftover dependence on the $\a_i$'s and $\b_i$'s is remained implies that the above two parameter family of couplings was just parameterizing different total derivative terms in (A)dS. Finally, putting together the various results one ends up with
\begin{multline}
\d\,\cK_{\sst a_1a_2a_3}\,=\,\d\,\exp\big[Y_1\partial_{\l_1}+Y_2\partial_{\l_2}+Y_3\partial_{\l_3}+G\partial_{\t}\big]\\\times\,\exp\big[- \tfrac{\hat\d}{L}\,\left(Z_1\,\partial_{\l_2}\partial_{\l_3}\,+\, Z_2Z_3\partial_{\l_1}\partial_\t\,+\,Z_1Z_2Z_3\,\partial_\t^2\,+\,\text{cyclic}\right)\big]\\\times\,\cK_{\sst a_1a_2a_3}\left(\l_i,\t\right)\Big|_{^{\l_i=0}_{\t=0}}\,.
\end{multline}
Now since the first exponential is just a translation operator one can eliminate the auxiliary variables ending up with
\begin{multline}
\d\,\cK_{\sst a_1a_2a_3}=\d\exp\big[- \tfrac{\hat\d}{L}\,\left(Z_1\,\bar\partial_{Y_2}\bar\partial_{Y_3}\,+\, Z_2Z_3\bar\partial_{Y_1}\bar\partial_G\,+\,Z_1Z_2Z_3\,\bar\partial_G^2\,+\,\text{cyclic}\right)\big]\\\times\,\cK_{\sst a_1a_2a_3}\left(Y_i,G\right)\,,
\end{multline}
whose form coincides with eq.~\eqref{explicit soluz}. With this techniques one can easily change the convention of the $Y_i$'s, that here have been chosen as in \eqref{Y and Z}, to any other convention. For instance in the antisymmetric convention
\be
Y_i^{\text{a}}\,=\,\partial_{U_i}\cdot\partial_{X_{i+1,i-1}}\,,
\ee
the corresponding solution in terms of $\hat{\delta}$ reads
\begin{multline}
\d\,\cK_{\sst a_1a_2a_3}\,=\,\d\,\exp\big[- \tfrac{\hat\d}{L}\,\big(3\,Z_1\,\bar\partial_{Y_2}\bar\partial_{Y_3}\,+\,2\, Z_2Z_3\bar\partial_{Y_1}\bar\partial_G\,+\,Z_1Z_2Z_3\,\bar\partial_G^2\\+\,\text{cyclic}\big)\big]\,\cK_{\sst a_1a_2a_3}\left(Y_i^{\text{a}},G^{\text{a}}\right)\,,
\end{multline}
while in the following convention
\be
Y^c_1\,=\,\tfrac{1}{2}\,\partial_{U_1}\cdot\partial_{X_{23}}\,,\qquad Y^c_2\,=\,\partial_{U_2}\cdot\partial_{X_3}\,,\qquad Y^c_3\,=\,-\,\partial_{U_3}\cdot\partial_{X_2}\,,
\ee
that is useful in order to extract the corresponding Noether currents the coupling looks like
\begin{multline}
\cK_{\sst a_1a_2a_3}(\hat\d,Y_i,Z_i)\\=\,\exp\left[-\,\tfrac{\hat\d}{L}\big(Z_2\,\bar\partial_{Y_3}\bar\partial_{Y_1}\,+\, Z_3\,\bar\partial_{Y_1}\bar\partial_{Y_2}\, +\,Z_{2}\,Z_3\,\bar\partial_{Y_1}\bar\partial_G\big)\right]\,\cK_{\sst a_1a_2a_3}\left(Y_i^{\text{c}},G^{\text{c}}\right)\,.\label{vertex currents}
\end{multline}
One can then extract the Ambient Noether currents as
\begin{multline}
J^a(X,U)\,=\,\exp\left[-\,\tfrac{\hat\d}{L}\big(Z_2\,\bar\partial_{Y_3}\bar\partial_{Y_1}\,+\, Z_3\,\bar\partial_{Y_1}\bar\partial_{Y_2}\, +\,Z_{2}\,Z_3\,\bar\partial_{Y_1}\bar\partial_G\big)\right]\\\times\,{\cK^a} _{\sst a_2a_3}\left(Y_i^{\text{c}},G^{\text{c}}\right)\,\Phi^{a_2}(X_2,U_2)\,
\Phi^{a_3}(X_3,U_3){\Big|}_{_{X_i=X}^{U_2=U_3=0}}^{\partial_{U_1}\rightarrow U^{\text{T}}}\,.
\end{multline}
The latter are conserved modulo the equations of motion while we have defined the tangent auxiliary variable
\be
U^{\text{T}}_{M}\,=\,U_M\,-\,\frac{U\cdot X}{X^2}\,X_M\,,
\ee
in order to translate the tangentiality constraint on the fields in terms of tangent currents.

As we have anticipated, before closing this section we want to describe a very simple class of solutions to the Noether procedure equations. Indeed, our strategy has been to explicitly solve the differential equations starting from the zero-th order solution or to deform the latter with total derivatives in order to control both the number of solutions and their non-triviality. On the other hand we could have solved the commutator equation \eqref{gaugeconscond1} directly obtaining in principle a subclass of the solutions. From this respect it can be interesting to notice that the following building blocks
\be
\tilde H_i\,=\,\partial_{U_{i-1}}\!\cdot\partial_{X_{i+1}}\,\partial_{U_{i+1}}\!\cdot\partial_{X_{i-1}}\,- \,\partial_{X_{i+1}}\!\cdot\partial_{X_{i-1}}\,Z_i\,,\label{H}
\ee
are tangent and identically gauge-invariant:
\be
\big[X_i\cdot\partial_{U_i},\tilde H_j\big]\,=\,0\,,\qquad \big[\tilde H_j,U_i\cdot\partial_{X_i}\big]\,=\,0\,.
\ee
Hence, the above identities actually imply that any function
\be
C_{\sst a_1a_2a_3}\,=\,\cK_{\sst a_1a_2a_3}(\tilde H_1,\tilde H_2,\tilde H_3)\,,
\ee
is a solution to the Noether procedure. This can be considered a trivial solution because it does not need to be completed further and does not require any TT constraint for the gauge invariance. On the other hand this set of couplings cover only a very limited class of the whole set that we have found above.

\section{Reduction to (A)dS-intrinsic expressions}

The cubic vertices \eqref{amb K} or \eqref{explicit soluz} constructed in the ambient-space formalism are given in terms of the Lorentz invariants \mt{\partial_{U_{i}}\!\!\cdot\partial_{X_{j}}} and \mt{\partial_{U_{i}}\!\!\cdot\partial_{U_{j}}}\,, and the expressions are compact but implicit with respect to (A)dS.
The explicit expressions in terms of (A)dS-intrinsic quantities can be obtained making use of the radial reduction formulas recovered in Section~\ref{(A)dS Geometry}. A convenient way for the reduction is to express the Lorentz invariants in terms of the following (A)dS-intrinsic bi-local quantities\footnote{In the following we shall concentrate on $\s=1$ remembering that the $\s=-1$ case can be obtained by analytic continuation $L\ra iL$.}:
\ba
	Z(x_{i},x_{j}) \edf \hat X_{\sst M}(x_{i})\,\hat X^{\sst M}(x_{j})\,,\\
	H_{\mu}(x_{i},x_{j}) \edf L\,\frac{\partial \hat X_{\sst M}(x_{i})}{\partial x_{i}^{\mu}}\,\hat X^{\sst M}(x_{j})\,,\\
	G_{\mu\nu}(x_{i},x_{j}) \edf L^{2}\,
	\frac{\partial \hat X_{\sst M}(x_{i})}{\partial x_{i}^{\mu}}
	\frac{\partial \hat X^{\sst M}(x_{j})}{\partial x_{j}^{\nu}}\,,
	\label{bilocal}
\ea
whose coincident-point limits are given by
\be
	Z(x,x)=1\,, \qquad H^{\mu}(x,x)=0\,, \qquad G^{\mu\nu}(x,x)=g^{\mu\nu}(x)\,.
	\label{prop bilocal}
\ee
Here the indices of the bi-local quantities are raised or lowered with the local metric tensor.
With these conventions, the ambient-space Lorentz invariant operators can be written as
\ba
	\partial_{U_{i}}\!\cdot\partial_{X_{j}}\eq
	\Big[\partial_{v_{i}}\,Z(x_{i},x_{j})+\partial_{u_{i}^{\mu}}\,H^{\mu}(x_{i},x_{j})\Big]\,
	\partial_{R_{j}} \nn
	&& \!\!\!\!\!\!\!\!\!\!\!\!\!\!\!\!\!\!+\,
	\Big[\partial_{v_{i}}\,H^{\nu}(x_{j},x_{i})+\partial_{u_{i}^{\mu}}\,G^{\mu\nu}(x_{i},x_{j})\Big]
	\Big[D_{j\,\nu}+\tfrac1L\,(u_{j\,\nu}\,\partial_{v_{j}}-v_{j}\,\partial_{u_{j}^{\nu}})\Big]
	\,\tfrac{L}{R_{j}}\,, \quad
	\label{div ij} \nn
	\partial_{U_{i}}\!\cdot\partial_{U_{j}}\eq
	\partial_{v_{i}}\,Z(x_{i},x_{j})\,\partial_{v_{j}}
	+ \partial_{u_{i}^{\mu}}\,H^{\mu}(x_{i},x_{j})\,\partial_{v_{j}}
	+ \partial_{v_{i}}\,H^{\nu}(x_{j},x_{i})\,\partial_{u_{j}^{\nu}} \nn
	&& +\,\partial_{u_{i}^{\mu}}\,G^{\mu\nu}(x_{i},x_{j})\,\partial_{u_{j}^{\nu}}\,,
	\label{tr ij}
\ea
where we have also introduced the compact notation
\be
	u^{\mu}_{i}:=u_{i}^{\alpha} \ e_{\alpha}^{\ \mu}(x_{i})\,,
	\qquad
	\partial_{u_{i}^{\mu}}:=\partial_{u_{i}^{\alpha}} \ e^{\alpha}_{\ \mu}(x_{i})\,.
\ee
These quantities are more convenient than the flat auxiliary variables for the explicit computations since they commute with the (A)dS-covariant derivative:
\be
	[\,D_{i\,\mu}\,,\, u_{j}^{\nu}\,]=0\,,\qquad	[\,D_{i\,\mu}\,,\, \partial_{u_{j}^{\nu}}\,]=0\,.
\ee
The advantage of the bi-local quantities \eqref{bilocal} rests on the fact that they are closed under the action of the (A)dS-covariant derivatives, as one can see by explicit computation:
\ba
	D_{i\,\mu}\,Z(x_{i},x_{j})=\tfrac1L\,H_{\mu}(x_{i},x_{j})\,,&
	&D_{i\,\nu}\,H_{\mu}(x_{i},x_{j})=-g_{\mu\nu}(x_{i})\,\tfrac1L\,Z(x_{i},x_{j})\,, \nn
	D_{j\,\nu}\,H_{\mu}(x_{i},x_{j})=\tfrac1L\,G_{\mu\nu}(x_{i},x_{j})\,,&
	&D_{i\,\rho}\,G_{\mu\nu}(x_{i},x_{j})=-g_{\rho\mu}(x_{i})\,\tfrac1L\,H_{\nu}(x_{j},x_{i})\,.\nn	
	\label{der bilocal}
\ea
Therefore, the ambient-space cubic vertices \eqref{amb K} can be reduced to the (A)dS-intrinsic ones with some algebra. Notice that the ambient-space derivatives $\partial_{X_{i}^{\sst M}}$ do not always reduce to
the (A)dS covariant ones $D_{i\,\mu}$\,, but they can produce some powers of $1/L^{2}$ either via the contractions between $\partial_{v_{i}}/L$'s and $v_{i}/L$'s or
via the actions on bi-local quantities.
Hence, an ambient-space vertex with a number $\Delta$
of ambient-space derivatives results in a tail of (A)dS vertices whose number of covariant derivatives
varies within the range \mt{\Delta,\, \Delta-2,\, \cdots ,1} (or $0$).
Whenever the number of derivatives decreases by two,
the corresponding mass-dimension is compensated by a factor $1/L^{2}$\,.

\subsection{Example: 3$-$3$-$2 vertex with lowest number of derivatives}

Let us deal with an explicit example in order to see how this radial reduction works.
We have chosen the 3$-$3$-$2 example with the least number of derivatives because it is both one of the simplest examples of HS interactions and one of the vertices constructed by FV in the frame-like formalism.
The 3$-$3$-$2 vertex was also obtained in  \cite{Boulanger:2008tg, Zinoviev:2008ck}
in terms of metric-like fields.

For simplicity, we leave aside the Chan-Paton factors and choose $\alpha$ and $\beta$ in a way\footnote{We take the \mt{\e\to0} limit with  \mt{\alpha=1-2\e} and \mt{\beta=1+\e}\,. Even though the second line's last factor in \eqref{amb K} diverges, it does not matter since we consider the case $n=s_{3}$\,.} that the cubic action has a symmetric form:
\ba
	S^{\sst (3)} \!\!\!\eq\!\!\! -\frac{2}{3}\, g^{\sst 332,2}\,
	\int d^{d+1}X\ \delta\Big(\sqrt{X^{2}}-L\Big)\
	 G^{2}\,
	\partial_{U_{1}}\!\!\cdot\partial_{X_{2}}\,\partial_{U_{2}}\!\!\cdot\partial_{X_{1}}
	\\
	&& \hspace{80pt}\times\,
	\Phi^{\sst (3)}(X_{\sst 1},U_{\sst 1})\
	\Phi^{\sst (3)}(X_{\sst 2},U_{\sst 2})\ \Phi^{\sst (2)}(X_{\sst 3},U_{\sst 3})\,
	\Big|_{\overset{X_{i}=X}{\sst U_{i}=0}}\,,\nonumber
	\label{ex 332}
\ea
where $G$ is given by
\be
	G=2\,\big[
	\partial_{U_{2}}\!\!\cdot\partial_{U_{3}} \partial_{U_{1}}\!\!\cdot\partial_{X_{2}}\,
	-\partial_{U_{1}}\!\!\cdot\partial_{U_{3}}\,\partial_{U_{2}}\!\!\cdot\partial_{X_{1}}
	+ \tfrac12\,\partial_{U_{1}}\!\!\cdot\partial_{U_{2}}\,\partial_{U_{3}}\!\!\cdot\partial_{X_{12}}\big]\,.
	\label{G11}
\ee
Expanding $G^{2}$ gives rise to six terms,
and in order to describe the procedure let us consider first the term
\be
	(\partial_{U_{2}}\!\!\cdot\partial_{U_{3}})^{2}\,(\partial_{U_{1}}\!\!\cdot\partial_{X_{2}})^{3}\,
	\partial_{U_{2}}\!\!\cdot\partial_{X_{1}}\,.
	\label{1 term}
\ee
Using \eqref{tr ij} and \eqref{div ij}, one gets
{\footnotesize
\ba
	&&
	\Big[ \partial_{v_{2}}\,Z_{\sst 23}\,\partial_{v_{3}}
	+\partial_{u_{2}}\!\!\cdot H^{\sst 2}_{\sst \ \,3}\,\partial_{v_{3}}\,
	+ \partial_{u_{3}}\!\!\cdot H^{\sst 3}_{\sst \ \,2}\,\partial_{v_{2}}\,
	+ \partial_{u_{2}}\!\!\cdot G^{\sst 23}\!\!\cdot \partial_{u_{3}} \Big]^{2}
	\\\ &&
	\times\,\Big[(\partial_{v_{1}}\,Z_{\sst 12}+
	\partial_{u_{1}}\!\!\cdot H^{\sst 1}_{\sst\ \,2})\,
	\partial_{R_{2}} +
	(\partial_{v_{1}}\,H_{\sst 1}^{\sst\ 2}+\partial_{u_{1}}\!\!\cdot G^{\sst 12})\!\cdot\!
	\big[D_{\sst 2}+\tfrac1L\,(u_{\sst 2}\,\partial_{v_{2}}-v_{\sst 2}\,\partial_{u_{2}})\big]
	\,\tfrac{L}{R_{2}}\Big]^{3}
	\nn &&
	\times\,\Big[(\partial_{v_{2}}\,Z_{\sst 21}+
	\partial_{u_{2}}\!\!\cdot H^{\sst 2}_{\sst\ \,1})\,
	\partial_{R_{1}} +
	(\partial_{v_{2}}\,H_{\sst 2}^{\sst\ 1}+\partial_{u_{2}}\!\!\cdot G^{\sst 21})\!\cdot\!
	\big[D_{\sst 1}+\tfrac1L\,(u_{\sst 1}\,\partial_{v_{1}}-v_{\sst 1}\,\partial_{u_{1}})\big]\,
	\tfrac{L}{R_{1}}\Big]\,,\nonumber
	\label{intm}
\ea}
\!\!where the subscripts or superscripts of $Z, H$ and $G$ encode the bi-local dependence on $(x_{i},x_{j})$\,, in particular,
\be
(H^{\sst i}_{\sst\ \,j})_{\mu}=H_{\mu}(x_{i},x_{j})\ ,\qquad (H_{\sst i}^{\sst\ j})_{\mu}=H_{\mu}(x_{j},x_{i})\,.
\ee
Even though eq.~\eqref{intm} has a rather complicated structure, many simplifications can be made. First, since
the operator \eqref{1 term} is acting on
\be
	 \tfrac{R_{1}}L\,\tfrac{R_{2}}L\
	 \varphi^{\sst (3)}(x_{1},u_{1})\,\varphi^{\sst (3)}(x_{2},u_{2})\,
	 \varphi^{\sst (2)}(x_{3},u_{3})\,\big|_{u_{i}=v_{i}=0}\,,
	 \label{hom cond}
\ee
the dependence in $R_{i}$ and $v_{i}$  can be removed
performing all possible contractions.
Second, the coincident limit \eqref{prop bilocal} simplifies some of the bi-local quantities, and the formula \eqref{intm} becomes
{\small
\ba
	&& \!\!\!\!\!\!\!\!\!\!\!\!(\partial_{u_{2}}\!\!\cdot\partial_{u_{3}})^{2}
	\,\Big[\, \Big\{ (\partial_{u_{1}}\!\!\cdot D_{\sst 2})^{2}
	(\partial_{u_{1}}\!\!\cdot G^{\sst 12}\!\!\cdot D_{\sst 2}+\tfrac1L\,\partial_{u_{1}}\!\!\cdot H^{\sst 1}_{\sst\ \,2})
	-\tfrac3{L^{2}}\,u_{2}\!\cdot\partial_{u_{1}}\,\partial_{u_{1}}\!\!\cdot\partial_{u_{2}}\,
	\partial_{u_{1}}\!\!\cdot D_{\sst 2}\Big\}\times\nn
	&&\hspace{65pt} \times\,
	(\partial_{u_{2}}\!\!\cdot G^{\sst 21}\!\!\cdot D_{\sst 1}
	+\tfrac1L\,\partial_{u_{2}}\!\!\cdot H^{\sst 2}_{\sst\ \,1}) \nn
	&&\hspace{40pt} +\,\tfrac1{L}\,\Big\{
	\tfrac1L\,(\partial_{u_{1}}\!\!\cdot D_{\sst 2})^{2}
	-\partial_{u_{1}}\!\!\cdot D_{\sst 2}\ H_{\sst 1}^{\sst\ 2}\!\cdot D_{\sst 2}\,
	(\partial_{u_{1}}\!\!\cdot G^{\sst 12}\!\!\cdot D_{\sst 2}+\tfrac1L\,\partial_{u_{1}}\!\!\cdot H^{\sst 1}_{\sst\ \,2})
	 \nn
	&&\hspace{50pt} +\,\tfrac1{L^{2}}\,\partial_{u_{1}}\!\!\cdot D_{\sst 2}\ u_{2}\!\cdot H^{\sst 2}_{\sst\ \,1}\,
	\partial_{u_{1}}\!\!\cdot G^{\sst 12}\!\!\cdot \partial_{u_{2}}
	-(\partial_{u_{1}}\!\!\cdot D_{\sst 2})^{2}(H_{\sst 1}^{\sst\ 2}\!\cdot D_{\sst 2}+\tfrac1L\,Z_{\sst 12}) \nn
	&& \hspace{50pt}+\,
	\tfrac2{L^{2}}\,u_{2}\!\cdot\partial_{u_{1}}\,\partial_{u_{1}}\!\!\cdot D_{\sst 2}\
	\partial_{u_{2}}\!\!\cdot H^{\sst 2}_{\sst\ \,1}\, \Big\}\,
	\partial_{u_{1}}\!\!\cdot G^{\sst 12}\!\!\cdot \partial_{u_{2}}\,\Big]\,.
	\label{332 1}
\ea}
\!\!Finally, the property \eqref{der bilocal} enables one to remove all bi-local quantities
replacing them with some powers of $L$\,. At the end, one obtains the (A)dS intrinsic expression for the operator \eqref{1 term} as
\be
	(\partial_{u_{2}}\!\!\cdot\partial_{u_{3}})^{2}\,
	(\partial_{u_{1}}\!\!\cdot D_{\sst 2})^{3}\,
	\partial_{u_{2}}\!\!\cdot D_{\sst1}
	-\tfrac6{L^{2}}\,
	\partial_{u_{1}}\!\!\cdot\partial_{u_{2}}\,
	\partial_{u_{1}}\!\!\cdot\partial_{u_{3}}\,
	\partial_{u_{2}}\!\!\cdot\partial_{u_{3}}\,
	\partial_{u_{1}}\!\!\cdot D_{\sst 2}\,\partial_{u_{2}}\!\!\cdot D_{\sst1}\,.
	\label{332 1 ds}
\ee
Notice that the first term has the same form of \eqref{1 term} with the replacement of $(\partial_{X_{i}}\,,\partial_{U_{i}})$ by $(D_{\sst i}\,,\,\partial_{u_{i}})$, but the second term has a lower number of derivatives and is proportional to $1/L^{2}$\,.

Five other terms in the expansion \eqref{ex 332} can be computed in a similar way (see Appendix \ref{sec:reduc} for more details),
and the cubic action \eqref{ex 332} can be finally expressed solely in terms of (A)dS intrinsic quantities as
{\small
\ba
	&& \!\!\!\!\!\!\!\!\!\!\!S^{\sst (3)} = -\frac{8}{3}\, g^{\sst 332,2}\,
	\int d^{d}x\,\sqrt{-g}\,\times \\
	&& \times\,\Big[
	\left(
	\partial_{u_{2}}\!\!\cdot\partial_{u_{3}}\,\partial_{u_{1}}\!\!\cdot D_{\sst 2}
	-\partial_{u_{1}}\!\!\cdot\partial_{u_{3}}\,\partial_{u_{2}}\!\!\cdot D_{\sst 1}
	+\tfrac12\,\partial_{u_{1}}\!\!\cdot\partial_{u_{2}}\,\partial_{u_{3}}
	\!\!\cdot D_{\sst 12}\right)^{2}
	\partial_{u_{1}}\!\!\cdot D_{\sst 2}\,
	\partial_{u_{2}}\!\!\cdot D_{\sst 1}\nn
	&& \quad\
	+\,\tfrac4{L^{2}}\,\partial_{u_{1}}\!\!\cdot\partial_{u_{2}}
	\left[
	(\partial_{u_{2}}\!\!\cdot\partial_{u_{3}})^{2}
	(\partial_{u_{1}}\!\!\cdot D_{\sst 2})^{2}
	+
	(\partial_{u_{1}}\!\!\cdot\partial_{u_{3}})^{2}
	(\partial_{u_{2}}\!\!\cdot D_{\sst 1})^{2}
	\right.\nn
    &&\left.\qquad\qquad\qquad\qquad\qquad\qquad\qquad\qquad-3\,\partial_{u_{1}}\!\!\cdot\partial_{u_{3}}\,\partial_{u_{2}}\!\!\cdot\partial_{u_{3}}
	\,\partial_{u_{1}}\!\!\cdot D_{\sst 2}\,\partial_{u_{2}}\!\!\cdot D_{\sst 1}
	\right] \nn
	&& \quad\
	+\,\tfrac3{L^{2}}\,(\partial_{u_{1}}\!\!\cdot\partial_{u_{2}})^{2}\,
	\left[	\partial_{u_{2}}\!\!\cdot\partial_{u_{3}}\,
	\partial_{u_{1}}\!\!\cdot D_{\sst 2}
	-\partial_{u_{1}}\!\!\cdot\partial_{u_{3}}\,
	\partial_{u_{2}}\!\!\cdot D_{\sst 1}
	\right.\nn
    &&\left.\qquad\qquad\qquad\qquad\qquad\qquad\qquad\qquad\qquad+\tfrac16\,\partial_{u_{1}}\!\!\cdot\partial_{u_{2}}\,
	\partial_{u_{3}}\!\!\cdot D_{\sst 12}
	\right] \partial_{u_{3}}\!\!\cdot D_{\sst 12}
	\nn
	&& \quad\
	-\,\tfrac5{4\,L^{2}}\,(\partial_{u_{1}}\!\!\cdot\partial_{u_{2}})^{2}\,
	\left[	\partial_{u_{2}}\!\!\cdot\partial_{u_{3}}\,
	\partial_{u_{1}}\!\!\cdot D_{\sst 2}\,\partial_{u_{3}}\!\!\cdot D_{\sst 1}
	+\partial_{u_{1}}\!\!\cdot\partial_{u_{3}}\,
	\partial_{u_{2}}\!\!\cdot D_{\sst 1}\,\partial_{u_{3}}\!\!\cdot D_{\sst 2}
	\right] 	\nn
	&&\quad\
	-\,\tfrac{7d+29}{2\,L^{4}}\,
	(\partial_{u_{1}}\!\!\cdot\partial_{u_{2}})^{2}
	\partial_{u_{1}}\!\!\cdot\partial_{u_{3}}\,\partial_{u_{2}}\!\!\cdot\partial_{u_{3}}
	\Big]\nn
    &&\qquad\qquad\qquad\qquad\qquad\times\,\varphi^{\sst (3)}(x_{\sst1},u_{\sst 1})\,
	\varphi^{\sst (3)}(x_{\sst 2},u_{\sst 2})\,\varphi^{\sst (2)}(x_{\sst 3},u_{\sst 3})\,
	\Big|_{\overset{x_{1}=x_{2}=x_{3}=x}{\sst u_{1}=u_{2}=u_{3}=0}}\,,\nonumber
	\label{ex 332 dS}
\ea}
\!\!where we organized the various contributions according to
the number of (A)dS covariant derivatives.


\section{Discussion}
\label{sec: discussions}

In this Chapter, we have obtained the TT part of the general solution to the cubic-interaction problem of HS gauge fields in (A)dS.
Interestingly, the structure of the vertices, when expressed in the ambient-space formalism, coincides with the flat-space ones up to non-trivial total-derivative terms whose form is completely constrained by the gauge consistency. This observation underlines the key role of the simpler YM couplings from which all possible HS interactions can be recovered systematically in terms of powers of the former. This resonates with the observations made in \cite{Bern:2010ue} in the simpler gravity case and follows here just as a consequence of the gauge principle behind the Noether procedure, as observed in \cite{Taronna:2011kt}. On the other hand, we have been able to explicitly carry out the integrations by parts relating together different total derivative deformations of the same fundamental building blocks modulo lower derivative pieces recovering in this way different but equivalent presentations of the same results.

\section{Relation to the Fradkin-Vasiliev vertices}

Let us consider the \mt{s\!-\!s\!-\!2} vertices, which
were originally constructed by FV.
They correspond to the case \mt{s_{1}=s_{2}=s} and \mt{s_{3}=n=2} in \eqref{amb K}, so that they are the vertices with lowest number of derivatives.
With the same choice of $\alpha$ and $\beta$ as in \eqref{ex 332},
they are given by
\ba
	S^{\sst (3)} \!\!\!\eq\!\!\! g^{ss{\sst 2,2}}\,
	\int {d^{d+1}X}\ \delta\Big(\sqrt{\s X^{2}}-L\Big)\
	 G^{\,2}\,
	(\partial_{U_{1}}\!\!\cdot\partial_{X_{2}}\,\partial_{U_{2}}\!\!\cdot\partial_{X_{1}})^{s-2}
	\nn
	&& \hspace{70pt}\times\,
	\Phi^{\sst (s)}(X_{\sst 1},U_{\sst 1})\
	\Phi^{\sst (s)}(X_{\sst 2},U_{\sst 2})\ \Phi^{\sst (2)}(X_{\sst 3},U_{\sst 3})\,
	\Big|_{\overset{X_{i}=X}{\sst U_{i}=0}}\,,\qquad
	\label{ss2}
\ea
where, for simplicity, we have absorbed a numerical factor into the definition of
the coupling constant. In the previous section, we have shown how to express ambient differential operators in terms of (A)dS-intrinsic quantities.
Likewise, expressing the operators
\mt{(\partial_{U_{1}}\!\!\cdot\partial_{X_{2}}\,\partial_{U_{2}}\!\!\cdot\partial_{X_{1}})^{s-2}}
 in the above formula using \eqref{div ij} yields an expression in terms of (A)dS-covariant derivatives, bi-local quantities and also the $v_{i}$'s. Taking the ordering where all
 (A)dS-covariant derivatives are placed on the RHS, one gets
\be
	(\partial_{U_{1}}\!\!\cdot\partial_{X_{2}}\,\partial_{U_{2}}\!\!\cdot\partial_{X_{1}})^{s-2}
	=\cA_{s-2}+\Lambda\,\cA_{s-3}+\cdots+\Lambda^{s-2}\,\cA_{0}\,,
	\label{A exp}
\ee
where $\cA_{r}$ is the portion containing the $2r$-th power of the (A)dS-covariant derivatives, or $\cA_{r}\propto D^{2r}$\,. Plugging \eqref{A exp} into \eqref{ss2}, the \mt{s\!-\!s\!-\!2} vertex admits a similar expansion given by
\be
	S^{\sst (3)} = g^{ss{\sst 2,2}}\left[
	A_{s}+\Lambda\,A_{s-1}+\cdots+\Lambda^{s-2}\,A_{2} \right],
	\label{S exp}
\ee
with
\be
	A_{r+2} = \int d^{d+1}X\ \delta\Big(\sqrt{\s X^{2}}-L\Big)\,
	 G^{\,2}\,\cA_{r}\
	\Phi^{\sst (s)}\, \Phi^{\sst (s)}\, \Phi^{\sst (2)}\,
	\Big|_{\overset{X_{i}=X}{\sst U_{i}=0}}\,.
\ee
Notice that each $A_{r}$ is separately gauge invariant under the spin 2 gauge transformations,
and this is due to the fact that the $\cA_{r}$'s trivially commute with the spin 2 gauge transformations.
Notice as well that $A_{r}$ involves $2(r-2)$ or $2(r-1)$ (A)dS covariant derivatives,
since the action of $G^{2}$ may or may not add two additional derivatives.

This expansion of the vertex is quite similar to the one obtained by FV, and in fact
one can make it as an expansion in inverse powers of $\Lambda$
by redefining the coupling constant and the fields as
\be
	g^{ss{\sst 2,2}}=\frac{\sqrt{G}}{\Lambda^{s-2}}\ \lambda_{s}\,, \qquad
	\varphi^{\sst (2)}=\tfrac1{\sqrt{G}}\,h\,,
	\qquad
	\varphi^{\sst (s)}=\tfrac1{\sqrt{G}}\,\phi^{\sst (s)}\,.
	\label{redef F}
\ee
The coupling constant $g^{ss{\sst 2,2}}$ has mass-dimension $(2-d)/2-2(s-2)$\,,
while with this redefinition the new coupling constant $\lambda_{s}$ together with the new fields
have vanishing mass-dimension.
Finally, the expansion \eqref{S exp} becomes
\be
	S^{\sst (3)} = \frac{\lambda_{s}}G\left[ \tilde A_{2}
	+\frac1\Lambda\, \tilde A_{3}+\cdots+\frac1{\Lambda^{s-2}}\, \tilde A_{s} \right],
\ee
where $\tilde A_{r}$'s are given schematically by
\be
\cL_r=D^{2(r-1)}h\,\phi^{(s)}\phi^{(s)}+\L\,D^{2(r-2)}h\,\phi^{(s)}\phi^{(s)}\,,
\ee
in terms of dimensionless fields $\phi^{\sst (s)}$ and $h$\,.

Two remarks are in order. First, the lowest-derivative part $\tilde A_{2}$ of the above expression
should involve the gravitational minimal coupling as well as non-minimal ones which do not
deform the gauge transformations.
Therefore, the simplest way to see this link is to analyze how the vertices here reviewed
deform the gauge transformations and the gauge algebra.
We leave this issue for future work.
Second, the highest-derivative part
(the so-called \emph{seed coupling}, according to \cite{Sagnotti:talk09})
has the same form as
the flat-space vertices with $\partial_{x_{i}^{\mu}}$'s replaced by $D_{i\,\mu}$'s.
The relation between the gravitational minimal coupling and the seed coupling was already noticed in \cite{Boulanger:2008tg},
and in the present Thesis we can see how
both lower-derivative and seed couplings come out at the same time from the ambient-space vertices.
From a more general perspective, it would be interesting to investigate the relation between the present construction (in metric-like approach) and the recent frame-like one of \cite{Vasiliev:2011xf}.

\paragraph{Boulanger-Leclercq-Sundell limit}

Since a curved space looks flat in the short-distance limit, the dominant term of the curved-space actions in the limit should correspond to flat-space ones. One may expect to obtain in this way the flat-space vertices from the FV ones, but because of the inverse power expansion in $\Lambda$
the dominant terms diverge in the limit. In \cite{Boulanger:2008tg}, the authors considered a particular limit of the FV system in order to extract flat-space information from AdS interactions. More precisely, they considered the limit where not only the cosmological constant but also the gravitational constant and the fields scale as
\ba
	&\Lambda=\epsilon\,\tilde\Lambda\,,\qquad
	&G=\epsilon^{2(s-2)}\,\tilde G\,,\nn
	& h=\epsilon^{s-2}\,\tilde h\,,\qquad
	& \phi^{\sst (s)}= \epsilon^{s-2}\,\tilde\phi^{\sst (s)}\,,
	\label{BLS}
\ea
with $\epsilon\to0$\,. Under this rescaling, the quadratic action remains invariant, but the cubic vertices scale in a way that only the seed coupling survives and one gets the flat-space vertices with $2s-2$ derivatives.

In our setting, this can be understood at the level of \eqref{S exp}, where the flat-space limit is not singular for fixed (or non-scaling) $g^{ss{\sst 2,2}}$ and $\varphi^{\sst (s)}$'s, and the flat-space cubic vertices are recovered. In this respect, the rescaling \eqref{BLS} can be viewed as a particular flat-space limit in \eqref{redef F} which holds $g^{ss{\sst 2,2}}$ and the $\varphi^{\sst (s)}$'s finite.

Let us conclude by summarizing our results and our strategy and by describing their possible applications from a more general perspective also in order to introduce the next Chapter where we are going to try and extend as much as possible the above analysis at higher orders.

First, we observed that the flat-space interactions play a key role, through the ambient-space formalism, in understanding and controlling cubic interactions in any constant curvature background. Second, the simplified TT (or \emph{on-shell}) system makes possible to identify and classify the consistent cubic interactions dividing the problem of finding them from the problem of computing their completion that can be studied later. From this respect we have in mind a direct application of these results to higher order amplitude computations in any constant curvature background as we shall see in the next Chapter. Indeed, the latter problem does not require in principle the aforementioned completions. We expect as well that many other key properties of the interactions can be appreciated already at this simpler level even thought we want to stress the importance of the completion in order to arrive at full consistent Lagrangians order by order in the number of fields.

The aforementioned perspectives open a new window for a systematic analysis of many other aspects of HS theory.
First of all, the issue of non-Abelian HS gauge algebras in (A)dS and flat space together with their relations might be addressed\footnote{
See  \cite{Bekaert:2010hp} for an analysis of flat-space gauge algebras.}.
In particular, it is interesting to draw some more lessons on the HS geometry\footnote{
See \cite{Francia:2002aa,Francia:2002pt} for the free HS geometric equations, and \cite{Manvelyan:2010jf} for a recent development on HS curvatures.}
from the relations between the \emph{minimal} (A)dS couplings and the \emph{non-minimal} ambient-space ones. Moreover, the nature of the non-localities, which appear in the flat-space Lagrangian starting from the quartic order, can be clarified from this point of view: flat-space non-localities might fit within the Vasiliev's system with the aid of the ambient-space formalism. If so, the strategy employed above can give an additional motivation for the flat-space HS gauge theory.

Further interesting applications of our results can be found in massive HS field theories,\footnote{See \cite{Francia:2007ee, Francia:2008ac} for the recent development of the free massive HS theory in the metric-like approach.} of which String Theory is the most important example.
Actually, the interactions of massive HS fields can be investigated with techniques similar to those used above and we refer to \cite{Joung:2012rv} for further details. It is indeed believed by many authors\footnote{See e.g. \cite{Zinoviev:2008jz, Zinoviev:2008ck, Zinoviev:2009hu} for some investigation along these lines.} that the masses of HS fields can play a role similar to that of the cosmological constant of massless HS theories, and the understanding of this relation can give more insights on the very nature of String Theory.\footnote{
See \cite{Bengtsson:1986ys, Heanault:book, Sagnotti:2003qa} for the triplet system which contains the same DoFs as
the massless limit of the first Regge trajectory of String Theory. See \cite{Porrati:2010hm} for the analysis of HS interactions in a constant electromagnetic background within the String Theory framework. See \cite{Polyakov:2009pk,Polyakov:2010qs,Polyakov:2010sk} for the construction of some cubic and quartic flat-space vertices of massless HS fields using vertex operators in String theory, and \cite{Polyakov:2011sm} for its recent extension to AdS.}

Moreover, other applications can be found in the AdS/CFT correspondence, which has been applied to HS theories starting from \cite{Sezgin:2002rt, Klebanov:2002ja}.\footnote{In the $AdS_{3}$ case, there has been considerable recent development after the works \cite{Campoleoni:2010zq, Henneaux:2010xg,Gaberdiel:2010pz}.} We expect that the ambient-space representation of interacting vertices
simplifies the computations of $n$-point functions. Moreover, loop computations might be addressed within this formalism,
shedding some light on the quantum aspects of HS gauge theories.

\clearpage{\pagestyle{empty}\cleardoublepage}

\rhead[\fancyplain{}{\bfseries
On four-point functions and beyond}]{\fancyplain{}{\bfseries\thepage}}
\lhead[\fancyplain{}{\bfseries\thepage}]{\fancyplain{}{\bfseries\rightmark}}
\chapter{On four-point functions and beyond}                

In this Chapter we are going to try and extend as much as possible the analysis of the Noether procedure to the quartic level. The discussion is based on \cite{Taronna:2011kt} and we are going to rewrite it in the ambient space formalism in order to make more clear its role in relation to the problem of finding order by order in the number of fields a consistent Lagrangian in any constant curvature background. We shall also discuss the role of locality arguing about the possibility of relaxing it. As before, the following analysis will be carried out in the TT setting, although mostly at the zero-th order level in $\hat\d$. The latter, as in the cubic case, is the leading part of the corresponding (A)dS couplings and coincides with their flat limit. The full result can be than recovered asking for a consistent lifting of the various zero-th order results and we leave this very interesting problem for future work. Moreover, for that regards the generalization of the following results to higher numbers of external legs we refer for brevity to the Appendix~A of \cite{Taronna:2011kt}.


\section{The Noether equations at quartic order}\label{sec:ternary}


In this section we analyze the quartic Noether procedure equation
\be
\int d^{d+1}X\ \delta\left(\sqrt{\s X^2}-L\right)\ \left[\delta_E^{(1)}\cL^{(3)}\,+\,\delta_E^{(0)}\cL^{(4)}\right]\,\approx\,0\,,\label{Noether quartic}
\ee
in order to recover the general form of a consistent quartic coupling. Let us recall that with $\approx$ we mean equality modulo the free EoMs while each identity is considered as before modulo traces and divergences in order to concentrate on a first instance on the TT portion of the Lagrangian.
As shown in Section~\ref{Noether Procedure}, some formal simplifications arise considering the corresponding Lagrangian coupling generating functions. Hence, as we have anticipated, we shall consider a generating function representation of the quartic coupling given by
\begin{multline}
\cL^{(4)}\,=\,\tfrac{1}{4!}\ C^{(4)}_{\sst a_1a_2a_3a_4}\left(\hat{\d};\partial_{X_i}\cdot\partial_{X_j};U_i\cdot\partial_{X_j},U_i\cdot U_j\right)\\\star_{\sst 1234} \Phi^{a_1}(X_1,U_1)\Phi^{a_2}(X_2,U_2)\Phi^{a_3}(X_3,U_3)\Phi^{a_4}(X_4,U_4)\Big|_{_{X_i=X}^{U_i=0}}\,,\label{quartic genfunc}
\end{multline}
whose labels $a_i$'s are associated to Chan-Paton factors. Henceforth, for convenience, we are going to consider the color ordered convention \eqref{coupling genfunc 3} thus splitting eq.~\eqref{Noether quartic} into independent color-ordered contributions. Restricting the attention to one of these, eq.~\eqref{Noether quartic} takes the form
\be
\partial_{U_i}\cdot\partial_{X_i}\ C^{(4)}_{\sst 1234}\,\approx\,-\,
[\delta_i^{(1)}\,C^{(3)}]_{\sst 1234}\,,\label{Consist2}
\ee
where we recall that the above equality is to be considered modulo traces, divergences and free EoMs while, for brevity, we have dropped the generating functions of the fields. Here, the $\d_i^{(1)}$ operator takes the form
\be
\delta_i^{(1)}\,=\,\delta_i^{(1)}\Phi^{a_j}(X_j,U_j)\,\star_j\,\frac{\d}{\d \Phi^{a_j}(X_j,U_j)}\,,
\ee
with $\delta_i^{(1)}\Phi^{a_j}(X_j,U_j)$ the deformation of the gauge transformations of the field $\Phi^{a_j}(X_j,U_j)$ induced from the cubic level\footnote{We are not distinguishing here between trivial and non-trivial deformations just because the following discussion is independent on this feature.}.
For convenience, we have introduced the functional derivative with respect to a generating function as
\be
\frac{\d}{\d \Phi^{a_i}(X_i,U_i)}\ \Phi^{a_j}(X_j,U_j)\,=\,\d_{a_ia_j}\delta(X_i-X_j)\ e^{(U_i\cdot U_j)_T}\,,\label{functional derivative}
\ee
where we have defined the tangent contraction
\be
(U_i\cdot U_j)_T\,=\,U_i\cdot U_j\,-\,\frac{U_i\cdot X_j\,U_j\cdot X_i}{X_i\cdot X_j}\,,
\ee
by consistency with the tangentiality constraints on the fields.
Therefore, exploiting the cubic Noether equation
\be
\int d^{d+1}X\,\d\left[\delta^{(1)}_i \cL^{(2)}\,+\,\delta^{(0)}_i\cL^{(3)}\right]\,=\,0\,,
\ee
one then ends up with
\be
\delta_i^{(1)}\Phi^{a_j}(X_j,U_j)\star_{\sst j} \square_j\Phi^{a_j}(X_j,U_j)\,=\,-\,\delta^{(0)}_i\cL^{(3)}\,,
\ee
from which it follows\footnote{Here we have used that $\delta_i^{(1)}\Phi^{a_j}(X_j,U_j)$ has the same degree of homogeneity of $\Phi^{a_j}(X_j,U_j)$.}
\be
\delta_i^{(1)}\Phi^{a_j}(X_j,U_j)\,=\,-\,\frac{1}{\square_j\!\!}\ \frac{\d}{\d\Phi^{a_j}(X_j,U_j)}\left(\delta^{\,(0)}_i\cL^{(3)}\right)\,.
\ee
Using this equation one then recovers
\begin{multline}
\delta^{(1)}_i\cL^{(3)}\,=\,-\,\left(\cL^{(3)}\right)\,\frac{\overleftarrow{\d}}{\d\Phi^{a_j}(X_j,U_j)}\
\frac{\star_{\sst j}\!\!}{\square_j\!\!}\ \frac{\overrightarrow{\d}}{\d\Phi^{a_j}(X_j,U_j)}\left(\delta^{\,(0)}_i\cL^{(3)}\right)\\= \,-\,\d_i^{(0)}\left[\frac{1}{2}\,\left(\cL^{(3)}\right)\,\frac{\overleftarrow{\d}}{\d\Phi^{a_j}(X_j,U_j)}\
\frac{\star_{\sst j}\!\!}{\square_j\!\!}\ \frac{\overrightarrow{\d}}{\d\Phi^{a_j}(X_{j},U_{j})}\,\left(\cL^{(3)}\right)\right]\,,
\end{multline}
where the sum runs over all possible $j$'s. The above expression can be also rewritten in terms of the corresponding color-ordered generating functions as
\begin{multline}
\delta^{(1)}_1\cL^{(3)}\,=\sum_{\s}\,\left\{-\,\partial_{U_i}\cdot\partial_{X_i}\, \left[C^{(3)}_{1\s(2)j}(\partial_{X_k},U_k)\,\tilde{\star}_{\sst j}\,C_{j'\s(3)\s(4)}^{(3)}(\partial_{X_k},U_k)\right.\right.\\\left.\left.+\,C^{(3)}_{4\s(1)j}(\partial_{X_k},U_k)\,\tilde{\star}_{\sst j}\,C_{j'\s(2)\s(3)}^{(3)}(\partial_{X_k},U_k)\right]\right\}\\\star_{\sst 1234}\,\text{Tr}\left[E_1(X_2,U_2)\,\Phi_{\s(2)}(X_{\s(2)},U_{\s(2)})\right.\\\left. \times\,\Phi_{\s(3)}(X_{\s(3)},U_{\s(3)})\,\Phi_{\s(4)}(X_{\s(4)},U_{\s(4)})\right]\,,
\end{multline}
where we have defined a new inner-product
\be
\tilde{\star}_{\sst j}\,=\,\left(\Phi(X_j,\overleftarrow{\partial}_{U_j})\ \frac{\overleftarrow{\d}}{\d\Phi(X_i,U_i)}\right)\
\frac{\!\star_{\sst i}\!\!}{\square_{\,i}\!\!}\ \left(\frac{\overrightarrow{\d}}{\d\Phi(X_{i},U_{i})}\ \Phi(X_{j'},\overrightarrow{\partial}_{U_{j'}})\right)\,.\label{star tilde}
\ee
The latter is naturally related to the propagator of the theory and rebuilds the corresponding current exchange being by definition proportional to the inverse Laplacian times projectors into the physical components of the fields\footnote{Recall that our fields are defined modulo traces and divergences and on-shell they become transverse and traceless. Hence compatibility of the Fierz system with the tangentiality constraint implies that the above expression is exactly the inverse of the Laplacian on the domain of tangent fields. Notice from this respect that the presence of two projectors into eq.~\eqref{star tilde} implies that the inverse Laplacian can be completely factorized either on the left or on the right just because the non-vanishing commutator is entirely tangent or transverse as one can see from eq.~\eqref{box completo}.}.

Finally, the Noether equations \eqref{Noether quartic} become the first order differential equations\footnote{The differential equation is to be considered with respect to the variables $U_i$.}
\begin{multline}
\partial_{U_i}\cdot\partial_{X_i}\, \left[C^{(3)}_{12j}(\partial_{X_k},U_k)\,\tilde{\star}_{\sst j}\,C_{j'34}^{(3)}(\partial_{X_k},U_k)\,+\,C^{(3)}_{41j}(\partial_{X_k},U_k)\,\tilde{\star}_{\sst j}\,C_{j'23}^{(3)}(\partial_{X_k},U_k)\right.\\\left.+\ C^{(4)}_{1234}(\partial_{X_k},U_k)\right]\,\approx\,0\,,\label{final diff eq}
\end{multline}
where $\approx$ means, as above, that the equality should hold modulo the free EoMs of the external fields $\square_i\Phi_i+\ldots\approx 0$ for $i=1,2,3,4$. Moreover, for brevity we have left implicit the $\hat\d$ dependence of the various generating functions. The above differential equations \eqref{final diff eq} encode precisely the content of the Noether procedure, that is thus equivalent to search for the most general solution $\tilde{C}_{1234}^{(4)}$ of
\be
\partial_{U_i}\cdot\partial_{X_i}\,\tilde{C}_{1234}^{(4)}(\partial_{X_k},U_k)\,\approx\,0\,,\label{homogeneous}
\ee
or equivalently for the most general solution of the commutator equation
\be
\left[\tilde{C}_{1234}^{(4)}(\partial_{X_k},\partial_{U_k}), U_i\cdot\partial_{X_i}\right]\,\approx\,0\,,
\ee
in the operator notation. Let us mention that the above equations have exactly the same form of the Noether procedure equations at the cubic level apart for the number of external legs.
Then, supposing to have recovered the most general solution to the above equations, the TT part of the quartic color-ordered Lagrangian coupling generating function can be obtained as
\begin{multline}
C^{(4)}_{1234}(\partial_{X_k},U_k)\,=\,\tilde{C}_{1234}^{(4)}(\partial_{X_k},U_k)\\-\,C^{(3)}_{12j}(\partial_{X_k},U_k)\,\tilde{\star}_{\sst j}\,C_{j'34}^{(3)}(\partial_{X_k},U_k)\,-\,C^{(3)}_{41j}(\partial_{X_k},U_k)\,\tilde{\star}_{\sst j}\,C_{j'23}^{(3)}(\partial_{X_k},U_k)\,.\label{quartic coupling}
\end{multline}
So far we have just addressed the problem of describing the general structure of the solution to the Noether procedure equations at the quartic order. More in detail we have solved a non-homogeneous differential equation finding a particular solution
\begin{multline}
C^{(4)\,\text{part.}}_{1234}(\partial_{X_k},U_k)\,=\,-\,C^{(3)}_{12j}(\partial_{X_k},U_k)\,\tilde{\star}_{\sst j}\,C_{j'34}^{(3)}(\partial_{X_k},U_k)\\-\,C^{(3)}_{41j}(\partial_{X_k},U_k)\,\tilde{\star}_{\sst j}\,C_{j'23}^{(3)}(\partial_{X_k},U_k)\,,
\end{multline}
that in our case has also a neat physical interpretation being exactly minus the color-ordered part of the current exchange. The complete solution has been given adding to the particular solution any solution of the corresponding homogeneous differential equation \eqref{homogeneous}
\be
{C}_{1234}^{(4)\,\text{hom}}(\partial_{X_k},U_k)\,=\,\tilde{C}_{1234}^{(4)}(\partial_{X_k},U_k)\,.
\ee
The particular solution so far obtained to the Noether procedure was recognized already in\footnote{See also \cite{Dempster:2012vw} for a derivation of the same particular solution at the quartic order in flat space.} \cite{Sagnotti:2010at} but without including the possible homogeneous solutions. It was than extended in the present form in \cite{Taronna:2011kt}.
We have mentioned above the physical interpretation of the particular solution. However, also the homogeneous solution has a neat interpretation from the physical viewpoint. Indeed, the homogeneous differential equation can be interpreted as enforcing the Ward identities necessary in order to end up with a consistent amplitude involving massless degrees of freedom. Hence, the role of the homogeneous solution is closely related to the possibility of recovering a non-vanishing four-point amplitude as we shall discuss in the following.

Summarizing, the most general form of the solution to \eqref{Noether quartic}, and hence to the Noether procedure at this order, is given in eq.~\eqref{quartic coupling}.
This form manifests the relations between $C^{(4)}_{1234}$, the current exchange part, quantities like $\tilde{C}^{(4)}_{1234}$ and the cubic couplings themselves $C^{(3)}_{123}$ that solve \emph{homogeneous} equations and entail the linearized gauge symmetries (tantamount to Ward identities) of the free system \eqref{(A)dS Fierz}. In this form the Lagrangian coupling can be, in principle, both local and non-local depending both on the structure of the cubic coupling and on the available solutions to the quartic homogeneous equations. In the next sections we shall analyze the key properties that one should exploit in order to put constraints on both the cubic and quartic couplings with the aim of recovering consistent unitary theories. Indeed, as we have anticipated in Section~\ref{Noether Procedure}, without supplementing the Noether procedure with further requirements, the result will not have any constraint apart from being of the general form above and, in general we expect it to be inconsistent\footnote{Notice in this respect that the same is true also for the cubic couplings that although have been found as solutions to the Noether procedure are not generally expected to be consistent at the full non-linear level. We expect only some particular combinations of them to be so and in the following we are going to analyze what can be the consistency constraint that should be enforced also in light of the general structure of the quartic solution to the Noether procedure}. Let us conclude this section mentioning that given the most general form of the solution to the cubic and to the quartic homogeneous equations as above, consistency lies behind the properties of the combination \eqref{quartic coupling} that cannot have an arbitrary structure but can in principle have a non-local nature.


\section{Four-point scattering amplitudes}\label{sec:four-point}


In this Section, starting from the Feynman rules associated to the couplings recovered above, we are going to compute the associated four-point amplitudes clarifying the physical interpretation of the generating function $\tilde{C}^{(4)}_{1234}$ introduced in the previous section.
As we have already anticipated, the propagator of the theory is given by
\be
\cP\,=\,\tilde{\star}\,,
\ee
where $\tilde\star$ has been defined in eq.~\eqref{star tilde}, and is nicely expressed in terms of the inner-product \eqref{star} up to some projectors that are needed by consistency with the tangentiality and TT constraints. The color-ordered Feynman rules for the cubic and quartic interactions associated to the color-ordered generating functions so far considered can be easily recovered from the previous definitions and read
\begin{align}
\cV^{(3)}_{123}\,=\,&+\,C^{(3)}_{123}\left(\hat\d,\partial_{X_i},U_i\right)\,,\\
\cV^{(4)}_{1234}\,=\,&-\,C_{12j}\left(\hat\d,\partial_{X_k},U_k\right)\tilde\star_{\sst j}\ \label{V4} C^{(3)}_{j34}\left(\hat\d,\partial_{X_k},U_k\right)\\&-\,C_{41j}\left(\hat\d,\partial_{X_k},U_k\right)\tilde\star_{\sst j}\ C^{(3)}_{j23}\left(\hat\d,\partial_{X_k},U_k\right)\,+\,\tilde{C}^{(4)}_{1234}\left(\hat\d,\partial_{X_k},U_k\right)\,.\nonumber
\end{align}
One can now compute the HS four-point amplitudes, ending up with
\begin{multline}
\cA_4\,=\,\sum_{\s}\tilde{C}^{(4)}_{1\s(2)\s(3)\s(4)}\left(\hat{\d},\partial_{X_k},U_k\right)\\ \star_{\sst1234}\,\text{Tr}\left[\Phi_{1}(X_1,U_1)\,\Phi_{\s(2)}(X_{\s(2)},U_{\s(2)})\,\Phi_{\s(3)}(X_{\s(3)},U_{\s(3)})\, \right.\\\left.\times\,\Phi_{\s(4)}(X_{\s(4)},U_{\s(4)})\right]\,,\label{A}
\end{multline}
where the sum is over the permutations, the trace is over the Chan-Paton factors and where only the contribution coming from $\tilde{C}^{(4)}_{1234}$ is still present, while the contribution given by the current exchanges is completely canceled by the first contribution to the quartic Feynmann rules of eq.~\eqref{V4}.

In some sense what we have recovered here is the equivalence between the Noether procedure at the Lagrangian level, and the decoupling of unphysical states at the amplitude level, that is intrinsically related to the linearized part of the gauge symmetry. Moreover, this construction clarifies the role of the homogeneous solution $\tilde{C}^{(4)}_{1234}$ that, indeed, coincides on-shell with the four-point amplitude generating function and from which, attaching the proper boundary to bulk propagator in AdS, one would recover the related CFT current correlators. In the following we shall provide the general \emph{tree-level} form for $\tilde{C}^{(4)}_{1234}$ extracting the \emph{Lagrangian} couplings and commenting on the issue of locality, also in relation to the content of Weinberg's theorem of \cite{Weinberg:1964ew}. Let us emphasize that we have split the TT part of the quartic Lagrangian couplings into portions that are non-local, putting on more general grounds their form along lines that are actually in the spirit of \cite{Barnich:1993vg}. From this point of view the TT part of the quartic couplings can be entirely characterized as the \emph{counterterm} canceling the non-vanishing linearized gauge variation of the current exchange amplitude. Hence, we see here the first instance in which the TT part itself of a coupling can be non-local canceling in this case the portion of the current exchange whose gauge variation \emph{cannot} be compensated by local terms. The quartic coupling $C^{(4)}_{1234}$ is indeed explicitly non-local if $\tilde{C}^{(4)}_{1234}$ does not factorize on \emph{all} possible current exchanges that one can construct starting from the cubic vertices, with the \emph{correct} relative coefficients. Let us mention although that the nature of the allowed non-localities is anyway very restricted even without enforcing consistency requirements like Unitarity but just imposing that the observable quantities are well-behaved on-shell. Indeed the only acceptable non-localities, as we have also commented in the previous Chapter, should not have a singular behavior on-shell\footnote{We reiterate here that this is the reason why at the cubic level non-localities can only enter the part proportional to traces and divergences. Let us stress, in order to avoid confusion, that non-localities are not all of the form $\tfrac{1}{0}\,\sim\,\infty$, like for instance the inverse of Mandelstam variables.}. Hence, they can be only of the type that is already allowed within the current exchange amplitudes where non-localities appear as inverses of the Mandelstam-like invariants. This feature is to be confronted with the cubic case in which no Mandelstam variable is available and indeed the TT part of the couplings has an intrinsic local, although higher derivative, nature. One could still require Locality as a consistency constraint at the quartic order but this constraint has proved to be at least too restrictive already at the quadratic and cubic levels, for that regards the \emph{completion} of the TT part. Hence, it would be interesting to better understand the role of this hypothesis in more general cases relaxing it and explicitly checking for more fundamental requirements like unitarity, as we have done for instance at the quadratic and cubic levels.

Having relaxed the standard \emph{locality} hypothesis, we need to find some physical alternative that has to give a rationale for the possibly non-local answer especially whenever \emph{no} local one is available. This is what we call \emph{minimal scheme}. It is defined via a number of constraints on the couplings, and hence also on the cubic coupling function together with the spectrum of the theory, that have to satisfy altogether the following prescriptions:
\begin{itemize}
\item any particle propagating within some exchange gives rise to non-vanishing four-point amplitudes where it plays the role of an external state\footnote{With this constraint we ensure that the S-matrix be non trivial and that all current exchanges be built from states present in the spectrum constraining both the latter and the coupling functions of the theory.},
\item no quartic coupling contains portions that are identically gauge invariant under the linearized gauge variation\footnote{With this requirement we avoid for simplicity local quartic couplings that are proportional to the amplitude itself multiplied with Mandelstam variables in order to get a local object. This requirement is not so strict and can be eliminated in the most general setting although it is suggested by the structure of the boundary CFT in the framework of AdS/CFT correspondence. Indeed, the full conformal algebra and all the correlation functions are entirely specified by the OPE of two operators and by their conformal dimensions and hence by the cubic couplings while no additional freedom comes entirely from higher orders.},
\item In standard constructions, the on-shell gauge invariance of four-point S-matrix elements is generally violated by current exchanges, but these violations can be eliminated by the contributions of contact terms, local quartic couplings. In this fashion, consistent four-point amplitudes result from the local three-point couplings that build the exchanges and from additional local quartic interactions, within the conventional framework of local field theories. With higher-spin interactions, however, this cancellation is in general impossible, as implied for instance by Weinberg's argument \cite{Weinberg:1964ew}. As a result, the only option that appears generally available is to eliminate the offending amplitudes altogether via non-local quartic couplings that cancel precisely the exchanges.
    More in general, given a current exchange $C_{1234}$ built from two cubic couplings with both external spins and propagating particle fixed, its gauge invariant completion has to be chosen local if a solution $\tilde{C}_{1234}$ of \eqref{homogeneous} factorizing on $C_{1234}$ exists.\footnote{This is equivalent to consider the maximal set of gauge invariant amplitudes enforcing a correspondence between the residues and the propagating particles present in the spectrum. More in detail if a residue is present in the amplitude, its form cannot be different from that of the corresponding current exchange. (See e.g. the YM example that we consider in the next section where $\a$ and $\b$ have to be chosen in order to recover a local result).}
\item only those non-vanishing current exchanges that \emph{do not} satisfy the above requirement have to be removed altogether by \emph{non-local} quartic couplings\footnote{Non-localities should never touch the correspondence between residues and exchanges.}.
\end{itemize}
At the end only the current exchanges of the latter type, if any, will not contribute to the amplitudes while any residue present in a given scattering amplitude will be associated with one of the propagating degrees of freedom present in the spectrum. This iterative procedure fixes the non-vanishing entries of the cubic coupling function and in principle also the relative coefficients between different cubic couplings. Moreover, it enforces constraints on the possible spectra playing the same job of locality in the lower-spin cases. Let us stress in this respect that the above requirements are \emph{equivalent} to locality whenever all possible exchanges satisfy the third condition. They depart from it only whenever there is no non-trivial gauge invariant amplitude factorizing on a certain exchange. However in those cases, as we shall see below, the amplitude must involve by consistency at least infinitely many propagating degrees of freedom contributing to the \emph{same} residue making much more difficult to verify unitarity and posing very interesting questions about the consistency of the above requirements in these unusual cases. For instance, whenever infinitely many degrees of freedom contribute to the same residue possible clashes with analyticity can in principle arise\footnote{See \emph{e.g.} \cite{Bekaert:2009ud,Taronna:2010qq,Sagnotti:2010at} for the analysis of the resummation of infinite exchanges contributing to the same residue}.

Other options are related to quartic local couplings that can arise as solutions to the homogeneous equations. This types of couplings are proportional to the amplitude itself but contain a sufficient number of Mandelstam variables in the numerator in order to give rise to local objects. Any coupling of this type can give in principle a consistent local field theory with no cubic coupling but since the tensorial structure is the same of $\tilde{C}^{(4)}_{1234}$, these kind of options are clearly encoded into choices of the relative functions of the Mandelstam variables that weight each different contribution to $\tilde{C}^{(4)}_{1234}$. In the following, we shall restrict our attention to the minimal scheme and pursuing this kind of program we are going to explore in which sense, if any, the usual notion of \emph{locality} at the Lagrangian level may be overcome, leaving though a systematic analysis of those constraints and of their solutions for the future\footnote{It can be useful to stress that for the lower-spin cases the solutions to the minimal scheme are all local.}. Moreover, we want to stress that in principle some of the solutions that comply to the minimal scheme can explicitly clash with many commonly accepted ideas about the structure of the S-matrix for massless particles. However, given the enormous difficulties present on this subject related to various assumptions whose origin cannot be proved in a rigorous sense starting from the usual causality and unitarity hypothesis \cite{Eden:1966}, we make the choice of exploring the minimal setting that can give rise to non-trivial HS interactions trying to understand in which sense their non-triviality makes them different from their lower-spin counterparts.

Furthermore, we leave for the future the important question of clarifying wheatear the minimal scheme here proposed is sufficient to imply global symmetries of higher spin, as it does for their lower-spin counterparts.


\section{The Yang-Mills example}\label{sec:YMfour}


In the previous sections we have obtained the general form of the quartic coupling stressing the role of linearized gauge symmetries related to the free system. In order to make the construction explicit let us apply these techniques to the familiar case of Yang-Mills theory coupled to scalar fields in the adjoint representation, whose cubic couplings have been recovered in the previous Chapter. For simplicity from now on we shall restrict the attention to the zero-th order in $\hat\d$ of the above generating functions, setting to zero any total derivative that otherwise would have acted on the $\d$-function inside the measure. Those solutions by the way can be interpreted as the flat limits of the corresponding (A)dS couplings, as we have observed at the cubic level.

In this framework we need to solve the equations
\be
\partial_{X_i}\cdot\partial_{U_{i}}\ \tilde{C}^{(4)}_{1234}(\partial_{X_j},U_{j})\,\approx\,0\,,\label{quarticS}
\ee
discarding any total derivative contribution while $\approx$ means that the result is to hold modulo the free EoMs for the external states and up to terms proportional to divergences and traces. Actually, as we have remarked above, the physical meaning of \eqref{quarticS}, and more generally of \eqref{gaugeconscond1}, amounts to the decoupling condition for unphysical polarizations (Ward identities) at the level of the amplitudes. The latter has indeed the same form of the linearized gauge symmetry of the free system. We emphasize here the analogy between these constraints and those satisfied by the cubic couplings. The only difference is that the solution of the decoupling condition is no more, in general, a Lagrangian coupling, simply because from the quartic order Lagrangian couplings do not coincide in general with the amplitudes but rather differ from them in the current exchange parts. What happens in the three-point case is actually an accident from this point of view.

Following our strategy, one can start from the current exchange amplitude that can be constructed from the following spin-1 cubic coupling\footnote{We introduce henceforth the notation $\partial_{X_{ij}}\,=\,\partial_{X_i}-\partial_{X_j}$.}
{\allowdisplaybreaks
\begin{multline}
\cG_{123}\,=\,(U_1\cdot U_2+1)\, U_3\cdot\partial_{X_{12}}\,+\,(U_2\cdot U_3 +1 )\,U_1\cdot\partial_{X_{23}}\\+\,(U_3\cdot U_1+1)\, U_2\cdot\partial_{X_{31}}\,,\label{G}
\end{multline}}
\!\!encoding precisely the YM cubic interactions together with the minimal coupling to a scalar.
Restricting the attention to a single color-ordered contribution, one is then left with the color-ordered current exchange
\be
\cA^{(exch.)}_{1234}\,=\,-\,\frac{1}{s}\ \cG_{12a}\ \star_{a}\ \cG_{a34}\,-\,\frac{1}{u}\ \cG_{41a}\ \star_{a}\ \cG_{a23}\,,\label{exch}
\ee
where only the $s$ and $u$ channels contribute while the Mandelstam variables are defined as\footnote{Notice that both sides of the current exchange have the same degree of homogeneity by consistency. Hence, by virtue of eq.~\eqref{box by parts}, one can integrate by parts inside the Mandelstam variables also when considering the subleading contributions in $\hat\d$.}
\be
s\,=\,-\,(\partial_{X_3}\,+\,\partial_{X_4})^2\ ,\quad t\,=\,-\,(\partial_{X_2}\,+\,\partial_{X_4})^2\,,\quad u\,=\,-\,(\partial_{X_2}\,+\,\partial_{X_3})^2\,.
\ee
This current exchange \eqref{exch} is not gauge invariant, and its linearized gauge variation reads
\begin{multline}
\d_4 \cA^{(c.e.)}_{1234}\,=\,\partial_{X_4}\cdot\partial_{U_4}\ \cA^{(c.e.)}_{1234}\,=\,-\,2\,U_2\cdot \partial_{X_4}\,+\,U_1\cdot \partial_{X_4}\\+\,U_3\cdot \partial_{X_4}\,-\,2\,U_1\cdot U_3\,U_2\cdot \partial_{X_4}\,+\,U_1\cdot U_2\,U_3\cdot \partial_{X_4}\,+\,U_2\cdot U_3\,U_1\cdot \partial_{X_4}\,,\label{gaugevar}
\end{multline}
so that the whole point of the Noether procedure, as we have stressed, is to produce a counterterm whose linearized gauge variation cancels this contribution. The totally cyclic counterterm can be worked out relatively easily in this case, it is \emph{local}, and is given by
\begin{multline}
C^{(4)\,\text{YM}}_{\,1234}\,=\,2\,U_1\cdot U_3\,U_2\cdot U_4\,- \,U_1\cdot U_4\,U_2\cdot U_3\,-\,U_1\cdot U_2\,U_3\cdot U_4\\+\,2\,\left(U_1\cdot U_3\,+\,U_2\cdot U_4\right)\,- \,U_1\cdot U_2\,-\,U_2\cdot U_3\,-\,U_3\cdot U_4\,-\,U_4\cdot U_1\,,\label{cV}
\end{multline}
so that it coincides precisely with the corresponding color-ordered contribution to the Yang-Mills quartic coupling generating function that can be deduced from the Yang-Mills Lagrangian coupled to scalar fields. Notice also that one would have ended with the same result starting from the full (A)dS result without dropping the $\hat\d$ contributions.

Returning to our discussion, we would like to reiterate that we have recovered the Yang-Mills quartic Lagrangian generating function $C^{(4)\,\text{YM}}_{1234}$ following the strategy outlined in the previous section and imposing the decoupling condition of unphysical degrees of freedom at the level of the amplitude. One then recovers the full generating function of Yang-Mills four-point amplitudes in color-ordered form, given by
\be
\cG_{1234}(\partial_{X_i},U_i)\,=\,-\,\frac{1}{s}\ \cG_{12a}\ \star_{a}\ \cG_{a34}\,-\,\frac{1}{u}\ \cG_{41a}\ \star_{a}\ \cG_{a23}\,+\,C^{(4)\,\text{YM}}_{1234}\,,\label{G1234}
\ee
that is actually the desired solution for $\tilde{C}^{(4)}_{1234}(\partial_{X_i},U_{i})$. In a similar fashion, any Lagrangian vertex can be recovered as a counterterm that guarantees the gauge invariance property of the corresponding amplitude while, as pointed out in the previous section, the key physical content of the Noether procedure is to produce amplitudes that decouple unphysical degrees of freedom. Moreover, we want to underline, as also stressed in \cite{Barnich:1993vg} and as remarked previously, that from Noether procedure alone there is no general argument forcing to choose local counterterms for $C^{(4)}_{1234}$. It is then interesting at least in principle to analyze the most general quartic coupling, studying also non-local solutions in this simple toy model. In this case, for instance, we can conceive to consider the solutions
\be
\tilde{C}^{(4)}_{1234}\,=\,\l\,\cG_{1234}\,,\label{ampl}
\ee
with $\l$ an overall coefficient that does not affect the defining property of eq.~\eqref{quarticS}. This choice would led to a \emph{non-local} quartic Lagrangian coupling of the form
\be
C^{(4)}_{1234}(\partial_{X_i},U_{i})\,=\,-\,\frac{\l\,-\,1}{s}\ \cG_{12a}\star_{\,a} \cG_{a34}\,-\,\frac{\l\,-\,1}{u}\ \cG_{41a}\star_{\,a} \cG_{a23}\,+\,\l\,\cV^{\,\text{YM}}_{1234}\,,\label{G1234non}
\ee
where we have subtracted the current exchange contribution in eq.~\eqref{exch}. Similarly, one could have also started from a more general current exchange of the form
\be
\cA^{(exch.)}_{1234}\,=\,-\,\frac{\a}{s}\ \cG_{12a}\,\star_{\,a} \cG_{a34}\,-\,\frac{\b}{u}\ \cG_{41a}\,\star_{\,a} \cG_{a23}\,,\label{exch}
\ee
weighted by different constants $\a$ and $\b$, that can be interpreted as parameterizing a violation of the Jacobi identity\footnote{This situation arise whenever one does not start directly with Chan-Paton factors but with arbitrary color factors that do not define a gauge group.}, and yielding to the \emph{non-local} Lagrangian quartic coupling
\be
C^{(4)}_{1234}(\partial_{X_i},U_{i})\,=\,-\,\frac{\l\,-\,\a}{s}\ \cG_{12a}\star_{\,a} \cG_{a34}\,-\,\frac{\l\,-\,\b}{u}\ \cG_{41a}\star_{\,a} \cG_{a23}\,+\,\l\,C^{(4)\,\text{YM}}_{1234}\,.\label{G1234non2}
\ee
The meaning, if any, of these class of solutions, that manifest themselves in this setting creates a sort of \emph{ambiguity} related to the various choices for the parameters $\l$, $\a$ and $\b$, so much so that if one wants to relax the \emph{locality} constraint one clearly needs to replace it with something else. Our observations move from the fact that, whatever the choice for these coefficients, the amplitude that one recovers at the end is always given by eq.~\eqref{ampl}, whose residues have a fixed form matching the coefficients of the current exchange part only if\footnote{Notice that, as required in the third point of the minimal scheme, the dependence of the Mandelstam variables inside the amplitude is the same as in the current exchange regardless of the choice of $\a$, $\b$ and $\l$.} $\l\,=\,\a\,=\,\b$. Hence, we are led to the conclusion that the only choice leading to a physically meaningful setting is exactly $\a\,=\,\b\,=\,\l$, in which the current exchange contributions extracted from \eqref{ampl} can be \emph{entirely} related to the cubic part of the theory via the current exchanges, as the \emph{minimal scheme} requires. The same argument can be given using tree-level unitarity since whenever a finite number of degrees of freedom is to contribute to the residue, arbitrary choices of $\a$, $\b$ and $\l$ would clearly violate it. This is true since the residue would not match the current exchange contribution of eq.~\eqref{exch} that counts the \emph{finite} number of degrees of freedom that ought to be propagating. The situation may well be different if an infinite number of degrees of freedom contributes to the same residue.

A different situation presents itself when the theory possesses cubic couplings leading to current exchanges whose violation of gauge invariance leads to amplitudes that \emph{cannot} factorize on the initial type of exchange. In such cases the quartic coupling becomes \emph{intrinsically} non-local and locality cannot be restored modifying the cubic couplings relative coefficients as above, as we shall see in the next sections\footnote{Notice that in the YM case just presented as well as in all standard low-spin examples, including classical gravity, the minimal scheme implies locality.}. In those cases either the cubic couplings cannot be present in a consistent theory because non-localities create further inconsistencies, or one needs to resort to the minimal scheme or to other similar frameworks. To reiterate, whenever the propagating multiplet contains a finite number of degrees of freedom the above discussion implies the necessity within a covariant approach of Lagrangian locality at the level of the TT part of the action, while the situation has to be still clarified when the multiplet contains infinitely many massless degrees of freedom. Let us anticipate that non-localities of this type, modulo all possible problems that can come together with them, require an infinite number of massless HS fields.

The full amplitude generating function is then obtained summing over all color orderings as
\begin{multline}
\cA(\Phi_1,\Phi_2,\Phi_3,\Phi_4)\,=\,\sum_\s \cG_{1\s(2)\s(3)\s(4)}(\partial_{X_i},U_{i})\\ \star_{\sst 1234}\ \text{Tr}\Big[\Phi_1(X_1,U_1)\,\Phi_{\s(2)}(X_{\s(2)},U_{\s(2)})\\\times\, \Phi_{\s(3)}(X_{\s(3)},U_{\s(3)})\, \Phi_{\s(4)}(X_{\s(4)},U_{\s(4)})\Big]\,,\label{SYangMills}
\end{multline}
where for simplicity in the YM case one can consider truncated matrix valued generating functions of the form
\be
\Phi(X,U)\,=\,\phi(X)\,+\,A(X)\cdot U\,,
\ee
encoding only the scalar wave-function and the polarization tensor of the gauge bosons. The sum is over all permutations of three elements, in order to recover the usual group theoretical factors
\be
f^{abe}f^{cde}\,\sim\,\text{Tr}\left([T^a, T^b][T^c,T^d]\right)\,,
\ee
together with a sum over the non-cyclic permutations of the external legs. Moreover, we have expressed the amplitude in terms of the star-contraction \eqref{star}. It is interesting to observe that $\cG_{1234}$ satisfies some simple relations like
\be
\cG_{1234}\,+\,\cG_{2134}\,+\,\cG_{2314}\,=\,0\,,\qquad\cG_{1234}\,=\,\cG_{4321}\,,\label{ide}
\ee
that, together with the cyclicity in the external legs, leave two independent objects, say for instance
\be
\cG_{1234}\,,\qquad\cG_{1243}\,.
\ee
Analyzing more in detail what we have gained, as for three-point amplitudes, one can look more closely at the various contributions to $\cG_{1234}$, distinguishing them by their order in the symbols $U_i$'s. In this case, for each of the independent terms $\cG_{1234}$ and $\cG_{1243}$, one can extract three different contributions of order $0$, $2$ and $4$ in the symbols, given by
\begin{align}
a_{-1}(s,t,u)\,G_{1234}^{(-1)}(\partial_{X_i})\,=\,-\,&\left(2\,t\,-\,2\,\frac{s\,u}{t}\right)\,G_{1234}^{(-1)}(\partial_{X_i})\\ &=\,\cG_{1234}(\partial_{X_i},\l\,U_i)\Big|_{\l\,=\,0}\,=\,\frac{t\,-\,u}{s} \,+\,\frac{t\,-\,s}{u}\,,\nn
G^{(1)}_{1234}(\partial_{X_i},U_i)\,&=\,\left(\frac{d}{d\l}\right)^{\,2}\ \left[ \cG_{1234}(\partial_{X_i},\l\,U_i)\right]\Big|_{\l\,=\,0}\,,\\
G^{(2)}_{1234}(\partial_{X_i},U_i)\,&=\,\left(\frac{d}{d\l}\right)^{\,4}\ \left[ \cG_{1234}(\partial_{X_i},\l\,U_i)\right]\Big|_{\l\,=\,0}\,.
\end{align}
The first is a function of the Mandelstam variables that is related to the four-scalar amplitude. Here, by convention we have defined $G^{(-1)}_{1234}$ as the factorized contribution with a scalar exchange, encoding in the function
\be
a_{-1}(s,t,u)\,=\,-\,\left(2\,t\,-\,2\,\frac{s\,u}{t}\right)\,,
\ee
the residue of the spin-$1$ exchange. On the other hand, the other two $G_{1234}^{(i)}$'s are related, respectively, to the two scalar -- two gauge boson amplitude and to the four gauge boson amplitude, whose structure has been recovered here enforcing \emph{linearized} gauge invariance.

To summarize, we have recovered the analogs of the three-point $Y$ and $G$ operators. They are:
\be
G^{(1)}_{1234}(\partial_{X_i},U_i)\,,\quad G^{(2)}_{1234}(\partial_{X_i},U_i)\,,
\ee
together with their independent non-cyclic permutations and can be related to the tree-level amplitudes
\begin{align}
\cA(\phi_{\,1},\phi_{\,2},A_3,A_4)\,&=\,\sum_\s \ G^{(1)}_{1\s(2)\s(3)\s(4)}(p_{\,i},\xi_{\,i})\\&\hspace{50pt}\star_{\,1234}\ \text{Tr}\Big[\phi_{\,1}\,\phi_{\,\s(2)}\,A_{\s(3)}\cdot U_{\s(3)}\, A_{\s(4)}\cdot U_{\s(4)}\Big]\nn
\cA(A_1,A_2,A_3,A_4)\,&=\,\sum_\s\ G^{(2)}_{1\s(2)\s(3)\s(4)}(\partial_{X_i},U_{i})\\&\hspace{30pt}\star_{\,1234} \text{Tr}\Big[A_1\cdot U_1\,A_{\s(2)}\cdot U_{\s(2)}\,A_{\s(3)}\cdot U_{\s(3)}\, A_{\s(4)}\cdot U_{\s(4)}\Big]\,.\nonumber
\end{align}
The role of the overall constant factor, that at the cubic level corresponds to the three-scalar coupling, is played here by the basic building block of a four-scalar amplitude given by
\be
G^{(-1)}_{1234}(\partial_{X_i})\,=\,-\,\frac{1}{s}\,-\,\frac{1}{u}\,=\,\frac{t}{s\,u}\,.\label{cG-1}
\ee
The other $G^{(i)}_{1234}$'s are on the contrary color-ordered amplitudes for the processes involving two or four gauge bosons. Explicitly
\begin{multline}
G^{(1)}_{1234}(\partial_{X_i},U_{i})\,=\,-\,\left[\frac{1}{s}\,\left(G^{(0)}_{12a}\,\star_{a}G^{(1)}_{a34}\,+\, G^{(1)}_{12a}\,\star_{a}G^{(0)}_{a34}\,+\,G^{(0)}_{12a}\,G^{(0)}_{a34}\Big|_{U_a\,=\,0}\right)\right.\\\left.
+\,\frac{1}{u}\,\left( G^{(0)}_{41a}\,\star_{a}G^{(1)}_{23a}\,+\, G^{(1)}_{41a}\,\star_{a}G^{(0)}_{23a}\,+\,G^{(0)}_{12a}\,G^{(0)}_{a34}\Big|_{\xi_a\,=\,0}\right)\right]\\ +\,2\,\left(U_1\cdot U_3\,+\,U_2\cdot U_4\right)\,- \,U_1\cdot U_2\,-\,U_2\cdot U_3\,-\,U_3\cdot U_4\,-\,U_4\cdot U_1\,,\label{G01}
\end{multline}
and
\begin{multline}
\cG^{(2)}_{1234}(\partial_{X_i},U_{i})\,=\,-\,\left[\frac{1}{s}\  G^{(1)}_{12a}\,\star_{a}G^{(1)}_{a34}\,+\,\frac{1}{u}\ G^{(1)}_{41a}\,\star_{a}G^{(1)}_{23a}\right]\\\,+\,2\,U_1\cdot U_3\,U_2\cdot U_4\,- \,U_1\cdot U_4\,U_2\cdot U_3\,-\,U_1\cdot U_2\,U_3\cdot U_4\,,\label{G02}
\end{multline}
where $G_{123}^{(0)}$ and $G_{123}^{(1)}$ are the pieces of $\cG_{123}$ in eq.~\eqref{G} of order one and three in the $U_i$'s, respectively.
The $G_{1234}^{(i)}$'s that we have found here can be used to find the most general solution for $\tilde{C}^{(4)}_{1234}$ satisfying \eqref{quarticS} in a theory with scalars and gauge bosons. The corresponding solution reads
\begin{multline}
\tilde{C}^{(4)}_{1234}\,=\,\cK^{(4)}_{1234}\left(\partial_{X_i}\cdot\partial_{X_j},\,G^{(i)}_{1234}\right)\,=\, a_{-1}(s,t,u)\,G^{(-1)}_{1234}\\+\,a_0(s,t,u)\,G^{(0)}_{1234}\,+\,a_1(s,t,u)\,G^{(1)}_{1234}\,+\,a_2(s,t,u)\,G^{(2)}_{1234}\,+\,\ldots\,,
\end{multline}
where the $a_i(s,t,u)$ are functions of the Mandelstam variables that do not introduce higher-order poles, laying the freedom left by Noether procedure in building a consistent theory\footnote{Notice that the structure recovered above is similar the one coming out from ST integrating out the massive modes. Again whenever a finite number of degrees of freedom is present higher-order derivatives would imply the violation of unitarity above a certain scale. Otherwise the situation may well be different whenever an infinite number of degrees of freedom is propagating.}. They encode the residues of the various processes as well as further local quartic couplings that are gauge invariant under the linearized gauge symmetry. In this case, by consistency, $\tilde{C}^{(4)}_{1234}$ does not contain exchanges with spin greater than one, while the $G^{(i)}_{1234}$'s are defined in eqs.~\eqref{cG-1}, \eqref{G01} and \eqref{G02}. Moreover, one can see the reason why we have left a free slot for $G_{1234}^{(0)}$. Indeed, one can consider a further contribution linear in the symbols $U_i$ and defined as\footnote{This additional building block can be interpreted as an amplitude involving 3 scalars and 1 gauge boson.}
\begin{multline}
G^{(0)}_{1234}\,=\,-\,\frac{1}{s}\,\left(U_{1}\cdot \partial_{X_{2a}}\,+\,U_2\cdot \partial_{X_{a1}}\,+\,U_3\cdot \partial_{X_{4\tilde{a}}}\,+\,U_4\cdot \partial_{X_{\tilde{a}3}}\right)\\-\,\frac{1}{u}\,\left(U_{1}\cdot \partial_{X_{b4}}\,+\,U_2\cdot \partial_{X_{3\tilde{b}}}\,+\,U_3\cdot \partial_{X_{\tilde{b}2}}\,+\,U_4\cdot \partial_{X_{1b}}\right)\,,
\end{multline}
where by convention
\begin{align}
\partial_{X_{a}}\,&=\,-\,\partial_{X_1}\,-\,\partial_{X_2}\,,\quad \partial_{X_{\tilde{a}}}\,=\,-\,\partial_{X_a}\,,\\ \partial_{X_b}\,&=\,-\,\partial_{X_1}\,-\,\partial_{X_4}\,,\quad \partial_{X_{\tilde{b}}}\,=\,-\,\partial_{X_b}\,.
\end{align}
Hence, the Yang-Mills plus scalar example is finally recovered with the choice
\be
\tilde{\cK}^{\text{YM}}_{1234}\,=\,-\,\left(2\,t\,-\,\frac{s\,u}{t}\right)\,G^{(-1)}_{1234}\,+\,G^{(1)}_{1234}\,+\,G^{(2)}_{1234}\,.
\ee
Other non-standard examples related to a theory with gauge bosons and scalars can arise whenever one takes into account the corresponding quartic amplitudes that can be extracted from higher powers of the building blocks
\be
\left[G_{1234}^{\,(i)}\right]^{\,n}\,.
\ee
In this case one can recover the local quartic couplings that are linked to the highest-derivative cubic couplings involving two or three gauge bosons that in generating function form read
\be
U_1\cdot \partial_{X_{23}}\,U_2\cdot \partial_{X_{31}}\,+\,\text{cyclic}\,,\qquad U_1\cdot \partial_{X_{23}}\,U_2\cdot \partial_{X_{31}}\,U_3\cdot \partial_{X_{12}}\,.\label{higher}
\ee
The corresponding four-point amplitudes are proportional to the generalized $H$ building blocks
\begin{align}
H^{(1)}_{ij}\,&=\,\frac{{U_j}\cdot\partial_{X_i} \, {U_i}\cdot\partial_{X_j}}{\partial_{X_i}\cdot\partial_{X_j}}\, -\,{U_i}\cdot{U_j}\,,\\
H^{(2)}_{ijk}\,&=\,\frac{U_i\cdot\partial_{X_j}}{\partial_{X_i}\cdot\partial_{X_j}}\,-\, \frac{U_i\cdot\partial_{X_k}}{\partial_{X_i}\cdot\partial_{X_k}}\,,
\end{align}
that are the analog of the $H$ operator \eqref{H} recovered in the cubic case with the difference that at the quartic or higher orders they can be non-local thanks to existence of the Mandelstam variables.
In the following we shall keep always in mind that the above amplitudes arise as particular combination of powers of the building blocks so far considered up to field redefinitions or, viceversa, that all $G_{1234}^{(i)}$ can be rewritten in terms of the $H^{(i)}$ operators. Indeed it is worth noticing that, contrary to the cubic case, starting from the quartic order the number of the above $H$ building blocks is enough in order to solve the full ambient-space Noether procedure in terms of a generic function of the form
\be
\tilde{C}_{1234}^{(4)}\,=\,\cK_{1234}^{(4)}\left(\partial_{X_i}\cdot\partial_{X_j},\,H^{(1)}_{ij},\,H^{(2)}_{ijk}\right)_{i\neq j\neq k}\,.\label{H soluz}
\ee
The key difference with respect to the cubic level, as we have anticipated, is the fact that starting from the quartic order the presence of the Mandelstam variables gives the possibility to define non-singular non-local structures. However, the above $H^{(i)}$ building blocks hide the important links between the quartic homogeneous solutions and the current exchanges. Hence, admitting some redundancy in the description, we shall consider generating functions of these operators together with the previous ones. Finally, it can be interesting to notice that the above solution \eqref{H soluz}, that we have recovered modulo traces and divergences, is actually complete because it is satisfying the Noether equations off-shell while what is left is just the precise relation between the $G^{(i)}$ and the $H^{(i)}$ building blocks.

To conclude this section, we want to emphasize the role of the scattering amplitudes in comparison with the Lagrangian couplings as we have recovered in \eqref{final diff eq}. As we have seen, it is possible to work directly at the amplitude level, where the gauge symmetry is realized linearly, extracting the quartic couplings as counterterms needed in order to guarantee the linearized gauge invariance. We want to emphasize here that, although the decoupling condition for the unphysical polarizations does not fix the relative functional coefficients between $G^{\,(0)}_{1234}$, $G^{\,(1)}_{1234}$ and $G^{\,(2)}_{1234}$, these can be completely fixed requiring the minimal scheme or the stronger locality condition, as we have seen.

In the following we will push forward these observations, generalizing the results to HS gauge fields, with special attention to the nature of the four-point Lagrangian couplings that for gauge bosons can be local, but in this approach result explicitly from subtractions between different non-local objects and can be, in general, non-local as well.


\section{The HS case}


In this section we are going to consider the general case of HS four-point couplings, extending the ideas of the previous section. In order to arrive at a systematic description of HS four-point amplitudes, we proceed as before, finding the solutions to the Noether procedure modulo divergences and traces. At the end, we shall comment on the Lagrangian couplings that arise after subtracting the current exchange portions. In order to achieve this goal we need to exhibit the general solution $\tilde{C}^{(4)}_{1234}$ satisfying
\be
\partial_{X_i}\cdot\partial_{U_{i}}\ \tilde{C}^{(4)}_{1234}(\partial_{X_j},U_{j})\,\approx\,0\,.\label{decoupling}
\ee
Actually, one can construct a general ansatz for a solution to eq.~\eqref{decoupling} starting from the results obtained in the previous section and generalizing what happens in the three-point case where, as we have discussed, the HS couplings are simply given by powers of the gauge-boson ones. Focusing on a \emph{generating function}, it suffices to exponentiate the $G^{(i)}$'s obtained in the previous section, so that a general class of solutions to eq.~\eqref{decoupling} can be given by
\be
\tilde{C}^{(4)}_{1234}(\partial_{X_j},U_{j})\,=\,-\,\frac{1}{su}\ \exp\left[-\,su\,\left(G^{(0)}_{1234}\,+\,G^{(1)}_{1234}\,+\,G^{(2)}_{1234}\right)\right]\,.\label{Opengenfunc}
\ee
Admitting some redundancy as anticipated in the previous section, we could also add to the exponent all other building blocks $H^{(i)}$.
We can call this result with a little abuse of language, of the \emph{open-string}-type since it is \emph{planar} and hence is naturally associated to Chan-Paton factors \cite{Paton:1969je}. Moreover, $\tilde{C}^{(4)}_{1234}$ should be considered modulo arbitrary relative functions of the Mandelstam variables, that are not constrained by eq.~\eqref{decoupling} and play the role of relative weights between the various totally cyclic terms in the expansion of \eqref{Opengenfunc}. Finally, we have considered a fixed ordering $1234$ of the external legs so that only the $G^{(i)}_{1234}$'s enter and contribute to the correct channels reproducing the HS exchanges. The corresponding current exchange amplitude generating function to which \eqref{Opengenfunc} should be confronted with is by the way
\be
\cA^{\text{c.e.}}\,=\,-\frac{1}{s}\ C^{\,(3)}_{12a}\star_{\sst a} C^{\,(3)}_{a34}\,-\,\frac{1}{u}\ C^{\,(3)}_{41a}\star_{\sst a} C^{\,(3)}_{a23}\,.
\ee
The full amplitude, that we can call again of the \emph{open-string}-type, is recovered as usual by
\begin{multline}
\cA(\Phi_1,\Phi_2,\Phi_3,\Phi_4)\,=\,\sum_\s\text{Tr}\Big[\Phi_1(\xi_1)\,\Phi_{\s(2)}(\xi_{\s(2)})\, \Phi_{\s(3)}(\xi_{\s(3)})\, \Phi_{\s(4)}(\xi_{\s(4)})\Big]\\\star_{\,1234} \tilde{C}^{(4)}_{1\s(2)\s(3)\s(4)}(p_{\,i},\xi_{\,i})\,,\label{Shs}
\end{multline}
where now $\Phi_i(\xi_i)$ is an arbitrary matrix valued generating function containing all totally symmetric HS polarization tensors, while the trace is over the color indices. We clearly recover the results of the previous section as soon as we restrict the attention to the linear part in the $G^{(i)}_{1234}$'s. Going ahead, we have chosen a dependence as
\be
su\,G^{(i)}_{1234}(\partial_{X_i},U_{i})\,,
\ee
multiplying with $(su)$ the $G^{(i)}$'s, in order to avoid higher-order poles as soon as one considers HS fields. For instance, the form of the above four-point scattering amplitude of the open-string-type in the case of four spin-$2$ fields becomes here
\begin{multline}
\tilde{C}^{(4)}_{1234}\,\sim\,-\,\frac{1}{su}\ \sum_{\a\,+\,2\b\,+\,4\g\,=\,8}\ a_{\a,\b,\g}(s,t,u)\left[-\,su\,G^{(0)}_{1234}\right]^\a\\\times\,\left[-\,su\,G^{(1)}_{1234}\right]^\b \left[-\,su\,G^{(2)}_{1234}\right]^\g \, ,\label{OFourspin2}
\end{multline}
where the $a_{\a,\b,\g}(s,t,u)$'s are some functions that do not introduce additional poles in the Mandelstam variables. Those are to be fixed, in our minimal scheme, in order to reproduce the current exchange amplitudes, and hence by comparison to the cubic coupling function, whose arbitrariness is in turn constrained by the minimal scheme itself.
The TT part of the quartic coupling generating function can be now extracted exploiting eq.~\eqref{quartic coupling} where in order to be explicit
\be
\tilde{C}^{(4)}_{1234}\,=\,-\,\frac{1}{su}\ \sum_{\a,\,\b,\,\g}\ a_{\a,\b,\g}(s,t,u)\left[-\,su\,G^{(0)}_{1234}\right]^\a\left[-\,su\,G^{(1)}_{1234}\right]^\b \left[-\,su\,G^{(2)}_{1234}\right]^\g \, .\label{OFourspingenfunc}
\ee
This form, as in the spin-$1$ case, encodes in principle also non-minimal choices, which reflect some freedom left by Noether procedure. The latter freedom is related to local quartic couplings that are gauge invariant under the linearized gauge variation and whose tensorial structure is exactly as in the amplitude, but multiplied by a sufficient number of Mandelstam variables that suffice to eliminate all poles. Moreover, although \eqref{OFourspin2} is a generic planar color-ordered expression consistent with gauge invariance that one can write for four spin-$2$ fields it does not exhaust all the possibilities, as it was the case for the cubic couplings. Indeed, we have at our disposal two independent building blocks $G_{1234}$ and $G_{1243}$ and another available option is to combine them together using $G_{1243}$ in place of the color factor. One ends up, in this way, with the following type of \emph{derived} generating function
\be
\tilde{C}^{(4)}(U_{i}\,,U^{\,\prime}_{\,i})\,=\,\left(\sum_\s\ \tilde{C}^{(4)}_{1\s(2)\s(3)\s(4)}(\partial_{X_i},U_{i})\ \tilde{C}^{(4)}_{1\s(2)\s(4)\s(3)}(\partial_{X_i},U^{\,\prime}_{i})\right)\,,\label{Shs3}
\ee
whose contributions of the form
\be
\left[G_{1243}\right]^{\,\a}\,\left[G_{1243}\right]^{\,\b}\,,
\ee
with neither $\a\,=\,0$ nor $\b\,=\,0$, could be called, with a little abuse of language, \emph{closed-string}-like amplitudes\footnote{To be precise, we can call in this way \emph{only} the contributions with $\a\,=\,\b$. The other contributions, that show up starting from spin-$3$, do not satisfy the analog of \emph{level matching} but are not ruled out here by gauge invariance. We cannot exclude at this stage that they are not ruled out by other arguments, but we leave a more detailed analysis of these potentially interesting options for the future.}. Here we have literally replaced the Chan-Paton contribution in eq.~\eqref{Shs} with the generating function $\tilde{C}^{(4)}_{1243}$, again defined in eq.~\eqref{Opengenfunc}, so that for tree-level scattering amplitudes involving totally symmetric fields, one recovers the generating function
\begin{multline}
\cA^{(4)}(U_i)\,=\,\frac{1}{stu}\ \\\times\,\sum_\s e^{-\,s_{1\s(2)}s_{1\s(3)}\,\left(G^{(0)}_{1\s(2)\s(4)\s(3)}(U_{i})\,+\,G^{(1)}_{1\s(2)\s(4)\s(3)}(U_{i}) \,+\,G^{(2)}_{1\s(2)\s(4)\s(3)}(U_{i})\right)} \\ \times e^{-\,s_{1\s(2)}s_{1\s(4)}\,\left(G^{(0)}_{1\s(2)\s(3)\s(4)}(U_{i})\,+\,G^{(1)}_{1\s(2)\s(3)\s(4)}(U_{i}) \,+\,G^{(2)}_{1\s(2)\s(3)\s(4)}(U_{i})\right)}\,.\label{Closedgenfunc}
\end{multline}
Here by definition
\be
s_{ij}\,=\,-\,(\partial_{X_i}\,+\,\partial_{X_j})^2\,,
\ee
the sum is over all permutation of the three elements $\{234\}$ and one has, again, the freedom to multiply each totally cyclic gauge-invariant term in the expansion of \eqref{Closedgenfunc} with arbitrary relative functions $a_i(s,t,u)$ of the Mandelstam variables that give rise to amplitude with single poles at most (see e.g. eq.~\eqref{OFourspingenfunc}). Moreover:
\begin{itemize}
 \item one can in principle constrain these functions relating any current exchange contribution belonging to $\tilde{C}$ to the corresponding contribution obtained from the cubic couplings of the theory via the minimal scheme,
 \item the only \emph{non-local} contributions to the quartic Lagrangian coupling are those related to the exchanged particles that cannot be made gauge invariant with the addition of a local counterterms and that, for this reason, can never belong to $\tilde{C}$.
\end{itemize}
We leave a more detailed analysis of these issues related to \emph{non-local} field theories for the future, trying to understand their eventual \emph{geometric} rationale and in which sense, if any, they can be consistent with unitarity, even though they clash with commonly accepted ideas about the structure of S-matrix poles like factorization.

Before going on with our discussion, it can be of interest to comment more in details on the nature of the couplings that we have obtained for spin-$2$ external particles as a toy model of more general cases, extracting the current exchange part and identifying the cubic couplings involved. It is also important to discuss the difference between the \emph{open-string}-like couplings of eq.~\eqref{Opengenfunc} and the \emph{closed-string}-like ones of eq.~\eqref{Closedgenfunc}. Let us begin considering the coupling in eq.~\eqref{Closedgenfunc} associated to
\be
\cA^{(4)} (U_i)\,=\,\ldots\,+\,\frac{1}{stu}\ \sum_\s \,s^{\,2}_{1\s(2)}s_{1\s(3)}s_{1\s(4)}\,G^{(2)}_{1\s(2)\s(4)\s(3)}\,G^{(2)}_{1\s(2)\s(3)\s(4)}\,+\,\ldots\,,
\ee
and contributing to the four spin-$2$ scattering amplitude. Explicitly this contribution is given by
\begin{multline}
\cA^{(4)}(U_i)\,=\,-\,\sum_\s\left[s\left(\frac{1}{s}\ G^{(1)}_{12a}\star_{\sst a} G^{(1)}_{a34}\,+\,\frac{1}{u}\ G_{41a}^{(1)}\star_{\sst a} G_{a23}^{(1)}\,-\,\cV^{\,\text{YM}}_{1234}\right)\right.\\\left.\times\,\left(\frac{1}{s}\ G_{12a}^{(1)}\star_{\sst a} G_{a34}^{(1)}\,+\,\frac{1}{t}\ G_{13a}^{(1)}\star_{\sst a} G_{a24}^{(1)}\,+\,\cV^{\,\text{YM}}_{1243}\right)\right]\,,\label{Grav}
\end{multline}
so that one recovers, as expected, a \emph{non-planar} amplitude and the various contributions conspire after some algebra to yield
\begin{multline}
\cA_{1234}\,=\,-\frac{1}{s}\ \left(G_{12a}^{(1)}\right)^{\,2}\star_{\sst a} \left(G_{a34}^{(1)}\right)^{2}\,-\frac{1}{t}\ \left(G_{13a}^{(1)}\right)^{\,2}\star_{\sst a} \left(G_{a42}^{(1)}\right)^{2}\\-\frac{1}{u}\ \left(G_{14a}^{(1)}\right)^{\,2}\star_{\sst a} \left(G_{a23}^{(1)}\right)^{2}\,+\,\ldots\,,
\end{multline}
where the ellipses represent local terms and where the current exchange amplitude have been completely reconstructed. Here, one can observe a four-point function involving the minimal coupling of two spin-$2$ fields with a propagating spin-$2$, since the number of derivatives entering the current exchange is precisely $4$. This result resonates with the fact that this particular four-point function \eqref{Grav} is exactly the standard ``four-graviton'' four-point function, written in a form analogous to that obtained in the field theory limit of ST in \cite{Kawai:1985xq}. On the contrary, let us now consider the \emph{open-string}-like amplitude
\be
\cA_{1234}\,=\,-\,\frac{1}{su}\,\left(-\,s\,u\,G^{\,(2)}_{1234}\right)^2\,,
\ee
that can be recovered from eq.~\eqref{OFourspin2}. In this case we see a different structure that is given explicitly by
\be
\cA_{1234}\,=\,-\,s\,u\left(\frac{1}{s}\ G_{12a}^{(1)}\star_{\,a} G_{a34}^{(1)}\,+\,\frac{1}{u}\ G_{41a}^{(1)}\star_{\sst a} G_{a23}^{(1)}\,-\,\cV^{\,\text{YM}}_{1234}\right)^2\,,
\ee
so that, extracting the pole part in order to read off the current exchange contribution, one recovers
\be
\cA_{1234}\,=\,-\,\frac{u}{s}\ \left(G_{12a}^{(1)}\right)^{\,2}\star_{\sst a} \left(G_{a34}^{(1)}\right)^{\,2}\,-\,\frac{s}{u}\ \left(G_{41a}^{(1)}\right)^{\,2}\star_{\sst a} \left(G_{a23}^{(1)}\right)^{\,2}\,+\,\ldots\,.
\ee
Here, as before, the ellipsis represent \emph{local} terms while, using functions $a_i(s,t,u)$ that do not introduce higher-order poles, one can only increase the power of the additional Mandelstam variables in the numerator. The latter are actually necessary in order to guarantee both the right pole structure and the decoupling of transverse \emph{unphysical} polarizations. This translates into the fact that the coupling in which the current exchange factorizes involves this time the exchange of a spin-$3$ excitation and is of the form
\be
C^{(3)}\,\sim\,\left[\,G^{\,(1)}_{ijk}\right]^{\,2}\left[\,G^{\,(0)}_{ijk}\right]\,,
\ee
as one can evince looking at the residue, that is of order six in the momenta. In principle, one could go ahead, considering higher powers of the Mandelstam variables that are associated with HS exchanges building a full overall function $a(s,t,u)$ that does not introduce additional poles. For instance, one possibility could be the following gauge-invariant amplitude
\be
\cA_{1234}\,=\,-\,\frac{1}{s\,u}\,e^{\,-t}\left(\vphantom{\frac{1}{2}}u\ \cG_{12a}\star_{\sst a} \cG_{a34}\,+\,s\ \cG_{41a}\star_{\sst a} \cG_{a23}\,-\,s\,u\,\cV^{\,\text{YM}}_{1234}\right)^2\,+\,\ldots\,,\label{fullampl}
\ee
where the exponential of $t$ accounts for an infinite number of exchanges. Also more complicated examples related to the results in \cite{Bekaert:2009ud,Taronna:2010qq,Sagnotti:2010at} can be available, while the ellipsis stand for terms containing also powers of $G^{(1)}_{1234}$, as in eqs.~\eqref{Opengenfunc} and \eqref{OFourspin2}. In principle, this kind of structure is needed if one wants to construct a consistent quartic amplitude that factorizes into an infinite number of exchanges. However, let us stress that here only spins different than two propagate, even if a cubic coupling with colored spin-two fields does in principle exist and in contrast to the previous case where the exchange of a spin-$2$ field was indeed present. The counterpart of this peculiar aspect turns out to be non-localities at the Lagrangian level, as expected from the result of \cite{Boulanger:2000rq}, since the corresponding spin-2 exchange \emph{do not} admit any local gauge-invariant completion. Moreover, this means that:
\begin{itemize}
\item a massless colored spin-$2$ field can have a charge of spin strictly higher than $2$,
\item a full theory producing such four-point functions has to contain an infinite tower of massless HS fields.
\end{itemize}
The latter can be inferred requiring the minimal scheme or, in more detail, requiring that any propagating HS particle brings non-trivial interactions. Therefore, as soon as a spin-$3$, say, propagates one can look at processes with also spin-$3$ external particles recovering again a propagating particle of higher spin and so on. Other possibilities are then available, since one can in principle consider also powers of $G^{(0)}_{1234}$ recovering exchanges where no minimal coupling is present, but at the price of increasing the exchanged particle minimum spin. Also other options are available and are related to the other $H^{(i)}$ building blocks, recovering in the four spin-$2$ case in flat space abelian self-interactions. For instance, one could build in this way the following amplitude
\be
\cA_{1234}\,=\,-\,\frac{1}{su}\left(-su\,G^{(2)}_{1234}\right)\left[-su\,\left(s\,H^{(1)}_{12}\,H^{(1)}_{34}+u \,H^{(1)}_{41}\,H^{(1)}_{23}\right)\right]\,,
\ee
containing the current exchanges built from the cubic couplings involving two spin-2 fields that are linear in the $G_{123}^{(1)}$ operator\footnote{Notice that this coupling exists only in dimensions strictly higher than 4.}. Let us remark that from this result it follows that the lowest spin propagating without introducing non-localities is now spin-$2$ together with a quartic coupling that is manifestly higher derivative. Indeed, the available spin-$1$ exchange does not admit a local gauge invariant completion, coming back to a situation similar to the one above for the spin-2 exchange. Along similar lines one can also construct the completion of the current exchanges involving the spin two cubic couplings with no $G^{(1)}_{123}$. In this case one ends up with the following expression
\be
\cA_{1234}\,=\,-\,\frac{1}{s}\,\left[s^2\,H^{(1)}_{12}\,H^{(1)}_{34}\right]^2- \frac{1}{u}\,\left[u^2\,H^{(1)}_{41}\,H^{(1)}_{23}\right]^2\,,
\ee
from which this time the lowest spin propagating is spin-0 but again with higher-derivative quartic couplings.

It is important to point out that, from our perspective, potential clashes between Lagrangian non-localities and \emph{tree-level unitarity} need to be analyzed taking into account that whenever an infinite number of exchanges is present on the same residue some subtleties can in principle arise. For instance, it is no more clear in this case how to disentangle all contributing residues outside the radius of convergence of the series of exchanges. Things would have been clearly different if only a finite number of exchanges had contributed to any given pole of the tree-level amplitude. Indeed, in the latter case the minimal scheme implies locality at least for that regards the TT part of the Lagrangian. Certainly, a deeper understanding of non-local field theories is needed in order to clarify such peculiar features that actually may be considered as the counterpart of an infinite number of higher-derivative cubic couplings contributing to the same residue and might well led to an inconsistent answer in a Minkowski background. The latter implications could be appreciated extending the analysis to higher order terms in $\hat{\d}$ and deforming the above results to full (A)dS results or to massive fields where concrete examples of this kind are available. Along the same lines, it can be interesting to study the tensionless limit of ST at the quartic order, from which one can expect to recover similar types of results. More information can also come solving for the most general theory that is consistent with the minimal scheme, and we plan to address these problems in the future.

The planar spin-$2$ example may also clarify the role of the spin-$2$ excitation present in the Vasiliev system. Indeed, the latter can be dressed in principle with Chan-Paton factors making its interpretation debatable. For some time the relation of such spin-$2$ excitation with gravity and/or with the \emph{massive} spin-$2$ excitation present in \emph{open} string theory was somehow unclear, as pointed out in \cite{Francia:2007qt}. Indeed, it was argued that although the interaction of two massive open-string spin-$2$ excitations with a graviton is forbidden by momentum conservation, this is not true in the tensionless limit, whenever one reaches a regime where the massive open string spin-$2$ becomes massless. This observation implies a potential mixing that can be already appreciated from the results of \cite{Taronna:2010qq,Sagnotti:2010at}. In fact, among the limiting cubic couplings of the massive spin-$2$, the lowest derivative one is exactly the same as that of the graviton. Our present discussion may clarify these issues, since at the quartic order two different possibilities show up distinguishing two options. One of these, given in eq.~\eqref{OFourspin2}, is naturally endowed with Chan-Paton factors while the other, eq.~\eqref{Closedgenfunc}, is closely related to gravity. They coexist in the amplitude of four spin-$2$ excitations and hence one is led to deduce that in the massless case the two options for spin-$2$ can give rise indeed to a \emph{Cabibbo-like} mixing between the combination of spin-$2$ fields that interacts as gravity and the singlet component interacting with \emph{open-string}-like four-point amplitudes, as anticipated in \cite{Francia:2007qt}. Obviously, the mixing so far considered is expected to disappear whenever the theory breaks the HS-symmetry. In this case, the colored spin-$2$ field, that brings about non-localities, becomes massive, while a combination of the massless spin-$2$ fields remains massless, playing the role of the graviton. Again, it is tempting to believe that behind the string structure of the interactions there are some field theory properties that have to be understood and that may be intimately related to HS theories.


\section{Weinberg's theorem revisited}\label{sec:Weinberg}


In this section we take a closer look, in light of the previous discussion, at a key no-go theorem on the subject, in order to understand as much as possible its assumptions and in which sense one can go beyond them, clarifying hopefully the meaning of the results proposed so far. Indeed, one of the strongest arguments that has been presented over the years is Weinberg's 1964 Theorem of \cite{Weinberg:1964ew} (see e.g. \cite{Bekaert:2010hw} for a review and also for an interesting discussion of its interpretation). It is an S-matrix argument based on the analysis of a would be flat-space S-matrix element with $N$ external particles with momenta $p_{\,i}$, $i\,=\,1,\ldots,N$ and a massless spin-$s$ particle of momentum $q$ and polarization tensor $\Phi_{\,\m_1\ldots\m_s}(q)$. In the following we shall review this argument explicitly in the case with arbitrary \emph{massless} particles entering the process and restricting the attention to the consistent cubic vertices studied in Chapter~3. The idea is to analyze the case in which the momentum $q$ of one of the particles participating in the scattering process tends to zero, called also \emph{soft limit}. This limit encodes the long distance behavior, if any, of the interactions, which is dominated by the pole part, and it is very interesting since it gives constraints coming from very general and model independent infrared (IR) properties. The dominant pole generates in this limit a resonance, so that one can factorize the amplitude, eliminating any \emph{local} contact interaction and leaving \emph{only} the contribution associated to the current exchange\footnote{We depart here from the original Weinberg proof that has been given in the S-matrix language assuming some commonly accepted ideas about the pole structure of the S-matrix. In this respect, the usual factorization property translates here into locality of the corresponding TT part of the Lagrangian.}. Actually, this is the contribution on which Weinberg concentrated in order to develop his argument, and in the following we shall study precisely the same contribution in our explicit setting recognizing what vertices contribute to long-distances and what vertices give instead a vanishing contribution in the same limit and reinterpreting Weinberg's conclusions. Going to momentum space, the explicit form of the S-matrix amplitude becomes in this limit
\begin{multline}
S(p_{\,1},\Phi_{\,1};\ldots;p_{\,N},\Phi_{\,N};q,\Phi)\,\approx\,\sum_{i=1}^N S(p_{\,1},\Phi_{\,1};\ldots;p_{\,i}+q,\tilde{U}_i;\ldots;P_{\,N},\phi_N) \\\star_{\,i}\frac{\hat{\cP}(\tilde{U}_i,U_1)}{2p_{\,i}\cdot q}\star_{\,1}\left[\exp\Big(G_{123}(U_1,U_2,U_3)\Big)\star_{\,2,3} \Phi_{\,i}(-p_{\,i},U_2)\,\Phi(-q,U_3)\right]\, ,\label{softlimit}
\end{multline}
where $\cP(\tilde{U}_i,U_1)$ is the propagator numerator and, apart from the pole factor, the dependence on $q$ has been completely factorized solely into $G_{123}$. Using for the momenta of the particles participating to the factorized scattering process the parametrization
\be
p_{\,1}\,=\,p_{\,i}\,+\,q\,,\qquad p_{\,2}\,=\,-p_{\,i}\,,\qquad p_{\,3}\,=\,-q\,,
\ee
one then recovers
\begin{multline}
G_{123}\,=\,2\,\left(1\,+\,U_1\cdot U_2\right)\,U_3\cdot p_{\,i}\,+\,\left(1\,+\,U_2\cdot U_3\right)\,U_1\cdot (q\,-\,p_{\,i})\\-\,2\,\left(1\,+\,U_3\cdot U_1\right)\,U_2\cdot q\,,\label{limiting}
\end{multline}
where we have made use of momentum conservation together with the transversality constraint
\be
p_{\,i}\cdot U_i\,=\,0\,.
\ee
First of all, from this form one can immediately conclude that for $s\,>\,3$ the relevant tensor structure contributing at long distances, whenever present, is always given by
\be
\frac{(U_3\cdot p_i)^{s}}{2p_i\cdot q}\,.
\ee
Second, we are now in a position to see whether or not the amplitude that we are recovering in this limit decouples the unphysical degrees of freedom and what are the cubic couplings that contribute. The latter physical requirement, as we have shown in Section~\ref{sec:ternary}, is precisely the content of the Noether procedure from a Lagrangian point of view. Hence, let us perform a linearized gauge transformation for the HS particle $\Phi$ whose momentum $q$ goes to zero. The unphysical polarizations are given by
\be
\d\Phi(-q,U_3)\, =\,-q\cdot U_3\,E(-q,U_3)\,,
\ee
and performing this substitution in \eqref{softlimit} one finally ends up with
\begin{multline}
\d\,S(p_{\,1},\Phi_{\,1};\ldots;p_{\,N},\Phi_{\,N};q,\L)\,\approx\,\sum_{i=1}^N S(p_{\,1},\Phi_{\,1};\ldots;p_{\,i}+q,\tilde{U}_i;\ldots;p_{\,N},U_N) \\\star_{\,i}\ \hat{\cP}(\tilde{U}_i,U_1)\ \star_{\,1}\left[\left(1\,+\,U_1\cdot U_2\right) \exp\Big(G_{123}(U_1,U_2,U_3)\Big)\right.\\\left.\star_{\,2,3} \Phi_{\,i}(-p_{\,i},U_2)\,E(-q,U_3)\right]\,,\label{softlimit2}
\end{multline}
where the offending pole has been canceled by the terms proportional to momentum squared produced by the $\star$-contraction of $q\cdot U_3$ with $G_{123}$. We can recognize here the most dangerous contribution\footnote{It is important to stress that in the limit $q\ra 0$ we recover only the order zero contribution in $q$ while, by consistency, \emph{all} contributions have to cancel identically. This underlines the \emph{no-go} character of this argument from which one can only extract information about possible obstructions.} in the limit $q\ra 0$, that is given by
\begin{multline}
\d\,S(q=0)\,\sim\,\sum_i\tilde{S}_i(U_i)\,\star_{\,i}\,\sum_{\a_i,\b_i}\left\{\Big(2\,U_3\cdot p_{\,i}\Big)^{\a_i}\right.\\\left.\left[1\,+\,U_i\cdot U_2\,\Big(2\,U_i\cdot U_2\,U_3\cdot p_{\,i}\Big)^{\b_i}\right]\,\star_{\,2,3} \Phi_{\,i}(-p_{\,i},U_2)\,E(-q,U_3)\vphantom{\left[1\,+\,U_i\cdot U_2\,\Big(2\,U_i\cdot U_2\,U_3\cdot p_{\,i}\Big)^\b\right]}\right\}\,,\label{softlimit3}
\end{multline}
where $\a_i\,\geq\, -1$ and $\b_i\,\geq\, -1$ are some integers\footnote{For $\a_i\,=\,-1$ or $\b_i\,=\,-1$ we simply define the corresponding contribution to be zero.}, we have called $\tilde{S}(U_i)$ the leftover part of the S-matrix together with the propagator numerator, and where the sum over $\a_i$ and $\b_i$ runs over all admissible values that are associated to consistent HS cubic couplings that can be generated from \eqref{softlimit2}. This contribution is \emph{dangerous} since it does not tend to zero when $q\ra 0$, and hence must vanish identically. Restricting the attention to the case in which all external fields but one are scalars and only one HS field $\Phi(-q,U)$ is present only the scalar exchange contributes to \eqref{softlimit3} and there is only one possible value for $\b$, $\b\,=\,-1$. Thus, in order to set to zero \eqref{softlimit3}, one recovers a non-trivial constraint given by
\be
\sum_i g^i\,p^{\,i}_{\,\m_1}\ldots p^{\,i}_{\,\m_{s-1}}\,=\,0\,,\label{WeinbergConstr}
\ee
where the $g^i$'s are the corresponding coupling constants. As pointed out by Weinberg, this equation does not admit non-trivial solutions unless in general $s=1$, and eq.~\eqref{WeinbergConstr} reduces to charge conservation
\be
\sum_i g^i\,=\,0\,,
\ee
or $s=2$, so that eq.~\eqref{WeinbergConstr} reduces to $g^{\,i}\,=\,\kappa$ for any $i$ since, by momentum conservation
\be
\sum_i p^{\,i}_{\,\m}\,\equiv\,0\,.
\ee
We then arrive at a potential inconsistency for HS interactions, since the argument explained so far forces
\be
g^{\,i}\,=\,0
\ee
for spin grater than $2$. Actually, considering a more general HS theory and referring again to \eqref{softlimit3}, we see that as soon as an insertion of a scalar field is present in the amplitude there can be similar obstructions. This happens since one can reiterate this argument, concentrating on the factorized amplitude in which a scalar field is exchanged and in which $\Phi_{\,i}$ in eq.~\eqref{softlimit3} is one of the scalar fields participating to the process. This conclusion has actually a deeper meaning, since it forbids the possibility of having an $s-0-0$ coupling whenever $s$ is greater than $2$ within the framework of \emph{local field theories}. This can be understood observing that as soon as such cubic couplings are present one generates automatically dangerous contributions to some current exchange amplitude. However this conclusion is true unless, by some mechanism, these dangerous exchanges are eliminated whenever they give rise to this kind of problems. Hence, the only possible way out is related to the fact that we have considered in the $q\ra 0$ limit only the current exchange contribution, so that one is led to a clash with perturbative locality on the Lagrangian side or with commonly accepted S-matrix properties like factorization on the S-matrix side. These anyway are possibly stronger statements than the fundamental unitarity and causality properties, and a closer look to them is potentially interesting in view of a better understanding of the tensionless limit of ST.

Let us now turn to see the implications of Weinberg's argument in more general examples. Concentrating on eq.~\eqref{softlimit3}, let us consider an external particle with arbitrary spin $s_i$. The factorization of the amplitude can give rise to problems in the soft limit only if a sufficient number of $U_2$ is contained in \eqref{softlimit3}, otherwise this offending contribution vanishes identically. Hence, we conclude that a dangerous term of the form \eqref{softlimit3} can be generated only if
\be
\b_i\,\geq\,s_i\,-\,1\,.\label{bound}
\ee
In order to analyze the most general case let us restrict the attention to a factorized process in which, referring to \eqref{softlimit3}, $\Phi_{\,i}$ is a spin-$s_{\,i}$ particle. If $s_i\,>\,s$ non-vanishing contributions cannot be generated and so, without loss of generality, we can concentrate on the cases in which $s_i\,\leq\,s$. In this case we recover a non-vanishing contribution to \eqref{softlimit3} whenever the bound \eqref{bound} is satisfied, but we also see that as soon as
\be
-1\,\leq\,\b_i\,<\,s_i\,-\,1\,,\label{bound2}
\ee
no contribution can be generated, so that no inconsistency follows by this argument. Moreover, since for $\b_i\,\geq\,s_i$ one gets simply zero in \eqref{softlimit3}, there is only one dangerous contribution given by $\b_i\,=\,s_i -1$ that is associated to an exchanged particle with spin
\be
s_{\,\text{exchanged}}\,=\,s_i\,.
\ee
We clearly recover the simplest case of before when $s_i\,=\,0$, since in this case there is no solution for $\b_i$, and as soon as $s_i\,\geq\,1$, one begins to recover non-trivial solutions to \eqref{bound2}.

Summarizing, one can convince oneself that the only possibly dangerous contributions come in this limit whenever one considers a current exchange built from a coupling of the form $s_i-s_i-s$ with $s$ derivatives when $s\,\geq s_i$ and the exchanged particle with spin $s_i$. As concluded by Weinberg, this argument poses strict restrictions on the long distance behavior of HS fields, that hence cannot interact at zero frequency. In particular all long-range couplings given by the \emph{minimal} ones, can be ruled out in a \emph{local field theory} while other multipolar couplings are not yet forbidden. The former actually entail exactly the leading contribution related to long distance physics on which Weinberg concentrated in \cite{Weinberg:1964ew}. Moreover, since for $s\,<\,s_i$ the number of derivatives for these couplings is given by
\be
2s_i\,-\,s\,>\,s\,,
\ee
we have explicitly shown that the content of Weinberg's argument together with the classification of all consistent cubic couplings completely forbids the minimal coupling for HS particles within the framework of \emph{local field theories}. It is now interesting to compare this result with the scattering amplitudes constructed in the previous sections. As we have remarked a possible solution to the problem can arise resorting to non-local quartic couplings whose job is to cancel the dangerous exchanges contributing in principle to the amplitudes\footnote{It is interesting to comment that as for what concerns the long-distance behavior of HS interactions Weinberg conclusions are still valid also if non-localities are introduced and one can easily check that the general solution to the Noether procedure that we have exhibited satisfies this property.} but without setting to zero all non-abelian cubic couplings. This circumstances of course would possibly violate causality or unitarity (see e.g. \cite{Adams:2006sv}) but at present we are not able to come up with a definite conclusion on this issue and we believe that more effort is needed in order to clarify the situation.

For instance the conclusion drawn so far are still valid in (A)dS if one restricts the attention to the (A)dS couplings whose leading terms are the problematic flat interactions above. This can be inferred because, assuming locality as we have done in this section, the eventual lower derivative tails of the interactions cannot cancel the dangerous pieces that we have obtained here. It is then interesting to see if the presence of an infinite multiplet can make the difference since in those cases, in particular in (A)dS, one cannot restrict the attention to a single propagating particle within the factorized exchange. We plan to address this as well as other related issues in the future.

Let us end this section with a simple observation about the consequences of our results. If we concentrate on String Theory we see that all these minimal couplings are indeed generated in the tension-less limit starting from the simplest one that concerns two scalars \cite{Taronna:2010qq,Sagnotti:2010at}. Hence, we see here a very severe obstruction if we insist to use the framework of \emph{local} field theories or the usual factorization properties at the level of the S-matrix in order to describe a would be \emph{tensionless string}. Similar considerations apply to the leading contribution of FV vertices, as observed in \cite{Boulanger:2008tg}. Hence, one can argue that if a background independent underlying theory exists it has to include non-local couplings or, possibly, non-local degrees of freedom, which motivate a closer look at unitarity and its general implications.

\clearpage{\pagestyle{empty}\cleardoublepage}
\chapter*{Conclusions}
\rhead[\fancyplain{}{\bfseries
CONCLUSIONS}]{\fancyplain{}{\bfseries\thepage}}
\lhead[\fancyplain{}{\bfseries\thepage}]{\fancyplain{}{\bfseries
CONCLUSIONS}}
\addcontentsline{toc}{chapter}{Conclusions}


In this Thesis we have studied the Noether procedure reconsidering it in the ambient space formalism.

We have described in detail the solutions to the cubic-interaction problem for massless HS fields in a constant-curvature background leaving aside for brevity the extension of the results to the cases of massive and partially-massless interactions which has been studied in \cite{Joung:2012rv,Joung:2012fv}.
This has been achieved through a dimensional reduction of a $(d+1)$-dimensional massless theory with a delta function insertion in the action.
For simplicity, the entire construction has been carried out focusing on the TT part of the Lagrangian. We expect that the completion of such vertices can be performed adding divergences and traces of the fields together with possible auxiliary fields, proceeding along the lines of what was done for the flat space vertices in \cite{Sagnotti:2010at}.

Our studies were mainly motivated by ST, whose very consistency rests on the presence of infinitely many HS fields. Conversely, string interactions may provide useful information on the systematics of the consistent HS couplings. In \cite{Taronna:2010qq,Sagnotti:2010at},
cubic vertices of totally-symmetric tensors belonging to the first Regge trajectory of the open bosonic string were investigated.
Those vertices are encoded in the generating function
\ba
\label{StringVer}\nonumber
	&& \tfrac1{\sqrt{G_{N}}}\,
	\cK_{\sst A_1 A_2 A_3}\\&&=
	i\,\frac{g_o}{\alpha{\sst'}}\,{\text{Tr}}\left[T_{\sst A_1}\,T_{\sst A_2}\,	
	T_{\sst A_3}\right]\,
	\exp\left[i\sqrt{2\alpha'}\,(y_1+y_2+y_3)+z_1+z_2+z_3 \right]\nn
	&&+\,i\,\frac{g_o}{\alpha{\sst'}}\,
	{\text{Tr}}\left[T_{\sst A_2}\,T_{\sst A_1}\,T_{\sst A_3}\right]\,
	\exp\left[-i\sqrt{2\alpha'}\,(y_1+y_2+y_3)+z_1+z_2+z_3 \right]\,,\nonumber
\ea
where $G_{N}$ denotes Newton's constant, $g_o$ the open string coupling constant and $\alpha'$ the inverse string tension related to the  masses of the string states as
\be
\label{spec st}
	M^2\,\varphi^{\sst (s)}=\frac{s-1}{\alpha'}\,\varphi^{\sst (s)}\,.\nonumber
\ee
Remarkably, the Taylor coefficients of the exponential function and the string spectrum combine nicely to reproduce consistent massive and massless vertices. In this respect, it would be interesting to understand how the exponential function above fits in with other ST properties and what its AdS counterpart may be from the point of view of the Noether procedure.
In particular, we have some reasons to believe that the choice of the exponential is crucial for the global symmetries as well as for the planar dualities of the theory. Let us mention however that in AdS an exponential coupling of the form
\be
	e^{i\,\sqrt{2\alpha'}\,(\tilde Y_{1}+\tilde Y_{2}+\tilde Y_{3})+Z_{1}+Z_{2}+Z_{3}}\,,\nonumber
\ee
where the $\tilde Y_{i}$'s are here any total-derivative deformations of the $Y_{i}$'s\,, is incompatible with any spectrum containing a massless spin 1 field together with massive fields, reflecting the difficulties encountered in quantizing ST on (A)dS backgrounds \cite{Tseytlin:2002gz,Bonelli:2003zu,Sagnotti:2003qa}. From this perspective it is conceivable that a better understanding of the global symmetries of ST and of their implementation at the interacting level may shed some light on this issue.
We have then extended the formalism to higher orders, reversing the usual perspective of focusing on four-point Lagrangian couplings. In this respect we have recovered directly a class of gauge invariant $4$-point functions involving \emph{massless} HS fields, as well as low-spin fields, from the \emph{linearized} gauge invariance of the free system, relating them in a relatively simple way to powers of the standard $4$-point functions in a theory with a scalar and a gauge boson. This generalizes the construction of \cite{Sagnotti:2010at}, making it possible to define similar color-ordered generating functions in the general case. Those include, as a particular example, the simpler cubic ones from which all cubic couplings originate. One is then able to extract, subtracting the current exchange parts, four-point and in general $n$-point couplings. These contain as a special case the familiar low-spin examples, together with an infinite set of local couplings, but manifest in general a \emph{non-local} nature as soon as one considers more exotic cases, as for instance a colored spin-$2$, or more generally HS fields. The non-local nature so far observed has an interesting and peculiar structure of the form pointed out in the Appendix of \cite{Sagnotti:2010at}. However, the meaning of non-localities is here to restrict the spins propagating within the amplitude to those whose violation of gauge invariance can be compensated by \emph{local} Lagrangian couplings\footnote{We are referring here to the fourth point of the minimal scheme so that the amplitude cannot factorize on the current exchanges that would require non-local quartic couplings.}. This fact entails the key obstruction that has been recognized long ago by Weinberg in \cite{Weinberg:1964ew}, as well as other inconsistencies at the level of Jacobi identity and so on \cite{Bekaert:2010hp}, that disappear as soon higher-derivative and explicitly non-local couplings are considered, as already noticed in \cite{Barnich:1993vg}. Of course a non-local solution to the problem, even if explicit, cannot be satisfactory without a full understanding of its implications and in particular of the status of the minimal scheme proposed here. In this respect the only thing we can say is that it is conceivable that potential clashes with the standard form of \emph{tree-level} unitarity, that can come together with the \emph{non-localities} allowed by the minimal scheme, do not materialize if an infinite number of degrees of freedom is present. Even considering the case in which the latter option does not hold, it can be interesting to extend the above analysis to constant curvature backgrounds, and we leave this for the near future. Nonetheless, let us stress that any residue of the set of amplitudes so far recovered can be related to lower-point couplings via exchange amplitudes if the minimal scheme is enforced in place of the stronger locality constraint. In this respect it can be interesting to ask what plays the role of locality in constant curvature backgrounds and whether the solution will still contain similar non-localities even if controlled by some expansion parameter, thinking to push forward our analysis in order to understand more clearly the possible need of resorting to this kind of picture. Finally, a deeper understanding of the peculiar features involved by HS interactions, that seem to imply a clash with commonly accepted ideas about the pole-structure of the S-matrix, can be hopefully related to the difficulties that are encountered in the definition of an S-matrix, and we believe that they deserve a better understanding motivated at least by their appearance within ST in its tensionless limit or within the Vasiliev system in its flat limit \cite{Boulanger:2008tg}. In this respect our aim is to investigate further these questions trying to gain some indication also from the AdS/CFT correspondence, considered from our perspective as a slightly different incarnation of the Noether procedure for theories defined in AdS.

Although the subject of non-local field theories is still a completely unexplored arena, the aforementioned properties of the amplitudes may open the way to a deeper understanding of Field Theory. In this respect, ST appears to contain the seeds for interesting generalizations, and hides, in our opinion, some key field theory properties that have surfaced in this Thesis. To wit, the remarkable construction of Closed String Field Theory in \cite{Zwiebach:1992ie} is very general in its starting point, but the mechanical model definition of the interactions hides somehow their non-local nature that has long been felt to be related to a broken phase of the HS symmetry. The mechanical model may hide somehow the non-local features that we have presented here by linking them to the string tension\footnote{See e.g. \cite{Sagnotti:1987tw,Pradisi:1988xd,Horava:1989vt,Horava:1989ga,Bianchi:1990yu,Bianchi:1990tb,Bianchi:1991eu,Sagnotti:1992qw,Dudas:2000bn,Angelantonj:2002ct,Sugimoto:1999tx,Antoniadis:1999xk,Angelantonj:1999jh,Aldazabal:1999jr,Angelantonj:1999ms,Dudas:2000nv,Pradisi:2001yv} where other examples in which the mechanical model appears to provide an incomplete description are discussed.}. A similar situation may arise in the Vasiliev system so that it is tempting to imagine that the intrinsic non-local form of the couplings of a colored spin-$2$ exhibited here may shed some light on the \emph{non-local} nature of the Vasiliev system itself, that seems to be obscured by the presence of the cosmological constant $\L$, whose role is similar to that of the string tension in ST and provides an expansion of perturbatively local terms in which non-local operators like $\frac{1}{\square}$ could split in terms of $\L$.

Other questions then arise in order to attain a meaningful quantization of systems of this kind, that at any rate can be naturally formulated in terms of the Batalin-Vilkoviski formalism \cite{Batalin:1981jr} or in terms of a usual loop expansion. Those can be recovered from the Feynman rules here considered or, alternatively, from recursion relations techniques \cite{Witten:2003nn,Cachazo:2004kj,Britto:2004ap,Britto:2005fq,Benincasa:2007xk,Benincasa:2011kn,Benincasa:2011pg}. Other issues regard the freedom in building a theory of massless HS that we have recursively related to the freedom of choosing a consistent cubic coupling function within what we have called minimal scheme. We leave this as well as other questions, like the extension of the quartic results to constant curvature backgrounds, for the near future.

\clearpage{\pagestyle{empty}\cleardoublepage}

\renewcommand{\chaptermark}[1]{\markright{\thechapter \ #1}{}}
\renewcommand{\theequation}{\thechapter.\arabic{equation}} \csname
@addtoreset\endcsname{equation}{section}
\lhead[\fancyplain{}{\bfseries\thepage}]{\fancyplain{}{\bfseries\rightmark}}

\appendix                               

\rhead[\fancyplain{}{\bfseries \:Appendix A}]
{\fancyplain{}{\bfseries\thepage}}
\chapter{Useful identities}\label{sec: identities}
This appendix contains some identities and mathematical tools used in our construction of the cubic vertices.
Basic commutation relations among the operators $\eqref{Y and Z}$ are
\ba
\label{commrelations}
\big[\,Y_i\,,\,U_j\!\cdot\partial_{X_j}\,\big] \eq
\delta_{ij} \
\partial_{X_{i}}\cdot\partial_{X_{i+1}}\,,\\
\big[\,Z_i\,,\,U_{i+1}\!\cdot\partial_{X_{i+1}}\,\big]
\eq \partial_{X}\cdot\partial_{U_{i-1}}-Y_{i-1}\,\,, \nonumber\\
 \big[\,Z_i\,,\,U_{i-1}\cdot\partial_{X_{i-1}}\big]
\eq Y_{i+1}\,,\nn
\big[\,X_i\!\cdot\partial_{U_i}\,,\,F(Y,Z)\,\big] \eq -Z_{i+1}\,\partial_{Y_{i-1}}\,F(Y,Z)\,,\nn
 \big[\,X_i\!\cdot\partial_{X_i}\,,\,F(Y,Z)\,\big] \eq -Y_{i-1}\,\partial_{Y_{i-1}}\,F(Y,Z)\,,\nn
 \big[\, F(Y,Z)\,,\,U_i\!\cdot\partial_{U_i}\,\big] \eq \left(Y_i\,\partial_{Y_i}+Z_{i+1}\,\partial_{Z_{i+1}}+Z_{i-1}\,\partial_{Z_{i-1}}\right) F(Y,Z)\,.\nonumber
\ea
Here $i, j$ are defined modulo 3: $(i,j)\cong (i+3,j+3)$.
Another identity used throughout all the Thesis concerns the commutator between an arbitrary function $f(A)$ of a linear operator $A$ and an other linear operator $B$\,:
\be
\label{fBcomm}
[\,f(A)\,,\,B\,]=\sum_{n=1}^\infty\, \frac{1}{n!}\, ({\text{ad}}_A)^{n}B\, f^{(n)}(A)\,,
\ee
where ${\text{ad}}_A\,B=[\,A\,,\,B\,]$ and $f^{(n)}(A)$ denotes the $n$-th derivative of $f$ with respect to $A$. In order to prove the latter formula, we represent $f(A)$ as a Fourier integral
so that the commutator appearing in \eqref{fBcomm} can be written as
\be
\label{fBcomm2}
[\,f(A)\,,\,B\,]=\int^\infty_{-\infty} dt\  [\,e^{itA}\,,\,B\,]\,f(t)\,.
\ee
Using the well-known identity
\be
e^{itA}\, B\, e^{-itA}=\sum_{n=0}^\infty\, \frac{(it)^n}{n!}\, ({\text{ad}}_A)^{n}\,B\,,
\ee
eq. \eqref{fBcomm2} becomes
\begin{multline}
[\,f(A)\,,\,B\,]=\sum_{n=1}^\infty\, \frac{1}{n!}\, ({\text{ad}}_A)^{n}\,B\,\int^\infty_{-\infty} dt\, (it)^n\, e^{itA}\,f(t)
\\=\sum_{n=1}^\infty\, \frac{1}{n!}\, ({\text{ad}}_A)^{n}\,B\, f^{(n)}(A)\,.
\end{multline}
Since our vertices are arbitrary functions of commuting operators, formula \eqref{fBcomm} applies independently to each of them.

\clearpage{\pagestyle{empty}\cleardoublepage}

\chapter{Proof at the $\delta^{\sst (2)}$ level}             
\rhead[\fancyplain{}{\bfseries \:Appendix B}]
{\fancyplain{}{\bfseries\thepage}}
\label{sec:C=0}

In this section, we prove that the total-derivative part $\cC$ in \eqref{gv end} does not impose additional conditions on the constants $\tilde\alpha_{i}$ and $\tilde\beta_{i}$\,. At the level of $\delta^{\sst (1)}$, $\cC$ does not vanish  with (\ref{B1} - \ref{B2}), but is simplified to
\ba
	&& \cC= ({\tilde\alpha_{1}}^{2}-1)\,\partial_{X}\!\cdot\partial_{U_{1}}
	+2(\tilde\alpha_{2}+1)\,\partial_{X}\!\cdot\partial_{U_{2}}
	-2(\tilde\alpha_{3}-1)\,\partial_{X}\!\cdot\partial_{U_{3}}
	\nn
	&&
    -\,(\tilde\alpha_{1}-1)(\tilde\beta_{3}-\tfrac12)\,\partial_{X}\!\cdot\partial_{U_{2}}\,\partial_{U_{3}}\!\!\cdot\partial_{U_{1}}
	-(\tilde\alpha_{1}+1)(\tilde\beta_{2}+\tfrac12)\,\partial_{X}\!\cdot\partial_{U_{3}}\,\partial_{U_{1}}\!\!\cdot\partial_{U_{2}}
	\nn
	&&
	+\,\big[\,
	2(\tilde\alpha_{1}\,\tilde\beta_{1}-1)\,\partial_{X}\!\cdot\partial_{U_{1}}
	+\tfrac32(\tilde\alpha_{2}+1)\,\partial_{X}\!\cdot\partial_{U_{2}}
	-\tfrac32(\tilde\alpha_{3}-1)\,\partial_{X}\!\cdot\partial_{U_{3}}\,\big]\,\partial_{U_{2}}\!\!\cdot\partial_{U_{3}}
	\nn
	&&
	+\,\big[\,
	({\tilde\beta_{1}}^{2}-1)\,\partial_{X}\!\cdot\partial_{U_{1}}\,\partial_{U_{2}}\!\!\cdot\partial_{U_{3}}
	-(\tilde\beta_{3}-\tfrac12)(\tilde\beta_{1}+\tilde\beta_{2})
	\,\partial_{X}\!\cdot\partial_{U_{2}}\,\partial_{U_{3}}\!\!\cdot\partial_{U_{1}} \nn
	&&\qquad
	-\,(\tilde\beta_{2}+\tfrac12)(\tilde\beta_{3}+\tilde\beta_{1})
	\,\partial_{X}\!\cdot\partial_{U_{3}}\,\partial_{U_{1}}\!\!\cdot\partial_{U_{2}}\,\big]\,
	\partial_{U_{2}}\!\!\cdot\partial_{U_{3}}\,.
\ea
We integrate by parts in order to replace \mt{\delta^{\sst (1)}\,\partial_{X}\!\cdot\partial_{U_{i}}}
with \mt{-\delta^{\sst(2)}\,X_{i}\!\cdot\partial_{U_{i}}/L^{2}}\,,
and then $\delta^{\sst (1)}\,\cC$ can be rewritten as $-\delta^{\sst (2)}\,\cD/L^{2}$
with $\cD$ some other differential operator.
We now push $\cD$ to the right hand side of $e^{L\,\cV}$ as
\be
	\int\ \delta^{\sst (2)}\ k\ \cD\ e^{L\,\cV}\ E_{1}\,\Phi_{2}\,\Phi_{3}\,\big|
	= \int\ \delta^{\sst (2)}\ k\ e^{L\,\cV}\ L\ \cE\ E_{1}\,\Phi_{2}\,\Phi_{3}\,\big|\,,
\ee
getting the following operator acting on the fields:
{\allowdisplaybreaks\ba
	&& \cE= \Big\{-\big[\,2(\tilde\alpha_{1}\,\tilde\beta_{1}-1)(\tilde\alpha_{2}-1)
	-(\tilde\alpha_{1}+1)(\tilde\alpha_{2}+1)(\tilde\beta_{2}-1)\nn
    &&\hspace{200pt}+2(\tilde\alpha_{2}+1)(\tilde\beta_{3}+\tilde\beta_{1})\,\big]\,
	\partial_{U_{1}}\!\!\cdot\partial_{U_{2}} \nn
	&&\hspace{25pt}
	-\,\big[\,2(\tilde\alpha_{1}\,\tilde\beta_{1}-1)(\tilde\alpha_{3}+1)-(\tilde\alpha_{3}-1)(\tilde\alpha_{1}-1)(\tilde\beta_{3}+1)\nn
	&&\hspace{200pt}+2(\tilde\alpha_{3}-1)(\tilde\beta_{1}+\tilde\beta_{2})\,\big]\,\partial_{U_{3}}\!\!\cdot\partial_{U_{1}}\nn
	&&\hspace{25pt}
	-\,\big[\,-(\tilde\alpha_{2}-1)({\tilde\beta_{1}}^{2}-1)\\
    &&\hspace{100pt}+(\tilde\alpha_{2}+1)(\tilde\beta_{2}-1)(\tilde\beta_{3}+\tilde\beta_{1})\,\big]\,
	\partial_{U_{1}}\!\!\cdot\partial_{U_{2}}\,\partial_{U_{2}}\!\!\cdot\partial_{U_{3}} \nn
	&& \hspace{25pt}
	-\,\big[\,-(\tilde\alpha_{3}+1)({\tilde\beta_{1}}^{2}-1)\nn
    &&\hspace{100pt}+(\tilde\alpha_{3}-1)(\tilde\beta_{3}+1)(\tilde\beta_{1}+\tilde\beta_{2})\,\big]\,
	\partial_{U_{3}}\!\!\cdot\partial_{U_{1}}\,\partial_{U_{2}}\!\!\cdot\partial_{U_{3}} \nn
	&&\hspace{25pt}
	+\,(\tilde\beta_{2}+\tilde\beta_{3})\,\big[\,
	\tilde\alpha_{1}(\tilde\beta_{2}
    +\tilde\beta_{3})+\tilde\beta_{2}-\tilde\beta_{3}+2\,\big]\,
	\partial_{U_{1}}\!\!\cdot\partial_{U_{2}}\,\partial_{U_{3}}\!\!\cdot\partial_{U_{1}} \Big\}
	\,\partial_{U_{2}}\!\!\cdot\partial_{U_{3}}\,.\nonumber
\ea}\!\!
None of these contributions can be compensated,
so that each coefficient in the above formula should vanish separately.
Using the general solutions \eqref{sol cnst} of (\ref{B1} - \ref{B3}), one can
verify that this is indeed the case.

\clearpage{\pagestyle{empty}\cleardoublepage}

\chapter{Radial reduction of the 3$-$3$-$2 vertex}             
\rhead[\fancyplain{}{\bfseries \:Appendix C}]
{\fancyplain{}{\bfseries\thepage}}
\label{sec:reduc}

In this Appendix we present more details of the reduction of the 3$-$3$-$2 vertex \eqref{ex 332} to the (A)dS-intrinsic expression \eqref{ex 332 dS}.
Expanding the operator in eq.~\eqref{ex 332}
gives altogether six terms:
\ba
	&&
	\big[ \partial_{U_{2}}\!\!\cdot\partial_{U_{3}}\,\partial_{U_{1}}\!\!\cdot\partial_{X_{2}}
	-\partial_{U_{1}}\!\!\cdot\partial_{U_{3}}\,\partial_{U_{2}}\!\!\cdot\partial_{X_{1}}\,
	+ \tfrac12\,\partial_{U_{1}}\!\!\cdot\partial_{U_{2}}
	\,\partial_{U_{3}}\!\!\cdot\partial_{X_{12}}\big]^{2}\nn &&\hspace{230pt}\times
	 \,\partial_{U_{1}}\!\!\cdot\partial_{X_{2}}\,\partial_{U_{2}}\!\!\cdot\partial_{X_{1}} \nn
	&& =\,\partial_{U_{2}}\!\!\cdot\partial_{U_{3}}\,
	(\partial_{U_{1}}\!\!\cdot\partial_{X_{2}})^{3}\,\partial_{U_{2}}\!\!\cdot\partial_{X_{1}}\,
	 + (1\leftrightarrow2)\nn
	&&\quad-\,2\,
	\partial_{U_{1}}\!\!\cdot\partial_{U_{3}}\,\partial_{U_{2}}\!\!\cdot\partial_{U_{3}}\,
	(\partial_{U_{1}}\!\!\cdot\partial_{X_{2}})^{2}\,(\partial_{U_{2}}\!\!\cdot\partial_{X_{1}})^{2}
	\nn
	&& \quad+\,
	\partial_{U_{1}}\!\!\cdot\partial_{U_{2}}\,
	\partial_{U_{2}}\!\!\cdot\partial_{U_{3}}\,
	(\partial_{U_{1}}\!\!\cdot\partial_{X_{2}})^{2}\,\partial_{X_{1}}\!\!\cdot\partial_{U_{2}}\,
	\partial_{U_{3}}\!\!\cdot\partial_{X_{12}} + (1\leftrightarrow2)\nn
	&& \quad+\,\tfrac14\,
	 (\partial_{U_{1}}\!\!\cdot\partial_{U_{2}})^{2}\,
	\partial_{U_{1}}\!\!\cdot\partial_{X_{2}}\,
	\partial_{U_{2}}\!\!\cdot\partial_{X_{1}}\,(\partial_{U_{3}}\!\!\cdot\partial_{X_{12}})^{2}\,,
	\label{332 exp}
\ea
but taking into account the symmetries under $1\leftrightarrow2$\,, one is left with four terms. One of such terms is \eqref{1 term}, and we have sketched how to get the corresponding (A)dS intrinsic expression \eqref{332 1 ds}. Applying the same techniques explained there, one can deal with the other three terms in the same manner.

We present the (A)dS intrinsic expression for each term. The third term in the expansion \eqref{332 exp} gives
\ba
	&& \partial_{U_{1}}\!\!\cdot\partial_{U_{3}}\,\partial_{U_{2}}\!\!\cdot\partial_{U_{3}}\,
	(\partial_{X_{2}}\!\!\cdot\partial_{U_{1}})^{2}\,
	(\partial_{X_{1}}\!\!\cdot\partial_{U_{2}})^{2} \nn
	 && \simeq\,
	 \partial_{u_{1}}\!\!\cdot\partial_{u_{3}}\,\partial_{u_{2}}\!\!\cdot\partial_{u_{3}}\,
	 (\partial_{u_{1}}\!\!\cdot D_{\sst 2})^{2} (\partial_{u_{2}}\!\!\cdot D_{\sst 1})^{2}\nn
    &&\quad+\tfrac1{L^{2}}\,  \partial_{u_{1}}\!\!\cdot\partial_{u_{2}}\,
	 \partial_{u_{1}}\!\!\cdot\partial_{u_{3}}\,\partial_{u_{2}}\!\!\cdot\partial_{u_{3}}\,
	  \partial_{u_{1}}\!\!\cdot D_{\sst 2}\,\partial_{u_{2}}\!\!\cdot D_{\sst 1}
	 \nn
	 &&\quad\
	 -\,\tfrac1{L^{2}}\,\partial_{u_{1}}\!\!\cdot\partial_{u_{2}}\,
	 (\partial_{u_{1}}\!\!\cdot\partial_{u_{3}})^{2}\,(\partial_{u_{2}}\!\!\cdot D_{\sst 1})^{2}
	 -\tfrac1{L^{2}}\,\partial_{u_{1}}\!\!\cdot\partial_{u_{2}}\,
	 (\partial_{u_{2}}\!\!\cdot\partial_{u_{3}})^{2}\,(\partial_{u_{1}}\!\!\cdot D_{\sst 2})^{2}
	 \nn
	 &&\quad\
	+\,\tfrac{d+4}{L^{4}}\,(\partial_{u_{1}}\!\!\cdot\partial_{u_{2}})^{2}\,
	 \partial_{u_{1}}\!\!\cdot\partial_{u_{3}}\,\partial_{u_{2}}\!\!\cdot\partial_{u_{3}}\,,\qquad
\ea
where $\simeq$ means equivalence of two operators under the condition \eqref{hom cond}. The fourth term gives
\ba	
	&& \partial_{U_{1}}\!\!\cdot\partial_{U_{2}}\,\partial_{U_{2}}\!\!\cdot\partial_{U_{3}}\,
	(\partial_{U_{1}}\!\!\cdot\partial_{X_{2}})^{2}\,
	\partial_{U_{2}}\!\!\cdot\partial_{X_{1}}\,\partial_{U_{3}}\!\!\cdot\partial_{X_{12}} \nn
	&&\simeq\,
	\partial_{u_{1}}\!\!\cdot\partial_{u_{2}}\,\partial_{u_{2}}\!\!\cdot\partial_{u_{3}}\,
	(\partial_{u_{1}}\!\!\cdot D_{\sst 2})^{2}\,\partial_{u_{2}}\!\!\cdot D_{\sst 1}\,
	\partial_{u_{3}}\!\!\cdot D_{\sst 12}\nn
	&&\quad+\tfrac1{L^{2}}\,
	\partial_{u_{1}}\!\!\cdot\partial_{u_{2}}\,(\partial_{u_{2}}\!\!\cdot\partial_{u_{3}})^{2}
	(\partial_{u_{1}}\!\!\cdot D_{\sst 2})^{2} \nn
	&&\quad	-\tfrac1{L^{2}}\,(\partial_{u_{1}}\!\!\cdot\partial_{u_{2}})^{2}\,
	\partial_{u_{1}}\!\!\cdot\partial_{u_{3}}\,
	\partial_{u_{2}}\!\!\cdot D_{\sst 1}\,\partial_{u_{3}}\!\!\cdot D_{\sst 12}\nn
    &&\quad+\tfrac2{L^{2}}\,(\partial_{u_{1}}\!\!\cdot\partial_{u_{2}})^{2}\,
	\partial_{u_{2}}\!\!\cdot\partial_{u_{3}}\,
	\partial_{u_{1}}\!\!\cdot D_{\sst 2}\,\partial_{u_{3}}\!\!\cdot D_{\sst 12} \nn
	&&\quad\
	 +\tfrac3{L^{2}}\,\partial_{u_{1}}\!\!\cdot\partial_{u_{2}}\,
	 \partial_{u_{1}}\!\!\cdot\partial_{u_{3}}\,\partial_{u_{2}}\!\!\cdot\partial_{u_{3}}\,
	\,\partial_{u_{1}}\!\!\cdot D_{\sst 2}\,\partial_{u_{2}}\!\!\cdot D_{\sst 1}\nn
    &&\quad-\tfrac{d+1}{L^{4}}\,(\partial_{u_{1}}\!\!\cdot\partial_{u_{2}})^{2}\,
	 \partial_{u_{1}}\!\!\cdot\partial_{u_{3}}\,\partial_{u_{2}}\!\!\cdot\partial_{u_{3}}\,,
\ea
and the fifth term can be obtained interchanging 1 and 2 in the above. The last term gives
\ba
	&& (\partial_{U_{1}}\!\!\cdot\partial_{U_{2}})^{2}\,
	\partial_{U_{1}}\!\!\cdot\partial_{X_{2}}\,
	\partial_{U_{2}}\!\!\cdot\partial_{X_{1}}\,(\partial_{U_{3}}\!\!\cdot\partial_{X_{12}})^{2} \nn
	&&\simeq\,
	\partial_{u_{1}}\!\!\cdot D_{\sst 2}\,(\partial_{u_{2}}\!\!\cdot D_{\sst 12})^{2}
	+\tfrac2{L^{2}}\,(\partial_{u_{1}}\!\!\cdot\partial_{u_{2}})^{3}\,
	(\partial_{u_{3}}\!\!\cdot D_{\sst 12})^{2}\nn
	&&\quad\
	-\tfrac{5}{L^{2}}\,\partial_{u_{1}}\!\!\cdot\partial_{u_{3}}\,(\partial_{u_{1}}\!\!\cdot\partial_{u_{2}})^{2}\,
	\partial_{u_{2}}\!\!\cdot D_{\sst 1}\,\partial_{u_{3}}\!\!\cdot D_{\sst 2}\nn
	&&\quad-\tfrac{5}{L^{2}}\,\partial_{u_{2}}\!\!\cdot\partial_{u_{3}}\,(\partial_{u_{1}}\!\!\cdot\partial_{u_{2}})^{2}\,
	\partial_{u_{1}}\!\!\cdot D_{\sst 2}\,\partial_{u_{3}}\!\!\cdot D_{\sst 1} \nn
	&&\quad\
	-\tfrac8{L^{2}}\,\partial_{u_{1}}\!\!\cdot\partial_{u_{2}}\,
	 \partial_{u_{1}}\!\!\cdot\partial_{u_{3}}\,\partial_{u_{2}}\!\!\cdot\partial_{u_{3}}\,
	\partial_{u_{1}}\!\!\cdot D_{\sst 2}\,\partial_{u_{\sst 2}}\!\!\cdot D_{\sst 1}\nn
	&&\quad+2\,\tfrac{d-9}{L^{4}}\,(\partial_{u_{1}}\!\!\cdot\partial_{u_{2}})^{2}\,
	 \partial_{u_{1}}\!\!\cdot\partial_{u_{3}}\,\partial_{u_{2}}\!\!\cdot\partial_{u_{3}}\,,\qquad
\ea
and collecting all these terms finally gives \eqref{ex 332 dS}\,.

\clearpage{\pagestyle{empty}\cleardoublepage}

\chapter{Completion of the TT part}             
\rhead[\fancyplain{}{\bfseries \:Appendix D}]
{\fancyplain{}{\bfseries\thepage}}
\label{Completion}


In this appendix we consider the completion of the flat cubic vertices in the Fronsdal formulation. In the following we shall start from the TT part of the flat vertices and we consider the same Noether procedure equation
\be
\left[\tilde{C}^{(3)}(\partial_{X_i},\partial_{U_i})\,,U_i\cdot\partial_{X_i}\right]\,\approx\,0\,,\label{decoupling2}
\ee
where now the symbol $\approx$ means that the above equation is satisfied on-shell modulo the full Fronsdal EoM's. The procedure is tedious but straightforward and rests on finding the needed counterterms proportional to traces and divergences that compensate the traces and divergences coming from the original TT result along similar lines as those used in Chapter~2 in order to find the free Lagrangians.

Restricting the attention without loss of generality to the completion of
\be
\tilde{C}^{(3)}\,=\,\exp\left(\ell\,\cV\right)\,,
\ee
with
\be
\cV\,=\,Y_1+Y_2+Y_3+G\,,
\ee
we can exploit the general setting of Chapter~2 considering the following completion of the EoMs
\be
\square\,\phi(X\,,U)\,\approx\,U\cdot\partial_X\,\cD(X\,,U)\,,\label{onshelleq}
\ee
Hence, evaluating the linearized gauge variation of the vertex on-shell in the sense of eq.~\eqref{onshelleq} the following commutators show up
\ba
&&
\big[\,\cV\,,\,U_{\sst 1}\!\cdot\partial_{X_1}\big]\,
\Phi_{\sst 1}\,\Phi_{\sst 2}\,E_{\sst 3}
\approx
\left(\cH_{23}\ U_{\sst 3}\!\cdot\partial_{X_3}-
U_{\sst 2}\!\cdot\partial_{X_2}\,\cH_{32}\right)
\Phi_{\sst 1}\,\Phi_{\sst 2}\,E_{\sst 3}\,,\nn
&&
\big[\,\cV\,,\,U_{\sst 2}\!\cdot\partial_{X_2}\big]\,
\Phi_{\sst 1}\,\Phi_{\sst 2}\,E_{\sst 3}
\approx
\left(\cH_{13}\ U_{\sst 1}\!\cdot\partial_{X_1}-
U_{\sst 3}\!\cdot\partial_{X_3}\,\cH_{31}\right)
\Phi_{\sst 1}\,\Phi_{\sst 2}\,E_{\sst 3}\,,\nn
&&
\big[\,\cV\,,\,U_{\sst 3}\!\cdot\partial_{X_3}\big]\,
\Phi_{\sst 1}\,\Phi_{\sst 2}\,E_{\sst 3}
\approx
\left(U_{\sst 2}\!\cdot\partial_{X_2}\ \cH_{12}-
U_{\sst 1}\!\cdot\partial_{X_1}\,\cH_{21}\right)
\Phi_{\sst 1}\,\Phi_{\sst 2}\,E_{\sst 3}\,,\nn
\ea
where
\be
	\cH_{ij}=(\partial_{U_{i}}\!\!\cdot\partial_{U_{j}}+1) \,\cD_{j}
	+\partial_{U_{i}}\!\!\cdot\partial_{X_{j}}\,\cA_{j}\,.
\ee
This set of commutators
encodes a recursive structure from which one can reconstruct the full \emph{off-shell} completion of the cubic vertex. The end result can be  expressed in the following compact form:
\be
{C}^{\sst (3)\,\text{Fronsdal}}=
e^{\ell\, \cV}\,
\Big[1+\ell^{2}\ {\cH}_{12}\,{\cH}_{13}
+\ell^{3}\,:{\cH}_{21}\,{\cH}_{32}\,{\cH}_{13}:\,
+\ \text{(cyclic\ perm.)} \Big]\,,
\label{genoffshell}
\ee
where $:\ :$ enforces an ordering in which the generalized de Donder operators
are to act directly on the fields and hence are to be put to the right:
\be
:\,{\cD}_{1}\ \cZ_{2}:\ =\cZ_{2}\ {\cD}_{1}\,.
\ee
In the Fronsdal setting one can also add Fermions and following \cite{Sagnotti:2010at} one ends up with the full complete result
\be
\begin{split}
{C}^{\sst (3)\,\text{Fronsdal}}\,&=\,\left(1\,+\,\slashed{\partial}^{\,23}_{U_1}\,+\,\slashed{\partial}^{\,31}_{U_2} \,+\,\slashed{\partial}^{\,12}_{U_3}\right) \exp\Big(\ell \cV\Big)\\&\times\,\left[1\,+\,\frac{\a^{\,\prime}\!\!}{2}\ \hat{\cH}_{\,12}\,\hat{\cH}_{\,13} +\,\left(\frac{\a^{\,\prime}\!\!}{2}\,\right)^{\frac{3}{2}}:\hat{\cH}_{21}\,\hat{\cH}_{\,32}\,\hat{\cH}_{\,13}:\,+\ \text{cyclic}\right]\\
&+\,\exp\Big(\ell \cV\Big)\ \left\{\slashed{\partial}_{U_1}^{\,12}\ \left[\sqrt{\frac{\a^{\,\prime}\!\!}{2}}\ \hat{\cH}_{\,23}\,+\,\frac{\a^{\,\prime}\!\!}{2}\ \hat{\cH}_{\,32}\,\hat{\cH}_{\,13}\right]\right.\\&\left.\,\qquad\quad\quad \qquad-\,\slashed{\partial}_{U_2}^{\,12}\ \left[\sqrt{\frac{\a^{\,\prime}\!\!}{2}}\ \hat{\cH}_{\,13}\,-\,\frac{\a^{\,\prime}\!\!}{2}\ \hat{\cH}_{\,31}\,\hat{\cH}_{\,23}\right]\,+\,\text{cyclic}\right\}\ ,\label{genoffshell}
\end{split}
\ee
where we have considered also fermionic labels that give zero contribution in the purely bosonic case, so that for instance $\slashed{\partial}^{ij}$ contracts the fermionic indices between the field $\Psi_{\,i}$ and $\Psi_{\,j}$ while the $1$ simply contracts the two fermionic indices together, whenever they are present.
The (A)dS off-shell completion is expected to work on very similar grounds. In this case one gets
the following schematic form of the gauge variation
\begin{multline}
	\delta\,S^{\sst (3)}=
	\int d^{d+1}X\ \sum_{n=1}^{\infty}\,
	\delta^{\sst (n)}\Big(\sqrt{X^{2}}-L\Big)\ \times\\
	\times\,
	\partial_{X_{i}}\!\!\cdot\partial_{U_{i}}\,\big( \cdots \big)\,
	E(X_{\sst1}, U_{\sst1})\,\Phi (X_{\sst 2}, U_{\sst 2})\,\Phi (X_{\sst 3}, U_{\sst 3})
	\Big|_{\overset{X_{1}=X_{2}=x_{3}=X}{\sst U_{1}=U_{2}=U_{3}=0}}\,,
\end{multline}
that is to be compensated adding further divergence and trace terms
at the $\delta^{\sst (1)}$-level.
This procedure is expected to work order by order, so leading eventually to
the off-shell form of the (A)dS cubic action.

\addcontentsline{toc}{chapter}{Bibliography}

\bibliographystyle{utphys}
\bibliography{ref}


\end{document}